\documentclass[12pt]{article}

\usepackage{algorithm}
\usepackage{algpseudocode}
\usepackage{booktabs}
\usepackage{amsmath} 
\usepackage{braket} 
\usepackage{rotating}
\usepackage{newtxtext,newtxmath}

\usepackage{graphicx}

\usepackage[letterpaper,margin=1in]{geometry}

\renewenvironment{abstract}
	{\quotation}
	{\endquotation}

\date{}
\newcommand\R[1]{\textcolor{black}{#1}}

\makeatletter
\renewcommand{\fnum@figure}{\textbf{Figure \thefigure}}
\renewcommand{\fnum@table}{\textbf{Table \thetable}}
\makeatother

\usepackage{scicite}

\usepackage{url}

\def\scititle{Quantum-Informed Machine Learning for Predicting Spatiotemporal Chaos with Practical Quantum Advantage}
\title{\bfseries \boldmath \scititle}

\author{
	Maida Wang$^{1\dagger}$,
	Xiao Xue$^{1\dagger}$,
    Mingyang Gao$^{1}$,
	Peter V. Coveney$^{1,2,3\ast}$\and
	\small$^{1}$Centre for Computational Science, University College London, London, UK\and
	\small$^{2}$Informatics Institute, University of Amsterdam, The Netherlands\and
    \small$^{3}$Centre for Advanced Research Computing, University College London, London, UK\and
	\small$^\dagger$These authors contributed equally to this work.\and
    \small$^\ast$Corresponding author. Email: p.v.coveney@ucl.ac.uk
}
\begin{document} 
\maketitle
\begin{abstract}\bfseries \boldmath
\noindent{\large Short Title:}

QIML for Predicting Chaos with Practical Quantum Advantage
\end{abstract}
\begin{abstract}\bfseries \boldmath
\noindent{\large Teaser:}

Quantum circuits efficiently learn invariant measures of chaotic systems to stabilize long-term predictions.
\end{abstract}

\begin{abstract}\bfseries \boldmath
\noindent{\large Abstract}

We introduce a quantum-informed machine learning (QIML) framework for modelling the long-term behaviour of high-dimensional chaotic systems. QIML combines a one-time, offline-trained quantum generative model with a classical autoregressive predictor for spatiotemporal field generation. The quantum model learns a quantum prior (Q-Prior) that guides the representation of small-scale interactions and improves the modelling of fine-scale dynamics. We evaluate QIML on the Kuramoto–Sivashinsky equation, two-dimensional Kolmogorov flow, and the three-dimensional turbulent channel flow used as a realistic inflow condition. Across these systems, QIML improves predictive distribution accuracy by up to 17.25\% and full-spectrum fidelity by up to 29.36\% relative to classical baselines. For turbulent channel inflow, the Q-Prior is trained on a superconducting quantum processor and proves essential: without it, predictions become unstable, whereas QIML produces physically consistent long-term forecasts that outperform leading PDE solvers. Beyond accuracy, QIML offers a memory advantage by compressing multi-megabyte datasets into a kilobyte-scale Q-Prior, enabling scalable integration of quantum resources into scientific modelling.

\end{abstract}

\maketitle

\section{INTRODUCTION}
Modelling high-dimensional dynamical systems remains one of the most persistent challenges in computational science. Partial differential equations (PDEs) provide the mathematical backbone for describing a wide range of nonlinear, spatiotemporal processes across scientific and engineering domains~\cite{biferale2004multifractal,galaktionov2012stability,long2018pde}. However, high-dimensional systems are notoriously sensitive to initial conditions and the floating-point numbers used to compute them~\cite{coveney2024sharkovskii,klower2023periodic,boghosian2019new,coveney2025molecular}, making it highly challenging to extract stable, predictive models from data. Modern machine learning techniques often struggle in this regime: while they may fit short-term trajectories, they fail to learn the invariant statistical properties that govern long-term system behaviour. These challenges are compounded in high-dimensional settings, where data are highly nonlinear and contain complex multi-scale spatio-temporal correlations. 

Machine learning has seen transformative success in domains such as large language models~\cite{chang2024survey,thirunavukarasu2023large}, computer vision~\cite{zhang2024vision,zhou2022learning}, and weather forecasting~\cite{price2025probabilistic,bi2023accurate,lam2023learning,cavaiola2024hybrid}, and it is increasingly being adopted in scientific disciplines under the umbrella of scientific machine learning~\cite{bae2022scientific}. In fluid mechanics, in particular, ML has been used to model complex flow phenomena, including wall modelling~\cite{yang2019predictive,xue2024physics}, subgrid-scale turbulence~\cite{maulik2019sub,pal2020deep}, and direct flow field generation~\cite{fukami2019synthetic,yousif2022physics}. Physics-informed neural networks~\cite{raissi2019physics,cheng2025machine} attempt to inject domain knowledge into the learning process, yet even these models struggle with the long-term stability and generalization issues that high-dimensional dynamical systems demand. To address this, generative models such as Generative Adversarial Networks (GAN)~\cite{goodfellow2020generative} and operator-learning architectures like DeepONet~\cite{lu2021learning} and Fourier Neural Operators (FNO)~\cite{li2020fourier} have been proposed. While neural operators offer discretization invariance and strong representational power for PDE-based systems, they still suffer from error accumulation and prediction divergence over long horizons, particularly in turbulent and other chaotic regimes~\cite{vanchurin2021toward,carleo2019machine}. Recent work, such as Dyslim~\cite{schiff2024dyslim}, enhances stability by leveraging invariant statistical measures. However, these methods depend on estimating such measures from trajectory samples, which can be computationally intensive and inaccurate in all forms of chaotic systems, especially in high-dimensional cases. These limitations have prompted exploration into alternative computational paradigms. Quantum machine learning (QML) has emerged as a possible candidate due to its ability to represent and manipulate high-dimensional probability distributions in Hilbert space~\cite{cerezo2022challenges}. Quantum circuits can exploit entanglement and interference to express rich, non-local statistical dependencies using fewer parameters than their promising counterparts, which makes them well-suited for capturing invariant measures in high-dimensional dynamical systems, where long-range correlations and multimodal distributions frequently arise~\cite{gao2022enhancing}.
QML and quantum inspired machine learning have already demonstrated potential in fields such as quantum chemistry~\cite{kandala2017hardware, o2016scalable}, combinatorial optimisation~\cite{stokes2020quantum,sanyal2022neuro}, and generative modelling~\cite{ghazi2025quantum,benedetti2019parameterized}. However, the field is constrained on two fronts: fully quantum approaches are limited by noisy intermediate-scale quantum (NISQ) hardware noise and scalability~\cite{Preskill_2018}, while quantum-inspired algorithms, being classical simulations, cannot natively leverage crucial quantum effects like entanglement to efficiently represent the complex, non-local correlations found in such systems. These challenges limit the standalone utility of QML in scientific applications today. Instead, hybrid quantum–classical models provide a promising compromise, where quantum submodules work together with classical learning pipelines to improve expressivity, data efficiency, and physical fidelity.  In quantum chemistry, this hybrid paradigm has proven feasible, notably through quantum mechanical/molecular mechanical (QM/MM) coupling~\cite{kubavr2023hybrid,böselt2021machine}, where classical force fields are augmented with quantum corrections. Within such frameworks, techniques such as quantum-selected configuration interaction (QSCI)~\cite{bickley2025extending} have been employed to enhance accuracy while keeping the quantum resource requirements tractable.
\R{In the broader landscape of quantum computational fluid dynamics, progress has been made towards developing full quantum solvers for nonlinear PDEs. Recent works by Liu et al.~\cite{liu2012three} and Sanavio et al.~\cite{sanavio2024lattice,sanavio2024three} have successfully applied Carleman linearization to the Lattice Boltzmann equation, offering a promising pathway for simulating fluid flows at moderate Reynolds numbers. These approaches, typically using algorithms like Harrow–Hassidim–Lloyd (HHL)~\cite{harrow2009quantum}, promise exponential speedups but generally necessitate deep circuits and fault-tolerant hardware.}

Quantum-enhanced machine learning (QEML) combines the representational richness of quantum models with the scalability of classical learning. By leveraging uniquely quantum properties such as superposition and entanglement, QEML can explore richer feature spaces and capture complex correlations that are challenging for purely classical models. Recent successes in quantum-enhanced drug discovery~\cite{ghazi2025quantum}, where hybrid quantum–classical generative models have produced experimentally validated candidates rivalling state-of-the-art classical methods, demonstrate the practical potential of QEML even before full quantum advantage is achieved. Despite these strengths, practical barriers remain. 
QEML pipelines require repeated quantum–classical communication during training and rely on costly quantum data-embedding and measurement steps, which slow computation and limit accessibility across research institutions. Moreover, most current QEML applications focus on microscopic problems, such as atomic and molecular systems in quantum chemistry, with little exploration of macroscopic phenomena like nonlinear PDEs and turbulent flows, where long-term stability and statistical fidelity pose distinct challenges~\cite{tennie2025quantum}.
\R{In parallel, the pursuit of quantum advantage in pure quantum computing faces closely related challenges. While theoretical speedups are well established for specific algorithms, identifying practical, application-level advantages remains difficult in realistic settings. This difficulty is exacerbated by the overhead associated with data encoding~\cite{cerezo2022challenges}, the impact of noise in near-term devices~\cite{Preskill_2018}, and fundamental limits on information extraction imposed by Holevo’s bound~\cite{holevo1973bounds}. As a result, many demonstrations of quantum advantage to date rely on analogue sampling tasks~\cite{zhong2020jiuzhang}, quantum-native data, or post-processing of quantum outputs~\cite{huang2022quantum}, rather than real-world, classically relevant problems.}

\begin{figure}
\centering
\includegraphics[width=0.8\textwidth]{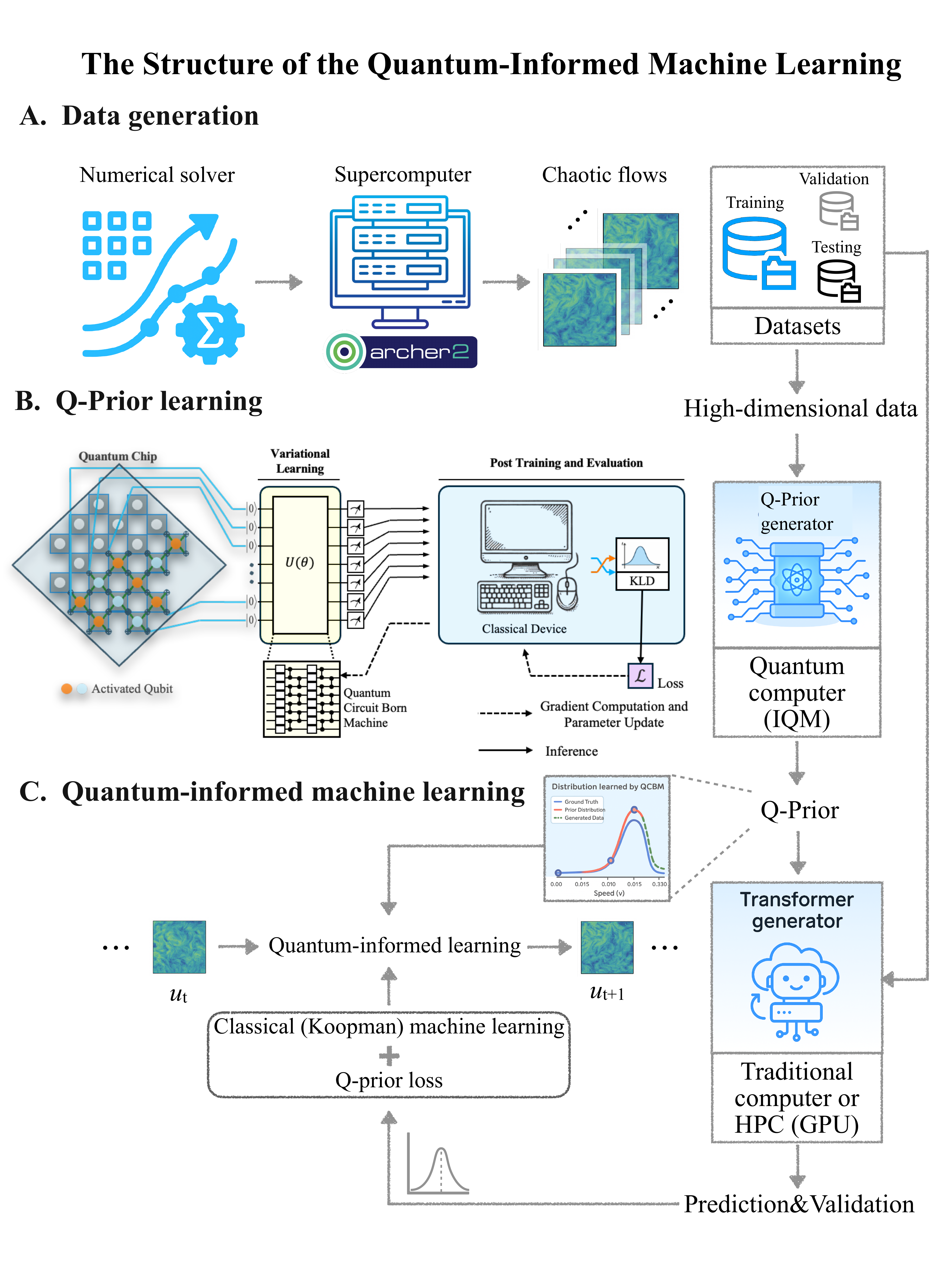}
\caption{\textbf{Architecture of the QIML framework.} 
\textbf{A.} High-dimensional dynamical flow fields are generated using high-resolution numerical solvers and used to construct training, validation and test datasets. 
\textbf{B.} A quantum circuit is trained on superconducting hardware to learn invariant statistical properties from the data (Q-Prior). 
\textbf{C.} The learned Q-Prior is integrated into a classical transformer machine learning, guiding it toward physically consistent predictions and improved long-term stability.
}
\label{fig:qimls}
\end{figure}

\R{To address these limitations, we propose a Quantum-Informed Machine Learning (QIML) architecture, as shown in Fig.~\ref{fig:qimls}m, a specialised hybrid quantum–classical framework for chaotic systems.} The quantum component is a sample-based quantum generator grounded on the quantum circuit Born machine~\cite{liu2018differentiable,benedetti2019generative}, trained once offline on quantum hardware to learn invariant statistical properties directly from observational data. The resulting compact quantum prior (Q-Prior) can be deployed broadly and efficiently without ongoing quantum–classical iteration, eliminating the need for repeated access to quantum hardware during training. This design is specifically tailored to learning the dynamics of nonlinear PDEs. These properties capture complex, high-dimensional distributions compactly by leveraging entangled quantum states. Unlike standard deep-learning surrogates, such as fully connected or convolutional networks that require large parameter counts to represent such structures, the quantum generators operate in exponentially large Hilbert spaces, enabling expressive modelling with as few as 10 to 15 qubits for the systems in this paper. Crucially, this framework circumvents the data-loading bottleneck common to many QML algorithms. The generator does not need to encode raw data onto quantum states; instead, it learns a compressed, generative model of the data's underlying statistical properties. The model is trained with a sample-based Maximum Mean Discrepancy (MMD) loss~\cite{gretton2012kernel}, enabling it to learn the quantum representation without knowing the data distribution. After the training of the quantum generator, the resulting Q-Prior is then integrated into an autoregressive, unitary-enhanced Koopman-based machine learning framework~\cite{budivsic2012applied}. This architecture is powerful for learning dynamical systems as it can linearize nonlinear dynamics in a lifted observable space, making it effective for capturing coherent structures~\cite{mezic2021koopman,brunton2021modern}. However, for high-dimensional chaotic systems, its ability to maintain predictive stability remains a significant challenge, often failing to preserve the systems' physical invariant measures over extended rollouts (i.e. recursive 1-in-1-out predictions where each forecasted state is fed back as input for the next step). This is particularly true when modelling data for which no closed-form governing equation is known—as is the case for a two-dimensional cross-section extracted from a three-dimensional turbulent domain.  We address this limitation by embedding the Q-Prior as a differentiable constraint directly within the loss function. This composite loss, which penalizes divergence from the Q-Prior, imposes a powerful statistical regularization on the evolution of the Koopman operator, guiding it to learn a physically consistent and stable long-term dynamic. We evaluate our QIML framework against its classical counterpart without Q-informed guidance, as well as two leading machine learning models: the FNO for general PDE-based systems and the Markov Neural Operator (MNO)~\cite{li2021learning} specifically designed for chaotic dynamics. Our QIML framework outperforms all baseline models while requiring orders of magnitude fewer trainable parameters.

\section{RESULTS}\label{sec:results}
\subsection{Numerical set up}

The selected test cases span increasing levels of physical and dynamical complexity. We begin with the Kuramoto–Sivashinsky (KS) equation, a PDE exhibiting spatiotemporal chaos. The solution is discretised into 512 spatial grid points, resulting in a 512-dimensional state vector at each time step. Next, we consider two-dimensional Kolmogorov flow on a $64 \times 64$ spatial grid (dimension $= 4096$), governed by the 2D incompressible Navier–Stokes equations in a periodic domain, where the quantum generator is trained in emulation to extract turbulent invariant measures. Finally, we examine cross-sections of a three-dimensional turbulent channel flow (TCF), simulated using the lattice Boltzmann method (LBM), which is numerically equivalent to solving the Navier–Stokes equations~\cite{succi2001lattice}. For the training process, the raw cross-sections are down-sampled and discretised onto a $64 \times 64$ spatial grid, resulting in a 4096-dimensional state vector. The Q-Prior for TCF is trained on superconducting quantum processors. While the full flow field is simulated numerically, the extracted two-dimensional cross-sectional datasets used in our analysis do not satisfy any closed-form governing equation. This distinction highlights a further step in complexity: moving from lower-dimensional systems with explicit dynamics to high-dimensional observational data without accessible evolution laws, thereby testing the capacity of QIML to generalize across different scenarios.
Throughout this study, `ground truth' or `raw data' refers to the high-resolution simulation data generated by established numerical solvers. While this raw simulation data has a large memory footprint, the quantum-informed priors offer parameter efficiency and memory advantage. They capture the flow's essential statistical features using fewer than 300 trainable parameters, occupying only a few megabytes of storage, which corresponds to a data storage reduction of over two orders of magnitude compared to the raw data. This approach enables a substantial reduction in data storage requirements compared to using the raw simulation data (see section Method E and Supplementary ~S2–S3). The classical components of the models were trained on a single NVIDIA A100 80GB GPU using default settings with 32-bit floating-point precision (for details, see Supplementary S4).

\begin{figure*}[htbp]
\centering
\includegraphics[width=0.85\textwidth]{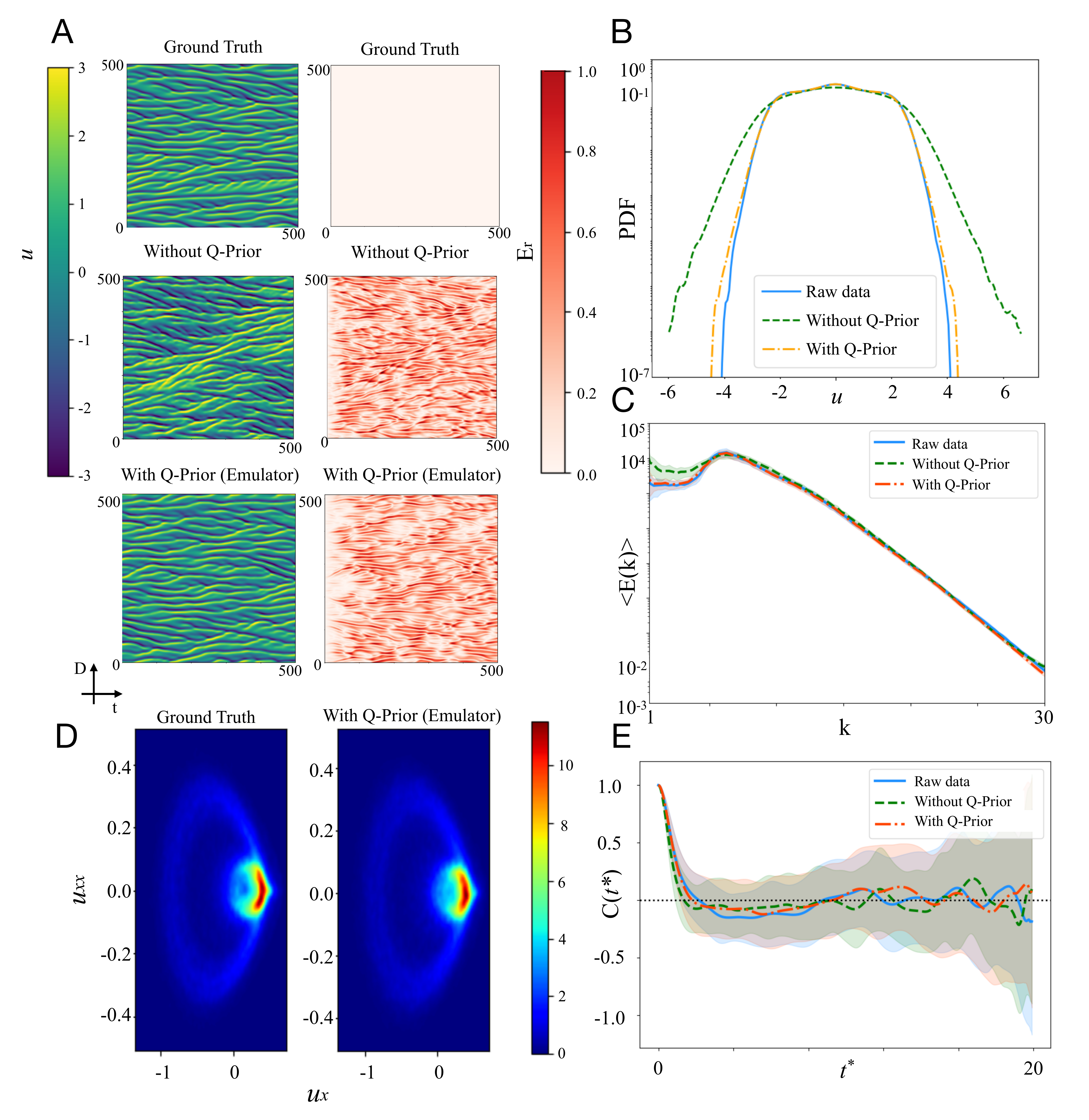   }
\caption{
\textbf{Evaluation of the QIML framework on the KS system. }
\textbf{A. }
This panel displays time-averaged mean velocity fields (left column) and corresponding relative error $E_r$ maps (right column) across the ground truth, the classical ML without Q-Prior, and the quantum-informed ML model with Q-Prior (from top to bottom).
\textbf{B.} 
This panel presents the probability distribution of $u$.
\textbf{C.} This panel shows the energy spectrum $\langle E(k) \rangle$ as a function of spatial wavenumber $k$, characterizing kinetic energy distribution across spectral modes. 
\textbf{D.} This panel visualizes the invariant density when dynamics are projected into the $(u_x, u_{xx})$ space, revealing the geometric support of the underlying invariant measure.
\textbf{E.} 
This panel plots the temporal autocorrelation, denoted by $C(t^*)$, computed from the time series of the field $u$ averaged over all spatial points. The correlation is shown as a function of the dimensionless \R{time lag $t^* = t/t_{Lyapunov}$.}} 
\label{fig:ks_all}
\end{figure*}



\subsection{Kuramoto-Sivashinsky equation}
We first apply our QIML framework to learn the solution of the Kuramoto-Sivashinsky equation, which is given by:
\begin{equation}\label{eq:kse}
\frac{\partial u}{\partial t} + u \frac{\partial u}{\partial x} + \nu \frac{\partial^2 u}{\partial x^2} + \mu \frac{\partial^4 u}{\partial x^4} = 0   
\end{equation}
where $u(x,t)$ denotes a scalar field representing the state of the system, with $u: [0, L] \times [0, \infty) \rightarrow \mathbb{R},$ $x \in [0, L]$ representing the spatial domain and $t \in [0, \infty)$ the time; $ \nu $ and $ \mu $ are constants which are set to 1 for this study. The spatial domain is discretised using $N = 512$ equidistant grid points. Raw data are generated as described in Supplementary S3. The dataset consists of $1200$ trajectories, where each trajectory is made up of $2,000$ time frames. The spatial domain for each frame is discretised using $512$ grid points.
We partition the trajectories chronologically into \(80\,\%\) training, \(10\,\%\) validation, and \(10\,\%\) test sets. 
For this system, both the classical Koopman machine learning (without Q-Prior) and the main QIML model (with Q-Prior) are trained for $500$ epochs.  The Q-Prior for the QIML configuration is generated by a sample-based quantum generator using $10$ qubits ($2^{10}=1\,024$ basis states) and \(120\) trainable parameters. 
Full architectural and training details for both configurations are provided in Supplementary S7.

Fig.~\ref{fig:ks_all} presents the QIML evaluation results on the Kuramoto–Sivashinsky system, comparing two configurations against the numerical simulation:  a classical machine learning model trained without Q-Priors (without Q-Prior), and a quantum-informed machine learning model incorporating a prior trained in simulation (with Q-Prior). Both machine learning models perform inference in an auto-regressive manner: starting from an initial condition at time step $t_0 = 0$, the model predicts the next frame at each subsequent step and recursively continues this process up to $t_{\text{end}} = 500$ time steps to ensure long-term statistics. Panel (A) shows the $u$ evolution alone the time: with 512 initial states and 500 prediction timesteps, and the corresponding relative error $E_r$ evolution. The error metric definition is presented in Supplementary~S1. From top to bottom, each row corresponds to the ground truth, the ML model without Q-Prior configuration, and the QIML model with Q-Prior configuration. These sub-panels are computed over the test dataset to visualize long-term statistical characteristics of the predictions. In the ``without Q-Prior" configuration, the model reproduces the general structure of the ground truth. However, over time, the predicted stripe patterns drift upward, revealing noticeable discrepancies in the associated error map. This indicates that the model accumulates errors during long-term prediction, likely due to the absence of prior information to guide the dynamics.
In contrast, the ``with Q-Prior" model demonstrates much closer alignment with the ground truth. The predicted fields preserve fine-scale features and capture the parallel stripe structures effectively. Additionally, the corresponding error maps exhibit substantially lower and more spatially diffuse errors. These results suggest that incorporating the Q-Prior enhances the model’s stability and accuracy in long-term forecasting. Panel (B) presents the probability density function (PDF) of $u$ for each configuration. The distribution is obtained via the entire spatiotemporal test domain (512 states in space and 500 prediction steps in time), allowing for a comparison of global statistical fidelity. The raw test data (solid blue line) serves as the ground truth. The model without Q-Prior (green dashed line) deviates from the raw data, particularly in the tails of the distribution, where it tends to overestimate the probability. In contrast, the model with Q-Prior (orange dash-dotted line) closely matches the ground truth across the full velocity range, especially in the tails. This improved statistical fidelity is confirmed by a quantitative analysis of the overall prediction error; the inclusion of the Q-Prior results in a $17.25\% \pm 5.25\%$ reduction in the Mean Squared Error (MSE) of the predicted fields with respect to the ground truth over the first 100 steps. This effect becomes even more pronounced in the distribution tails (events where $|u| > 4$), where the error is suppressed by nearly two orders of magnitude. Moreover, an analysis of the spectrum demonstrates superior performance across all energy scales, achieving a $29.36\%\pm 9.01\%$ reduction in MSE for this metric. This demonstrates that incorporating the Q-Prior allows the machine learning model to more accurately reproduce the statistical characteristics of the system, including rare events. Panel (C) shows the energy spectrum $\langle E(k) \rangle$ as a function of the wave number $k$, comparing the raw data, the machine learning model with and without Q-Prior. The energy spectrum of the raw data follows a smooth decay with increasing $k$, reflecting the distribution of energy across spatial frequencies. The model without Q-Prior overestimates the energy in the large-scale, deviating from the ground truth. However, the model with Q-Prior closely aligns with the raw data across all wave numbers, able to capture both large-scale and small-scale (high-$k$) behaviour of the system.  To further visualize the invariant structures underlying the KS system, we project both the ground truth and the Q-Prior configuration into the $(u_x, u_{xx})$ phase space, as shown in Panel (D). The resulting density plots depict the support of the system's invariant measure, providing insight into its long-term statistical behaviour. Over an extended prediction, the model with Q-Prior closely aligns with the ground truth in phase space, accurately capturing the geometry and concentration of the invariant set. This qualitative agreement highlights the Q-Prior model’s ability to preserve the system's underlying statistical structures over time.
Panel (E) illustrates the temporal autocorrelation $C(t^*)$ of the velocity field, averaged over all spatial points, as a function of the dimensionless \R{time lag $t^*=t/t_{Lyapunov}$}. Autocorrelation measures how strongly the system retains correlation with its past states over time. Specifically, its calculation is presented in Supplementary ~S1. In this plot, all cases exhibit a decay in autocorrelation, eventually showing no correlation with the initial state. The configuration with Q-Prior demonstrates stronger short-term correlation (the same as the ground truth data) compared to the model without Q-Prior, indicating better preservation of temporal dynamics in the early prediction period.
These outcomes demonstrate that our designed Q-Prior can regularize the classical ML model, preserving both long-scale and small-scale structures, and improving long-term statistical accuracy. 

\subsection{2D Kolmogorov flow}
Next, we apply our QIML framework to learn the dynamics of the 2D Kolmogorov flow, a canonical example of incompressible Navier-Stokes dynamics under sinusoidal forcing. The governing equations are:
\begin{equation}
\frac{\partial \mathbf{u}}{\partial t} + (\mathbf{u} \cdot \nabla) \mathbf{u} = -\nabla p + \nu \nabla^2 \mathbf{u} + \mathbf{F},
\end{equation}

\begin{equation}
\nabla \cdot \mathbf{u} = 0,
\end{equation}
where $\mathbf{u}$ is the velocity field, $p$ represents the pressure. Here $\mathbf{u}$ denotes the velocity field on the 2D cases. In the algorithmic expressions, we revert to the simpler symbol $u$, using it as a generic placeholder for both 1D scalar and 2D velocities, since the learning procedure applies uniformly across dimensionalities.
$\nu$ is the kinematic viscosity, and the forcing term is given by $\mathbf{F}(x, y) = (F \sin(k y), 0)$ for some amplitude $F$ and wave number $k$. 
The domain is periodic in both $x$ and $y$ directions: $(x, y) \in [0, L_x] \times [0, L_y]$ with $L_x=L_y=64$ grids. The data source is presented in Supplementary ~S4. We use $80\%$ for training, $10\%$ for validation, and $10\%$ for testing.
For this system, both the classical Koopman machine learning (without Q-Prior) and the main QIML model (with Q-Prior) are trained for $500$ epochs.
For QIML, the quantum generator uses $n=10$ qubits ($2^{10}=1\,024$ basis states) and \(180\) trainable parameters. Detailed network architectures, training hyperparameters, and data generation procedures are provided in Supplementary S7.

\begin{figure*}[htbp]
\centering
\includegraphics[width=0.95\textwidth]{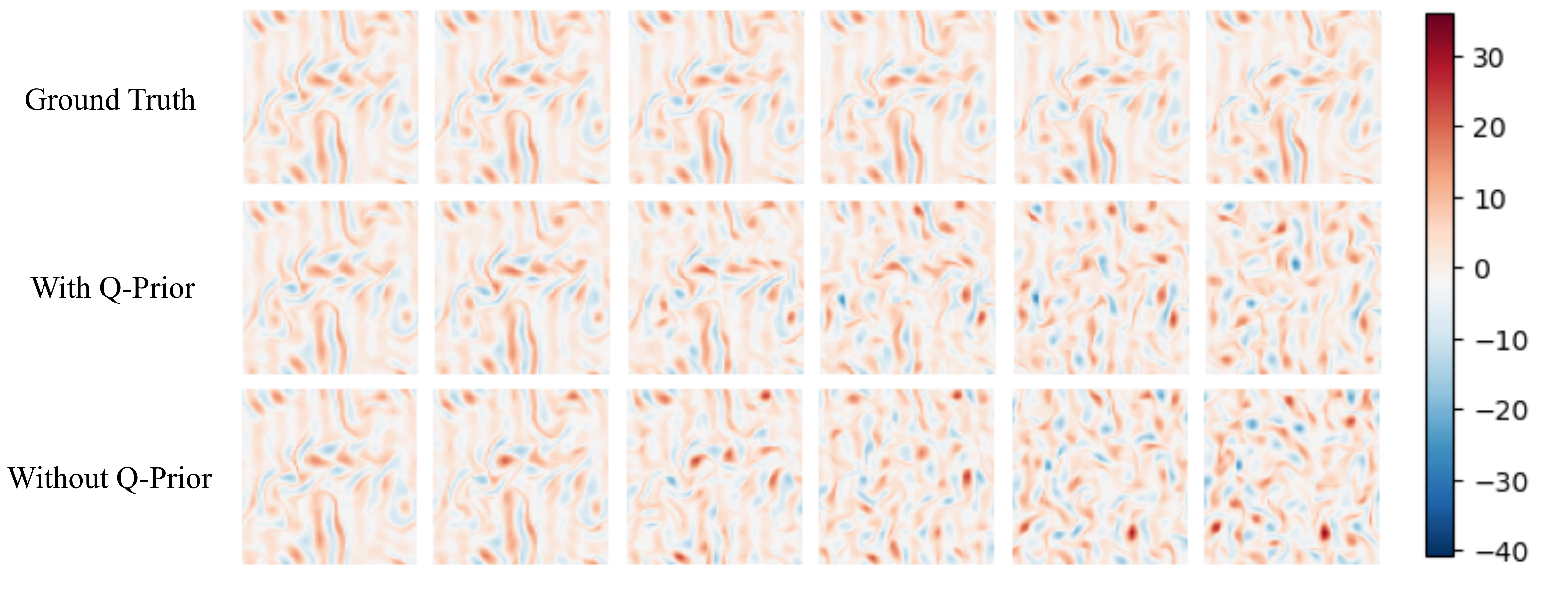  }
\caption{\textbf{Early-time chaotic evolution of streamwise velocity fields.} The streamwise velocity rollout of ground truth (top), the QIML with Q-Prior (middle), and the classical model without Q-Prior (bottom) given the same initial state at $\hat{t}$ = 0.0, 0.2, 0.4, 0.6, 0.8, 1.0 (from left to right), \R{corresponding to Lyapunov times $t^{*} = 0.0, 0.04, 0.08, 0.12, 0.16,$ and $0.20$, respectively.}
}
\label{fig:kf_pic}
\end{figure*}

\begin{figure*}[htbp]
\centering
\includegraphics[width=0.9\textwidth]{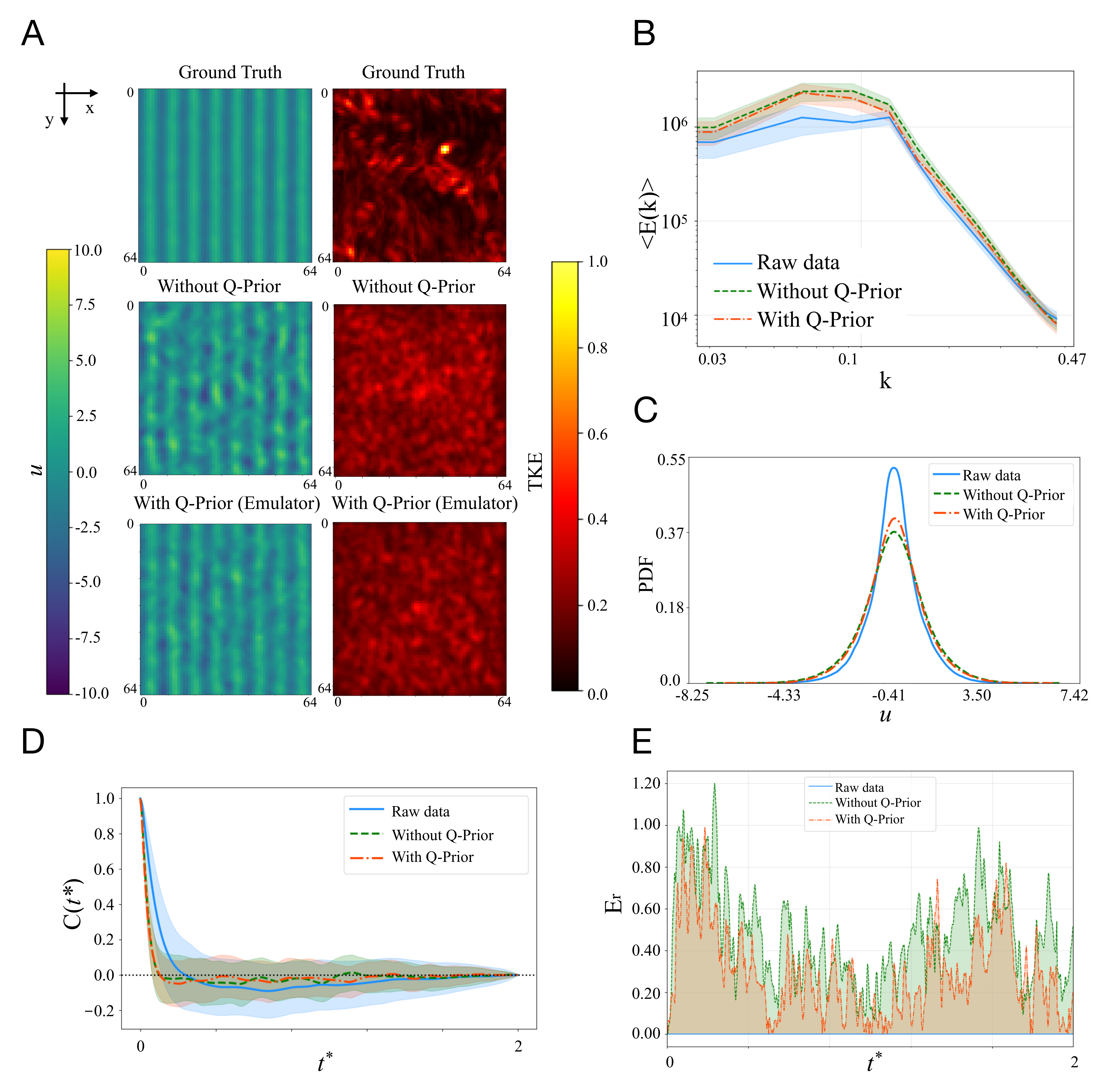   }
\caption{\textbf{Evaluation of the QIML framework on the 2D Kolmogorov flow. }
\textbf{A.} This panel displays time-averaged mean $u$ fields (left column) and corresponding TKE fields (right column) for the ground truth, the classical model without Q-Prior, and the quantum-informed model with Q-Prior. These maps are evaluated on the test set, highlighting long-term flow patterns and energy distribution. 
\textbf{B.} 
This panel presents the ensemble-averaged energy spectrum $\langle E(k) \rangle$ as a function of spatial wavenumber $k$. 
\textbf{C.} 
This panel shows the PDF of the field variable $u$, comparing the distribution from the raw test data with the PDFs predicted by the models with and without the Q-Prior.
\textbf{D.} 
This panel plots the temporal autocorrelation $C(t^*)$ of the velocity field as a function of dimensionless time $t^*$.  
\textbf{E.} 
This panel reports the normalized relative error, $E_r$, with respect to the ground truth, calculated over time on the same validation trajectory for a 1000-step rollout. 
}
\label{fig:kf_all}
\end{figure*}

Fig.~\ref{fig:kf_pic} presents the predicted streamwise velocity fields over a 20-step autoregressive rollout ($t_{\text{window}}=20$), where the \R{time is normalized as $\hat{t}=t/t_{\text{window}}$}. The snapshots qualitatively demonstrate that the QIML model preserves the coherent vortical structures of the ground truth with higher fidelity and for a longer duration compared to its classical counterpart, in which these features decay more rapidly into a disorganized state. Fig.~\ref{fig:kf_all} reports the comparative
performance of the baseline ML model (without Q-Prior) and the quantum-informed ML model (with Q-Prior) on the two-dimensional Kolmogorov flow dataset. Panel~(A) shows the time-averaged velocity field and the corresponding turbulent kinetic energy (TKE) field across three configurations: the ground truth from simulation, the classical model trained without Q-Prior, and the quantum-informed model trained with Q-Prior. All models are evaluated on the same test data. Both the ``with Q-Prior" and \R{``without Q-Prior" models perform auto-regressive (one frame in one frame out) inference over a horizon of $1000$ which is around 2 Lyapunov times $t_{Lyapunov}=500$~\cite{Kochkov2021-ML-CFD}. Beyond O(1) Lyapunov time, trajectory-wise prediction is not expected; therefore we assess long-horizon performance via statistical fidelity.} The visualizations span a two-dimensional domain of size $64 \times 64$ in the $x$–$y$ plane. The time-averaged velocity field is plotted in the left column, and the corresponding TKE field is shown in the right column. The ground truth velocity field (top left) exhibits coherent, stripe-like structures, and its TKE field (top right) reveals concentrated regions of high energy, indicating the presence of organized flow features. The model without Q-Prior (middle row) shows degraded spatial coherence in the velocity field, with disrupted patterns and more noise-like structures. In contrast, the model with Q-Prior (bottom row) recovers a more organized velocity field, better preserving the stripe patterns observed in the ground truth. Its associated TKE field also aligns more closely with the ground truth.
Panel~(B) presents the ensemble-averaged energy spectrum $\langle E(k) \rangle$ as a function of the wavenumber $k$ for the raw data (solid blue line), the machine learning model without Q-Prior (green dashed line), and the model with Q-Prior (orange dash-dotted line). The spectrum from the model without Q-Prior overestimates the energy spectrum at low and intermediate wave numbers. However, the Q-Prior aligns closer to the raw data across the whole spectrum. Then, we further examine the overall velocity distributions. As illustrated in panel~(C), we compare the predicted velocity probability density functions from both machine learning models. The raw data exhibit a sharply peaked distribution with heavy tails, characteristic of the system’s inherent velocity fluctuations. Both machine learning models capture the general symmetric shape of the distribution, but tend to slightly underestimate the peak and overestimate the spread. Among them, the model with Q-Prior more closely matches the raw data, particularly around the peak and in the tails, indicating improved accuracy in reproducing the system’s statistical behaviour. This is quantitatively supported by a $6.57\% \pm 3.68\%$ reduction in the MSE for the Q-Prior model's predictions compared to the classical model without the prior over the first 100 steps. This advantage is particularly pronounced in the high-probability peak of the distribution (where $|u| < 2$), where the MSE reduction for the Q-prior model increases to $10.39\% \pm 4.23\%$. On a global scale, this superior performance is also reflected in the energy spectrum, where the Q-Prior achieves a 14.16\% MSE reduction. These results further support the conclusion that incorporating the Q-Prior enhances the machine learning model's ability to recover velocity statistics. We also examined the autocorrelation of the field over time. Panel~(D) plots the autocorrelation $C(t^*)$ of the velocity field as a function of dimensionless \R{time $t^*=t/t_{Lyapunov}$}. The correlation is computed over 100 randomly sampled locations, with shaded regions indicating one standard deviation. Both models reproduce long-term decay in autocorrelation followed by fluctuations around zero, consistent with the expected loss of memory in a chaotic or turbulent system. To further analyze the model, panel~(E) reports the normalized relative auto-correlation error $E_r$ with respect to the ground truth. The raw data serves as the reference baseline with zero error. Across the entire temporal range, the QIML model incorporating the Q-Prior consistently exhibits lower autocorrelation error than the classical model trained without the Q-Prior.

These observations confirm that the quantum-informed Q-Prior can provide additional information that improves classical ML fidelity and velocity statistics, and enhances long-horizon accuracy.


\subsection{Turbulent channel flow inflow generation}

\begin{figure*}[htbp!]
\centering
\includegraphics[width=1\textwidth]{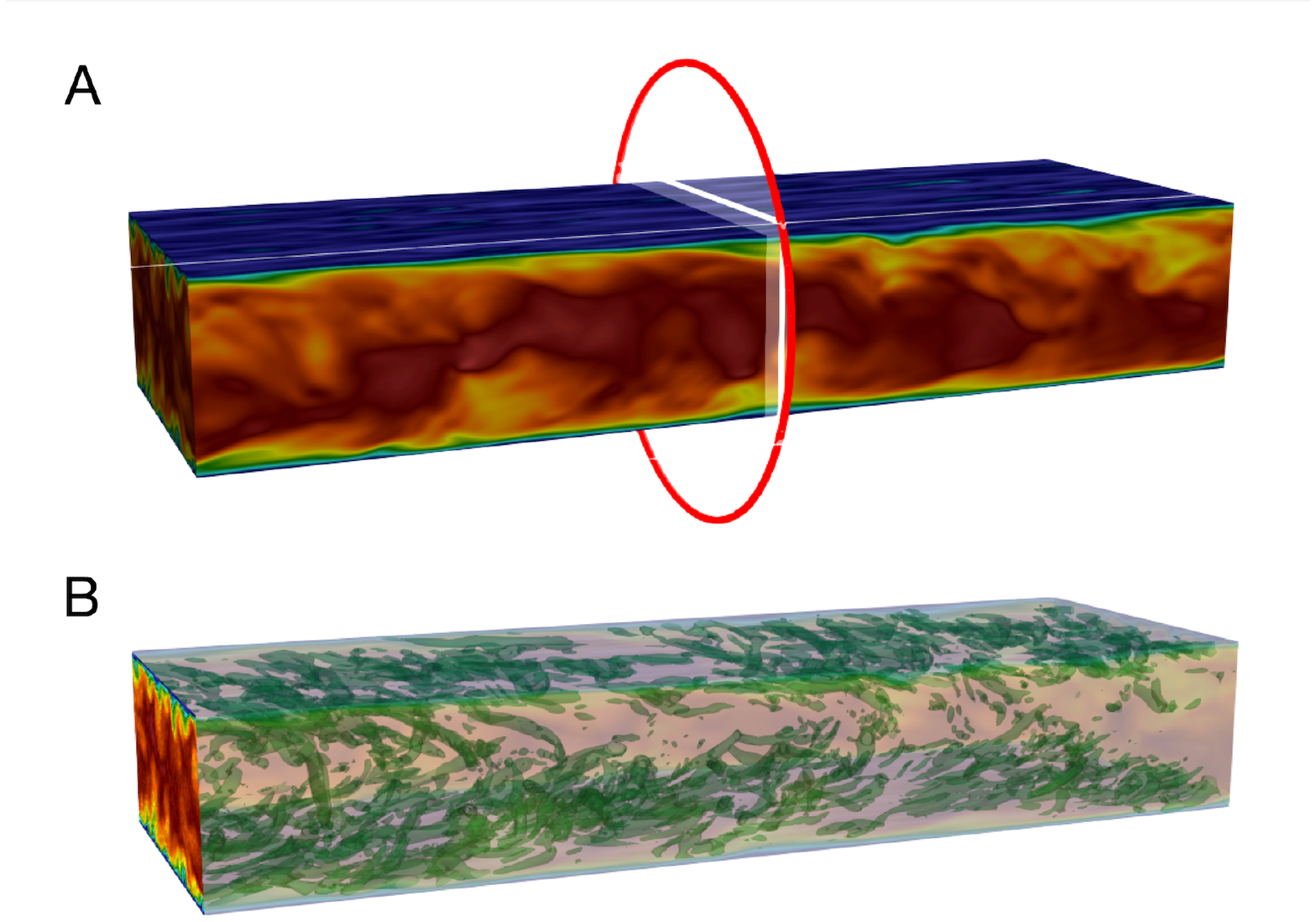   }
\caption{
\textbf{Generation of Synthetic Turbulent Inflow from Turbulent Channel Flow Data.} \textbf{A.} Instantaneous flow field from a turbulent channel flow simulation, representing the raw data used for turbulent inflow generation. The red ellipse indicates the cross-section where velocity field data are extracted. \textbf{B.} The QIML generative model can be used as turbulent inflow conditions in 3D turbulent channel flow for large eddy simulations. 
}
\label{fig:channel_in}
\end{figure*}

We further evaluate our QIML framework on turbulent channel flow, a representative case of wall-bounded turbulence. In this setting, the training data is obtained from the mid-plane cross-section of a fully developed periodic turbulent channel flow at friction Reynolds number $Re_{\tau}=180$. Note that the reference data has been validated with the DNS data~\cite{Moser1999} (See Supplementary S5). The reference data is generated using the LBM solver, which serves as an efficient alternative to directly solving the Navier-Stokes equations~\cite{succi2001lattice,chen1998lattice}. A large eddy simulation (LES) formulation is employed to capture the large-scale turbulent structures~\cite{xue2022synthetic,xue2024physics}. Fig.~\ref{fig:channel_in} \textbf{a} shows the instantaneous velocity field extracted from the high-fidelity turbulent channel flow simulation, with the red ellipse marking the cross-section used as the source data for inflow synthesis. The governing equations used for generating the data are provided in Supplementary S5. The detailed description of the data generation setup is in Supplementary S8. The channel flow snapshots are split $80{:}10{:}10$ into training, validation, and test sets. For this system, both the classical Koopman machine learning model (without Q-Prior) and the main QIML model (with Q-Prior) are trained for $500$ epochs. For QIML, we employ a \(15\)-qubit quantum generator (\(2^{15}=32\,768\) basis states) with \(240\) trainable parameters. Detailed network architectures, training hyperparameters, and data generation procedures are provided in Supplementary S7. Fig.~\ref{fig:channel_in} \textbf{b} depicts the corresponding synthetic turbulent inflow generated by the QIML model, demonstrating its capability to produce physically consistent inflow conditions for large-eddy simulations~\cite{xue2022synthetic}.

\begin{figure*}[htbp!]
\centering
\includegraphics[width=0.9\textwidth]{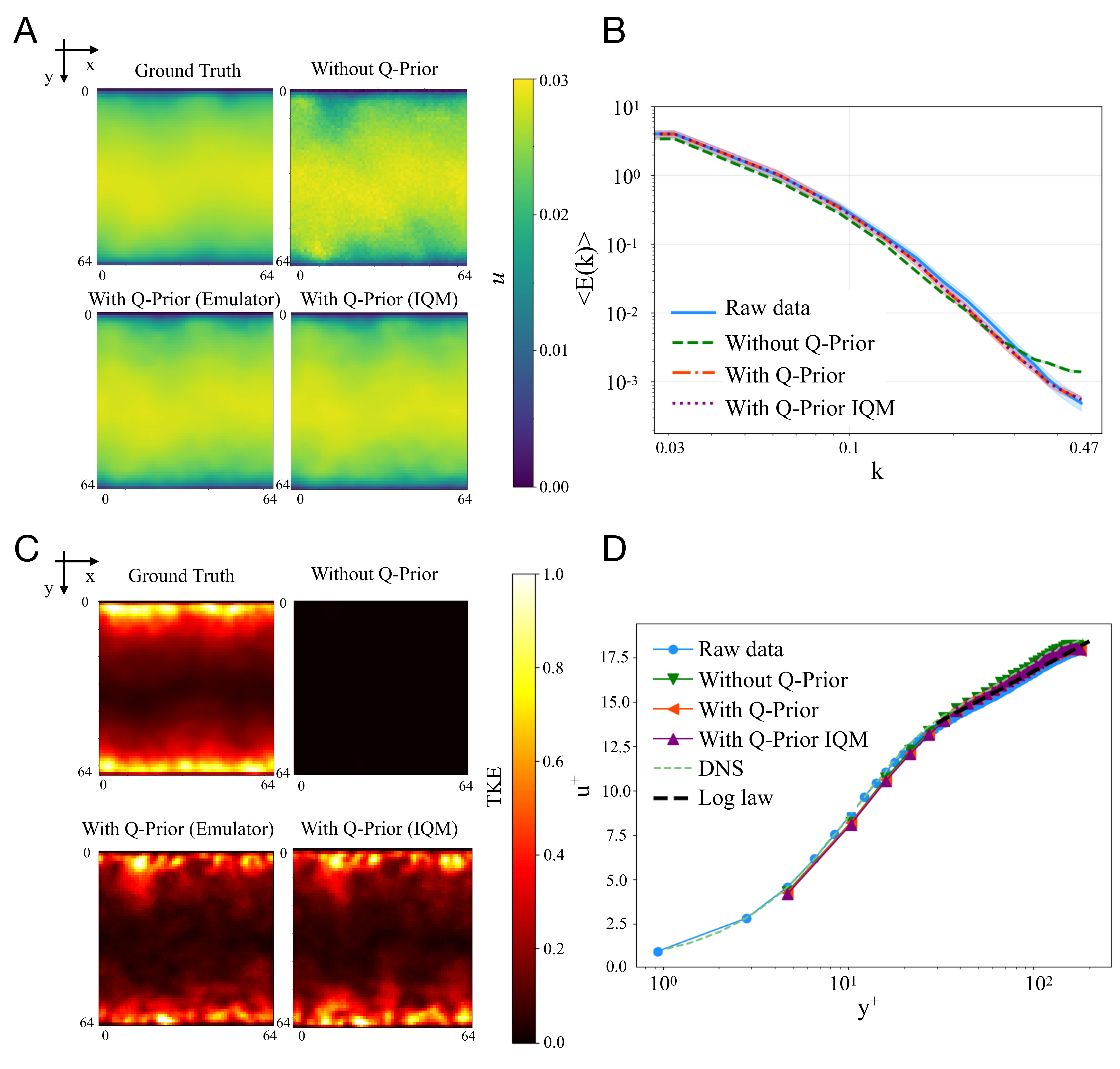   }
\caption{
\textbf{Evaluation of the QIML framework on the TCF system. }
\textbf{A.}
This panel displays time-averaged streamwise velocity fields across different models. Each row corresponds to a different model output: the top row displays the ground truth obtained from DNS, followed by results from the classical model without Q-Prior, the QIML model using Q-Prior trained on the emulator, and the model using a Q-Prior trained on real superconducting hardware (IQM).
\textbf{B.}
This panel plots the ensemble energy spectrum $\langle E(k) \rangle$ as a function of the spatial wavenumber $k$.
\textbf{C.} 
This panel displays the time-averaged TKE field for the same four configurations as in panel a, using a normalized color scale to enhance visual comparability. 
\textbf{D.}
This panel shows the non-dimensional velocity profile $u^+$ as a function of the wall-normal coordinate $y^+$ on a logarithmic scale. Results are compared with the empirical log-law and DNS reference, using the same validation data as in previous panels. 
}
\label{fig:tcf_all}
\end{figure*}

Fig.~\ref{fig:tcf_all} presents the evaluation of the quantum-informed machine learning framework on the turbulent channel flow dataset. Our ML models perform auto-regressive (one frame in one frame out) inference over a horizon of \R{$1000$ time steps}. The visualizations span a two-dimensional domain of size $64 \times 64$ in the $x$–$y$ plane. Panel (A) shows a comparison of time-averaged velocity magnitudes $u$ across a $64 \times 64$ domain for the ground truth, the machine learning model without Q-Prior, and two Q-Prior-enhanced variants: the emulator and real superconducting IQM device. The colorbar on the right indicates the magnitude of $u$, with higher values in yellow and lower values in blue. The ground truth (top left) displays smooth, large-scale variations in the velocity field. The model without Q-Prior (top right) introduces noticeable small-scale noise and fails to capture the smooth gradients seen in the true field. In contrast, both Q-Prior (with emulator and with IQM device) models (bottom row) closely match the ground truth, preserving the dominant spatial structures and exhibiting lower spurious variability. Panel~(B) plots the ensemble energy spectrum $\langle E(k) \rangle$ as a function of the wavenumber $k$ for the raw data (solid blue line), the machine learning model without Q-Prior (green dashed line), the model with Q-Prior (orange dash-dotted line), and the model with Q-Prior IQM (purple dotted line). The model without Q-Prior deviates from the ground truth at higher wavenumbers, with discrepancies arising at smaller spatial scales, which indicates a failure to capture the essential small-scale physical dynamics of the system. As we will demonstrate in the following section~\ref{sec:comparison}, this is a common failure mode for state-of-the-art classical architectures. In contrast, both Q-Prior variants better capture the energy distribution across all scales. Panel~(C) compares the normalized TKE fields across a $64 \times 64$ spatial domain for the ground truth, the model without Q-Prior, and two Q-Prior-based configurations: the emulator and IQM. The ground truth (top left) displays spatially structured regions of elevated TKE, particularly near the domain boundaries, reflecting characteristic flow features. The model without Q-Prior (top right) fails to capture any meaningful TKE patterns, instead producing a near-zero field, which indicates an almost static prediction over time. In contrast, both Q-Prior-based models (bottom row) recover structured TKE fields that closely resemble the ground truth. These models preserve the dominant energy distribution and capture the spatial organization of turbulent structures more accurately. Panel~(D) shows the dimensionless velocity profile $u^+$ as a function of the dimensionless wall-normal coordinate $y^+$. Our results are compared with the raw data, log-law and DNS reference. All machine learning models reproduce the canonical log-law behaviour and match well with the reference data, DNS data and log law.

Overall, the Q-Prior provides a clear improvement in the spatiotemporal prediction of cross-sections in turbulent channel flow, where governing equations for the cross-sectional dynamics are not explicitly defined. This makes it a potentially valuable tool for engineering applications that rely on coupling Reynolds-Averaged Navier–Stokes (RANS) and LES approaches~\cite{xue2022synthetic}.


Collectively, these results demonstrate that embedding Q-Priors significantly improves long-term prediction results and statistical alignment of classic machine learning models in high-dimensional dynamical systems. Even with modest quantum resources, that is, no more than 15 qubits and under 300 parameters (specifically shown in method~F, Supplementary S2), our framework captures essential dynamics of high-dimensional flows and enables practical hybrid learning pipelines for scientific applications. By anchoring predictions to Q-Prior, the QIML model improves statistical fidelity and stability, laying the groundwork for practical quantum–classical synergies in the modelling of turbulent and other chaotic physical systems.

\section{DISCUSSION}
\subsection{Comparative analysis of QIML against machine learning baselines}\label{sec:comparison}

To rigorously quantify the contribution of our quantum-informed machine learning framework, we conduct a comparative analysis of our full QIML architecture against key baseline models. The aim is to show that the stability and statistical fidelity observed in our results arise from the Q-Prior guidance, rather than being solely attributable to the autoregressive classical ML model’s architecture. Both the QIML framework and the classical baselines are trained on exactly the same raw dataset. The fundamental difference lies in the architectural approach instead of unequal access to data. \R{Here, we consider three primary classes of baselines: (i) a classical state-of-the-art Koopman-based autoregressive machine learning model trained without the Q-Prior, (ii) neural operators for nonlinear PDEs and chaotic systems such as the Fourier neural operator and the Markov neural operator~\cite{li2021learning}, and (iii) a classical-informed machine learning (CIML) variant in which the same architecture is guided by a classical generative prior~\cite{kingma2013auto}.}

\begin{figure*}[htbp]
\centering
\includegraphics[width=0.95\textwidth]{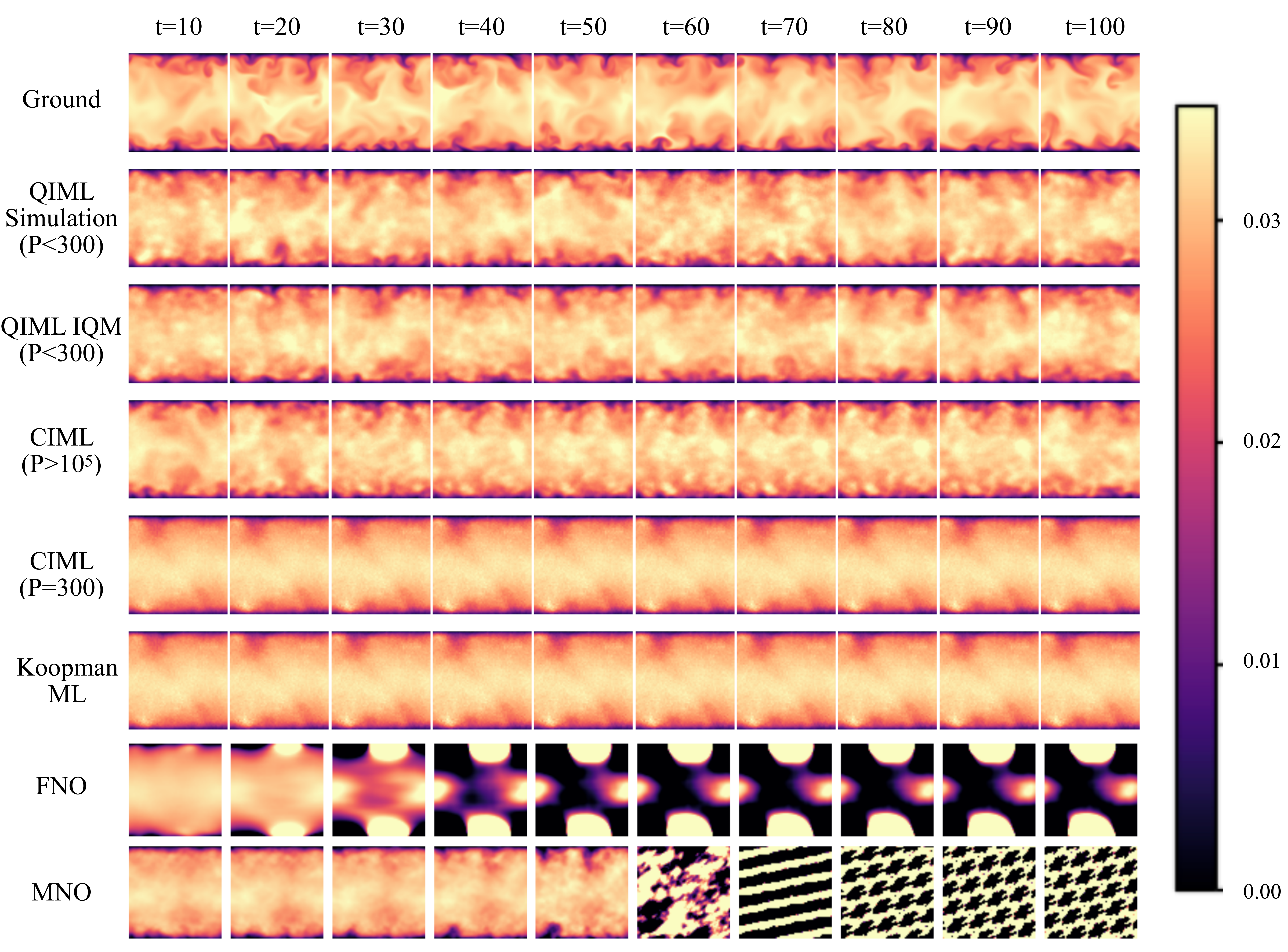   }
\caption{\R{\textbf{Qualitative comparison of long-term autoregressive predictions for the turbulent channel inflow, with all models evaluated under an identical one-step-in, one-step-out autoregressive rollout.} Snapshots of the velocity field are shown at representative predictive time steps $t$ from $10$ to $100$ (from left to right) for the Ground Truth and for predictions from the QIML (IQM), CIML (VAE), classical Koopman ML, FNO, and MNO models. $P$ represent the number of parameters used in Q-Prior or C-Prior. The comparison demonstrates that QIML most consistently preserves the structural features of the ground-truth flow over long prediction horizons. In contrast, most classical baselines either relax toward an overly smooth, time-averaged state or develop non-physical artefacts during rollout. While the CIML model can reproduce comparable large-scale structures when equipped with a substantially larger parameter budget, it requires more than 400 times as many parameters to approach the performance of QIML. Moreover, after $t = 20$, the CIML predictions display a pronounced locking behaviour, in which the flow field exhibits reduced temporal variability and gradually settles into a nearly static configuration. By contrast, QIML shows a more physically consistent temporal evolution, with sustained variability and evolving flow structures over time.}}
\label{fig:comparison}
\end{figure*}

As demonstrated in Fig.~\ref{fig:comparison}, our classical model without the Q-Prior consistently fails to maintain long-term stability. While capable of accurate short-term predictions, it inevitably succumbs to error accumulation, leading to prediction divergence and a failure to reproduce the system's invariant statistical measures. The comparison with neural operators provides a more stringent test. FNOs and MNOs are known for their strong performance in learning resolution-independent solutions to nonlinear PDEs and chaotic systems. We trained both models on the same datasets used for our framework and ensured that both baselines were evaluated under the same autoregressive setting, using a one-step-in, one-step-out rollout in the spatiotemporal domain. Under this regime, however, the FNO and MNO models exhibit the same weakness: their predictions diverge progressively from the ground truth, failing to preserve the correct long-term spatio-temporal evolution of the system (see Fig.~\ref{fig:comparison}). This behaviour is particularly striking given their substantial size; both the FNO and MNO employ an order of magnitude more trainable parameters than our QIML framework, as detailed in Table S2 of the Supplementary Information. The underlying issue is well documented~\cite{mccabe2023towards}: standard neural operators are typically trained in a teacher-forcing or multi-step-output setting, but when forced into strict one-step autoregressive rollouts, small local errors compound at each iteration. An important observation is that existing FNO and MNO models have demonstrated strong performance in equation-driven PDE systems~\cite{lippe2023pde}, such as 2D Kolmogorov flow, where the governing equations are explicitly known. Nonetheless, in settings like our cross-section of three-dimensional turbulent channel flow, where no closed-form PDE governs the evolution of the extracted two-dimensional data, these operator-learning models face intrinsic challenges. This highlights that while FNOs and MNOs excel in PDE-constrained environments, their generalization to more complex, data-driven scenarios still needs to be improved.

\R{Furthermore, to rigorously assess the role of the Q-Prior, we conducted a controlled comparison against a high-capacity classical generative baseline. Specifically, a deep Variational Autoencoder (VAE)~\cite{kingma2013auto} was trained on the identical TCF dataset to construct a classical prior (C-Prior) with the same output dimensionality as the Q-Prior.
When scaled to more than $1.2 \times 10^{5}$ trainable parameters, the VAE-based C-Prior is able to reproduce the coarse-grained energy distribution and large-scale statistics at a level comparable to the Q-Prior (Fig.~\ref{fig:comparison}). However, beyond intermediate prediction times, the CIML rollouts exhibit a pronounced tendency toward dynamical saturation, with the predicted flow becoming progressively locked into a quasi-stationary pattern and displaying reduced temporal variability. These effects accumulate during long-term autoregressive prediction and lead to increased dynamical drift relative to QIML. In addition, systematic deviations emerge in the high-wavenumber regime, where the classical prior underestimates the tail of the energy spectrum and fails to faithfully capture fine-scale, high-frequency structures (see Supplementary S7). 
A more direct comparison emerges when the C-Prior is constrained to a parameter budget comparable to that of the Q-Prior (approximately $300$ parameters). In this regime, the model collapses to an overly smooth, mean-field representation and fails to recover meaningful small-scale statistics.}

\R{This comparative analysis leads to a clear conclusion: the key enabling component of our framework is the Q-Prior integrated into the machine learning. The long-term instability is a fundamental challenge for even powerful classical architectures like FNOs, MNOs and even CIML. By providing a physically grounded, memory-efficient statistical regularizer, the Q-Prior confers the necessary stability and long-term fidelity that are otherwise absent. This result underscores the unique and practical advantage of our QIML architecture.}

\R{This observation raises a fundamental question: \textit{why does the Q-Prior provide a distinct benefit in modelling chaotic systems?} We hypothesize that the answer lies in the intrinsic structure of chaotic data, which often exhibits complex, non-local statistical dependencies---structures that may be functionally analogous to quantum entanglement or contextuality. These global correlations, inherent to the invariant measure of a chaotic attractor, are notoriously difficult for classical generative models to capture efficiently due to their reliance on factorized or locally conditioned distributions~\cite{gao2022enhancing}. In contrast, quantum circuits naturally encode such correlations through entangled amplitudes, enabling the Q-Prior to compactly represent statistical features that span multiple scales and degrees of freedom. We posit that quantum generative models may thus offer a provable advantage in learning the invariant measure of systems with these characteristics. A theoretical motivation and more discussions are presented in Supplementary Section~S8 and S10.
}

\subsection{Q-Prior memory efficiency}\label{sec:efficiency}
\label{sec:parameterandmemory}

A defining feature of the proposed framework is its capacity to extract high-dimensional physical information using significantly fewer trainable parameters than conventional deep learning models. The Q-Prior, operating on only 10–15 qubits and comprising fewer than 300 trainable parameters, successfully captures invariant distributions over spatial domains with dimensionality exceeding $10\textsuperscript{4}$. In contrast, a classical neural network capable of learning an equally expressive distribution over the same domain would require at least $\mathcal{O}(10^4)$ parameters, assuming a minimal fully connected architecture with comparable granularity.   For example, on a grid of \(D=2^{10}=1{,}024\) spatial points, the lightest fully-connected (FC) network that can approximate an arbitrary probability map with standard universal-approximation accuracy must allocate \(\mathcal{O}(D^{2})\) weights~\cite{barron2002universal,Goodfellow2016}.  
Even with an aggressively thinned hidden layer, this still amounts to \(\sim\!2.5\times10^{5}\) tunable parameters—three orders of magnitude more than the \(300\) parameters used by our ten-qubit Q-Prior (see Supplementary S2).  
Moreover, classical models lack access to quantum entanglement, limiting their ability to compactly encode non-local correlations and higher-order statistical dependencies prevalent in turbulent or chaotic systems. 

However, possessing a parameter-efficient quantum representation is only the first step; leveraging it to guide a classical machine learning model is a distinct and significant challenge. We address this by designing a composite loss function that seamlessly integrates the statistical information from the Q-Prior into the classical model's training process. This enables a true ``quantum-informed'' learning paradigm where the quantum component actively regularizes the classical dynamics, leading to superior results as demonstrated in this work. As shown in the Table~\ref{tab:compression_simple}, the successful implementation of this hybrid integration unlocks a further, highly practical potential of quantum computing: memory efficiency in data storage and transfer. 

Specifically, for the turbulent–channel dataset, retaining all mid-plane snapshots in double precision requires roughly \(5.0\times10^{2}\,\text{MB}\).
Archiving the corresponding Q-Prior checkpoint for each snapshot
(\(\,4.3\,\text{kB}\) per file) occupies only
\(595\times4.3\,\text{kB}\approx 2.5\,\text{MB}\),
a compression greater than two orders of magnitude while preserving
the relevant flow statistics.  Comparable savings are obtained for
Kolmogorov flow (\(\sim\!180\) parameters, \(\,\sim1.4\,\text{kB}\) per file, \(\,\sim0.44\,\text{MB}\) in total) and Kuramoto–Sivashinsky data (\(\sim\!120\) parameters, \(\sim1\,\text{kB}\) per file, \(\,\sim1.4\,\text{MB}\) in total).
Full storage calculations are shown and reported in Supplementary S8.
Such dramatic savings make the Q-Prior attractive whenever data are expensive to generate or archive, while still providing a statistically faithful, low-parameter regularizer for the classical Koopman machine learning. 

Crucially, this provides a blueprint for future hybrid computing infrastructures where the quantum processors might be co-located with every classical computer. Instead of transferring voluminous raw datasets, which can amount to hundreds of gigabytes or more, one would only need to transfer the compact Q-Prior. As our results show, this corresponds to a data storage reduction of over two orders of magnitude.
Furthermore, this vision is made practical by a key finding that directly addresses the number of measurement shots, a widely scrutinized issue for NISQ-era algorithms. While the results presented in this work are based on a high shot count (20,000 per iteration) to demonstrate the framework's maximum potential, we found that the classical model does not require a perfectly resolved, exponentially large distribution to be effective. In fact, reducing the measurement count to 8,000 shots had a negligible impact (less than 5\%) on the final model’s performance, showing that an approximate prior is sufficient to act as a powerful guiding signal. Crucially, the Q-Prior is trained only once in an offline stage, rather than being re-evaluated during every rollout or trained iteratively within the ML framework like QEML. \R{This one-time training greatly reduces the overall computational burden and differentiates our approach from iterative hybrid quantum–classical schemes that require repeated quantum calls throughout the learning process. Ideally, the sample complexity for estimating gradients of Born machines scales polynomially with the number of qubits~\cite{huang2025vast}, ensuring that the shot overhead remains manageable even as we scale to larger systems.} Thus, QIML establishes a viable and economically attractive framework: a central, high-fidelity quantum processor performs the intensive, one-time training of a master Q-Prior, whose compact parameters are then distributed at low cost to remote systems. This paradigm may serve as a template for future cloud quantum platforms, where the quantum processors provide powerful statistical guidance and data support for large-scale classical models.

\begin{table}[h!]
\centering
\caption{\textbf{Quantum resources and storage compression achieved by the Q-Prior.}}
\label{tab:compression_simple}
\begin{tabular}{lccccc}
\toprule
\textbf{Dynamical System} &
\textbf{Raw data} &
\textbf{Q-Prior} &
\textbf{Compression} \\
 & (full set) & (full set) & (ratio)\\
\midrule
Kuramoto–Sivashinsky & \(300\;\mathrm{MB}\) & \(0.25\;\mathrm{MB}\) & \(\approx 1.2\times10^{3}\!:\!1\) \\
2D Kolmogorov flow   & \(400\;\mathrm{MB}\) & \(0.40\;\mathrm{MB}\) & \(\approx   10^{3}\!:\!1\) \\
Turbulent channel flow &  \(500\;\mathrm{MB}\) & \(2.3\;\mathrm{MB}\) & \(\approx        2.2\times10^{2}\!:\!1\) \\
\bottomrule
\end{tabular}
\end{table}

\subsection{On the nature of practical quantum advantage in QIML}\label{sec:advantage}

The term ``quantum advantage" must be approached with caution, particularly in the context of quantum computing and quantum machine learning~\cite{huang2025vast,jerbi2024shadows}. As recent analyses have highlighted, many claims of quantum advantage are later matched or surpassed by sophisticated classical algorithms, revealing that the initial gap was due to the immaturity of classical methods rather than a fundamental quantum superiority~\cite{huang2025vast}. With the exception of algorithms like Shor's~\cite{shor1999polynomial}, which leverage the fundamental nature of quantum mechanics in a provably superior way, many claims of advantage are eventually challenged. Even powerful algorithms such as the HHL algorithm have faced scrutiny~\cite{harrow2009quantum}, with new classical works emerging that achieve comparable performance in specific contexts~\cite{huang2025vast}.  We do not assert that QIML holds an unassailable advantage. Instead, we posit that its effectiveness stems from a foundational quantum principle that, in this application, has not been classically replicated. \R{There is growing evidence that certain probability distributions are exponentially hard to represent or learn classically if they embody forms of contextuality or nonlocal correlations analogous to those in quantum states~\cite{gao2022enhancing}.}

The foundational advantage exploited by our Q-Prior stems from the ability of an $n$-qubit quantum state to represent a classical vector in a $2^n$ dimensional space, which is considered a genuine quantum
advantage~\cite{huang2022quantum}. This provides an exponential advantage in information storage capacity. This type of advantage has found recent applications in quantum machine learning, such as for computing quadratic functions of data~\cite{gilboa2024exponential}. However, this potential is famously constrained by Holevo's bound, which dictates that only $O(n)$ classical bits of information can be reliably extracted from an n-qubit state through measurement~\cite{holevo1973bounds}. If a full reconstruction of the classical information were required, the need for an exponential number of measurements would negate the storage advantage entirely.

In ergodic theory, the invariant measure governs the long-time statistical behaviour of chaotic dynamics, even though individual trajectories diverge exponentially~\cite{cornfeld2012ergodic}. The central insight of the QIML framework is to circumvent Holevo’s bound by extracting only this essential invariant measure without reconstructing the full dataset, yielding a compressed yet sufficient representation of the system. This low-dimensional representation is precisely what is needed to regularize the classical ML, guiding it away from non-physical trajectories without needing to know every detail of the system's state. Therefore, the advantage demonstrated here is a representational and memory advantage that translates directly into a performance advantage. \R{The theoretical basis for this advantage is detailed in Supplementary Section~S8.}

\textbf{Memory Efficiency}:  The Q-Prior leverages quantum entanglement to capture complex, non-local correlations within the invariant measure using a few parameters. This translates to a dramatic, orders-of-magnitude reduction in data storage, as quantified in Table~\ref{tab:compression_simple}.

\textbf{Performance Gain}: This compact and physically rich representation provides the crucial regularization that leads to long-horizon stability and statistical fidelity, a task at which larger, more complex classical models like FNO and MNO demonstrably fail for this class of problems.

While it is an open question whether other classical models could be adapted to leverage the Q-Prior as effectively, our results suggest that the co-design of the QIML architecture is key. We anticipate that our results may inspire new classical approaches. For instance, using large generative models to learn the invariant measure could achieve comparable predictive performance. However, we argue that such methods are unlikely to replicate the foundational source of the advantage demonstrated here: the physical ability to logarithmically compress high-dimensional data and extract the necessary guiding information without an exponential measurement cost. This is possible because the invariant measure appears to be a sufficiently rich representation of the essential physics of chaos. Capturing its key features provides the necessary guidance to unlock the observed gains in performance and, crucially, in memory efficiency. This efficiency is a direct, physical consequence of the quantum representation, and while our theoretical understanding of the measure's precise role is still developing, the empirical results provide compelling evidence for this paradigm's value.

Another notable point is that much of that research begins with a search for theoretical advantage, often struggling to connect with practical, ``useful" applications~\cite{huang2025vast}. In contrast, our investigation was born from a practical necessity: to solve the long-horizon prediction problem for chaotic systems, where leading classical methods fail. By showing that a 10-qubit quantum device can provide a unique resource, this work provides a compelling example of a practical quantum advantage and motivates the further development of both quantum and classical algorithms for scientific machine learning.

\subsection{Limitations, immediate and future impact of QIML}\label{sec:discussion}


To reduce the impact of hardware noise and enable stable training, the quantum generator is optimized separately and remains fixed during the training of the classical model. This architectural constraint arises from the limited scalability and noise of NISQ superconducting devices, which make full end-to-end quantum training infeasible at present. For near-term work, further improvements are possible within this offline training paradigm. While current Q-Priors are limited by qubit count, circuit depth, and hardware noise, deeper architectures with structured entanglement (e.g. hardware-efficient ans\"atze) may further improve expressivity for highly anisotropic distributions.  Besides, while this work demonstrates improvements in learning from digitally generated high-dimensional dynamical system data, we acknowledge the inherent limitations of the digital computing part of our method in fully capturing the true continuum behaviour of high-dimensional dynamical systems~\cite{coveney2024sharkovskii,klower2023periodic,boghosian2019new,coveney2025molecular}. Our ground truth here is, by definition, a 64-bit floating-point approximation. Future work will explore the use of analogue computation to provide a more direct comparison with the true invariant measures, thus offering a more rigorous validation of our approach against the ultimate ground truth.

Although there are limitations from NISQ hardware, in contrast to prevailing scepticism about the immediate applicability of quantum learning algorithms~\cite{cerezo2022challenges,jerbi2024shadows}, QIML demonstrates that even current quantum devices can contribute meaningfully to the modelling of high-dimensional dynamical systems. Without resorting to fault-tolerant quantum computing (FTQC) hardware, lightweight NISQ-based quantum modules enhance classical machine learning models by providing data-driven Q-Priors that efficiently capture complex statistical structures and non-local correlations—features that are otherwise challenging for classical methods to acquire with comparable memory efficiency. The resulting gains in statistical fidelity and long-horizon stability, coupled with a substantial reduction in data storage requirements, point to an immediate and practical role for near-term quantum resources in scientific machine learning. The compact 10- to 50-qubit Q-Prior presented here exemplifies how such modules can be effectively integrated today.

\begin{figure}[t]
  \centering
  \includegraphics[width=\linewidth]{ImpactofQIML.pdf}
  \caption[Conceptual roadmap of QIML]{
    \R{\textbf{Conceptual roadmap of the QIML paradigm.} 
    Stage~1 shows the current NISQ-era QIML framework in which an offline-trained quantum generator produces a memory-efficient Q-Prior that guides the classical model. 
    Stage~2 presents a hybrid co-learning QIML framework with shared gradients, structured entanglement, and noise-aware training. 
    Stage~3 outlines quantum-native machine learning with a quantum generative backbone and analogue quantum simulation that enable universal Q-Prior transfer for distributed communication. 
    The diagram highlights the immediate impact on scientific and engineering applications and future breakthroughs through compact, kB-scale Q-Prior transfer between supercomputers and quantum computers.}
  }
  \label{fig:qiml_roadmap}
\end{figure}

\R{Looking forward, the QIML architecture establishes a strategic roadmap bridging the NISQ era and the future of FTQC devices. In the near term, we identify a practical operational envelope for QIML involving approximately 50 qubits with circuit depths kept below 20 layers; this configuration balances the representational capacity required for complex flows against the coherence limits of current hardware. This defines our immediate trajectory for scaling up to high-dimensional geophysical datasets, specifically the European center for Medium-Range Weather Forecasts data, as delineated in Stage 2 of Fig.~\ref{fig:qiml_roadmap}. Beyond this, we envision a unified computational paradigm that integrates digital QIML with analogue quantum verification mechanisms to tackle industrial-scale turbulence. As illustrated in Stage 3 of Fig.~\ref{fig:qiml_roadmap}, this hybrid ecosystem, combining quantum generative priors with classical high-performance computing, serves as a blueprint for realizing practical quantum advantage in computational fluid dynamics and climate science.} 

Furthermore, this work opens up more exciting avenues for future research. A key question is how the framework performs with sparse or partial datasets, a common scenario in real-world applications where obtaining a complete ground-truth solution is impossible. Our current results are already promising in this regard, as the Q-Priors were successfully trained using only a subset of the available data (see section~\ref{sec:results}) while still providing effective guidance.
This suggests the potential for a more profound, out-of-distribution generalization~\cite{caro2023out}. For instance, future work could explore whether a single Q-Prior, trained on the invariant measure of one system, could successfully regularize a model for a different system that shares similar statistical properties but has different underlying dynamics. The successful demonstration of such a ``universal" prior would have significant implications, potentially allowing one Q-Prior to guide multiple machine learning models, further enhancing the practical value and efficiency of the QIML paradigm.

As quantum hardware continues to advance, in terms of qubit count, coherence time, and error mitigation, the prospect of replacing classical generative backbones with fully quantum circuits becomes increasingly viable. Also, this hybrid architecture lays the groundwork for a future paradigm of highly efficient information transfer within large-scale infrastructures. As quantum devices advance, one can envision a scenario where highly compressed Q-Priors, containing essential physical statistics, are transferred between compute nodes in place of voluminous classical datasets, which are riddled with spurious correlations~\cite{calude2017deluge}. This mechanism offers a scalable solution for future large-data training and provides a foundation for progressively more capable quantum-native machine learning models as the hardware matures. Thus, the parameter and memory efficiency demonstrated in this work suggest a pathway for Q-Priors to ultimately influence entire sections of large classical models.

\section{MATERIALS AND METHODS}
\subsection{Quantum-informed machine learning}\label{sec:framework}

\R{In this section,  we present the QIML architecture that leverages a quantum-classical hybrid approach for machine learning-based modelling of high-dimensional dynamical systems. Our QIML scheme is a specialized form of hybrid quantum–classical machine learning, which is illustrated in Fig.~\ref{fig:qimls} and Fig.~\ref{fig:qiml_trainingloop}.  Rather than combining quantum and classical modules through layered architectures or repeated variational optimisation, QIML employs a quantum generative model as a statistical prior, exploiting the structure of Hilbert space to constrain and regularise a classical predictor.} The workflow commences with the data generation stage (Fig.~\ref{fig:qimls}a), where we employ a numerical solver on a classical computer to simulate complex high-dimensional flows, producing the necessary training, validation, and testing datasets. Subsequently, in the quantum machine learning phase (Fig.~\ref{fig:qimls}b), a quantum generator is implemented on a quantum computer to learn the underlying physical invariant distribution, referred to as the Q-Prior. Finally, this Q-Prior is used to inform and regularize the classical machine learning model during training (Fig.~\ref{fig:qimls}c), guiding it toward long-term statistical consistency with the target dynamical system.  We begin by introducing the machine learning component, which is based on a Koopman operator formulation. The Koopman operator offers a linear representation of nonlinear dynamical systems by acting on observables instead of directly on the state variables. In this framework, the evolution of the system is captured not in the original state space, but in a lifted function space where the dynamics become linear. Consider a discrete-time dynamical system:

\begin{equation}
    x_{t+1} = f(x_t),
\end{equation}
where $x_t \in \mathbb{R}^n$ with $n$ denotes the dimension of the state space at time $t$, and $f$ is a generally nonlinear transition function. To enable auto-regressive prediction and long-term rollouts, we aim to find a mapping from the high-dimensional space to a latent space where a linear operator governs the time evolution. The measure-preserving property in the latent space is particularly important for capturing the statistical properties and invariant measure $\mu$ of high-dimensional dynamical systems.
The Koopman operator $K$ acts on observables $g: \mathbb{R}^n \to \mathbb{C}$ such that:

\begin{equation}
    (Kg)(x) = g(f(x)).
\end{equation}
While $K$ is linear, our implementations seek finite-dimensional approximations by learning a set of observables whose evolution under $K$ is closed and linear.
Next, we approximate the Koopman operator using a neural network; we define an encoder $\phi: \mathbb{R}^n \to \mathbb{R}^l$, with $l$ denotes the latent dimension, that maps the system state into a latent (feature) space, and a decoder $\psi: \mathbb{R}^l \to \mathbb{R}^n$ that reconstructs the original state. 
Unlike generic autoencoders primarily focused on data reconstruction, our Koopman operator neural network is specifically designed to learn a linear, norm-preserving evolution in the latent space, which is crucial for maintaining the long-term stability and physical fidelity of high-dimensional dynamical systems. 
Our detailed autoencoder-decoder architecture description can be found in Supplementary S7 for 1D and 2D cases. The dynamics in the latent space are modeled by a linear operator $K \in \mathbb{R}^{l \times l}$, such that:

\begin{equation}
    z_t = \phi(x_t), \quad z_{t+1} = K z_t, \quad \hat{x}_t = \psi(z_t),
\end{equation}
where $z_t$ is the encoded latent state and $\hat{x}_t$ is the reconstructed state. To preserve the long-term statistical behaviour of the system and approximate the spectral properties of a measure-preserving dynamical system, we constrain the operator $K$ to be unitary, i.e., $K^\top K = I$~\cite{petersen1989ergodic,budivsic2012applied,cheng2025learning}. This ensures that the Koopman dynamics in the latent space are norm-preserving and stable under long-term rollouts. In this work, unitarity is imposed by adding a regularization term to the loss function of the form $\mathcal{L}_{\text{unitary}} = \|K^\top K - I\|^2$. The Koopman operator $K$ evolves the system in a latent Hilbert space of encoded observables, enabling a linear and spectrally coherent representation of the dynamics. However, the preservation of long-term statistical properties in this latent space does not automatically guarantee consistency with the statistical structure of the original physical space. 

To address this, we integrate a sample-based quantum generator, which approximates the invariant distribution Q-Prior $p_\theta(x)$ of the system. This quantum-learned prior is used as a statistical regularizer and physically informed guidance during the classical machine learning training, guiding the model to produce predictions that are not only temporally coherent but also statistically consistent with the system's long-term behaviour (See Method C). This integration is achieved by incorporating the Q-Prior into a composite loss function that combines reconstruction fidelity with statistical alignment metrics, which is shown below. 

While the Koopman operator is introduced using a generic state variable, in this work, we focus on predicting physical states ${u}$. We use the Koopman operator to predict the next time frame of the physical states $\hat{{u}}_{t+1}$. The training of our model is guided by a composite loss function. We define its components below.

First, to ensure predictive accuracy, we use a standard reconstruction loss $\mathcal{L}_{\text{recon}}$ which measures the mean squared error between the predicted state $\hat{{u}}_{t+1}$ and the ground truth ${u}_{t+1}$:
\begin{equation}
\mathcal{L}_{\text{recon}} = \left\| \hat{{u}}_{t+1} - {u}_{t+1} \right\|^2.
\end{equation}
Second, we include the unitary regularization term $\mathcal{L}_{\text{unitary}}$ as previously discussed:
\begin{equation}
\mathcal{L}_{\text{unitary}} = \left\| K^{\top}K - I \right\|^2_F,
\end{equation}
where $I$ is the identity matrix and $\|\cdot\|_F$ denotes the Frobenius norm.
Finally, we incorporate the Q-prior, $p_{\theta}(x)$, as a statistical constraint. This is achieved through two complementary metrics. The Kullback-Leibler (KL) divergence $\mathcal{L}_{\text{KL}}$ captures low-order, information-theoretic discrepancies between the empirical distribution of predicted states $\hat{q}(x)$, and the Q-Prior:
\begin{equation}
\mathcal{L}_{\text{KL}} = D_{\text{KL}}\left( \hat{q}(x) \,\|\, p_{\theta}(x) \right).
\end{equation}
To align higher-order statistical moments, we also include an MMD term $\mathcal{L}_{\text{MMD}}$:
\begin{equation}
\mathcal{L}_{\text{MMD}} = \left\| \mathbb{E}_{x \sim \hat{q}(x)}[\phi(x)] - \mathbb{E}_{x \sim p_{\theta}(x)}[\phi(x)] \right\|^2_{\mathcal{H}}.
\end{equation}

These individual components are combined into a single training objective, where the Q-Prior terms are weighted by empirically selected hyperparameters $\lambda_{\text{KL}}$ and $\lambda_{\text{MMD}}$:
\begin{equation}
\mathcal{L}_{\text{total}} = \mathcal{L}_{\text{recon}} + \mathcal{L}_{\text{unitary}} + \lambda_{\text{KL}} \mathcal{L}_{\text{KL}} + \lambda_{\text{MMD}} \mathcal{L}_{\text{MMD}}.
\end{equation}
The full QIML framework is demonstrated in Algorithm~\ref{alg:surrogate}. All models are implemented in PyTorch with fully differentiable integration of Q-Priors. Architectural details for predicting 1D and 2D systems are provided in Supplementary S7.

\begin{algorithm}[H]
\caption{Training of the Quantum-Informed Machine Learning Model}
\label{alg:surrogate}
\begin{algorithmic}[1]
\Require Dataset $\{{u}_{t}, {u}_{t+1}\}$, Q-Prior distribution $p_\theta(x)$
\State Initialise machine learning model parameters $\phi$
\State Set loss weights $\lambda_{\text{KL}}, \lambda_{\text{MMD}}$ 
\For{each training epoch}
    \For{each mini-batch $({u}_{t}, {u}_{t+1})$}
        \State Predict next-step field: $\hat{{u}}_{t+1} \gets f_\phi({u}_{t}, \dots)$
        \State Compute empirical distribution $\hat{q}(x)$ from $\hat{{u}}_{t+1}$
        \State Compute reconstruction loss: 
        \[
        \mathcal{L}_{\text{recon}} \gets \|\hat{{u}}_{t+1} - {u}_{t+1}\|^2
        \]
        \State Compute unitary regularization loss:
        \[
        \mathcal{L}_{\text{unitary}} \gets \|K^{\top}K - I\|^2_F
        \]
        \State Compute KL divergence: 
        \[
        \mathcal{L}_{\text{KL}} \gets D_{\text{KL}}(\hat{q}(x) \,\|\, p_\theta(x))
        \]
        \State Compute MMD loss: 
        \[
        \mathcal{L}_{\text{MMD}} \gets \left\|\mathbb{E}_{x \sim \hat{q}(x)}[\phi(x)] - \mathbb{E}_{x \sim p_\theta(x)}[\phi(x)]\right\|_{\mathcal{H}}^2
        \]
        \State Compute total loss: 
        \[
            \mathcal{L}_{\text{total}} = \mathcal{L}_{\text{recon}} + \mathcal{L}_{\text{unitary}} + \lambda_{\text{KL}} \mathcal{L}_{\text{KL}} + \lambda_{\text{MMD}} \mathcal{L}_{\text{MMD}}
        \]
        \State Backpropagate and update $\phi$ via Adam optimizer
    \EndFor
\EndFor
\State \Return Trained model $f_\phi$
\end{algorithmic}
\end{algorithm}


\subsection{Invariant measure for high-dimensional dynamical systems}
The concept of an invariant measure originates from the field of ergodic theory and has become a fundamental tool in the study of dynamical systems. First formalized in the context of measure-preserving transformations by Birkhoff and von Neumann in the early 20th century, invariant measures capture the long-term statistical behaviour of systems evolving under deterministic rules \cite{birkhoff1931proof,neumann1932proof}.

An invariant measure describes how a dynamical system spends its time in different regions of its state space over the long term. For a system evolving under a map $T$, the invariant measure $\mu(x)$ satisfies:
\begin{equation}
    \mu(T^{-1}(A)) = \mu(A),
\end{equation}
which means that the total probability of the system being in a region $A$ remains the same after one step of evolution. This property reflects a kind of statistical balance: although individual trajectories may change, the overall distribution stays stable. In our work, we aim to capture this long-term behaviour by ensuring that the model’s predictions remain consistent with a reference distribution, which we learn using a quantum generator. 

In turbulent flows, such invariant measures typically manifest as conserved distributions, such as velocity probability, turbulent kinetic energy, or energy spectra that encode physical long-term steady states. Capturing these invariant features is essential for ensuring robustness in extrapolative prediction regimes, where a classical machine learning method often suffers from gradient explosion, mode collapse, or long-term error accumulation~\cite{pascanu2013difficulty}.

\subsection{The quantum generator to learn the Q-Prior}\label{sec:method-qcbm}

\begin{figure}[t]
\centering
\includegraphics[width=0.85\textwidth]{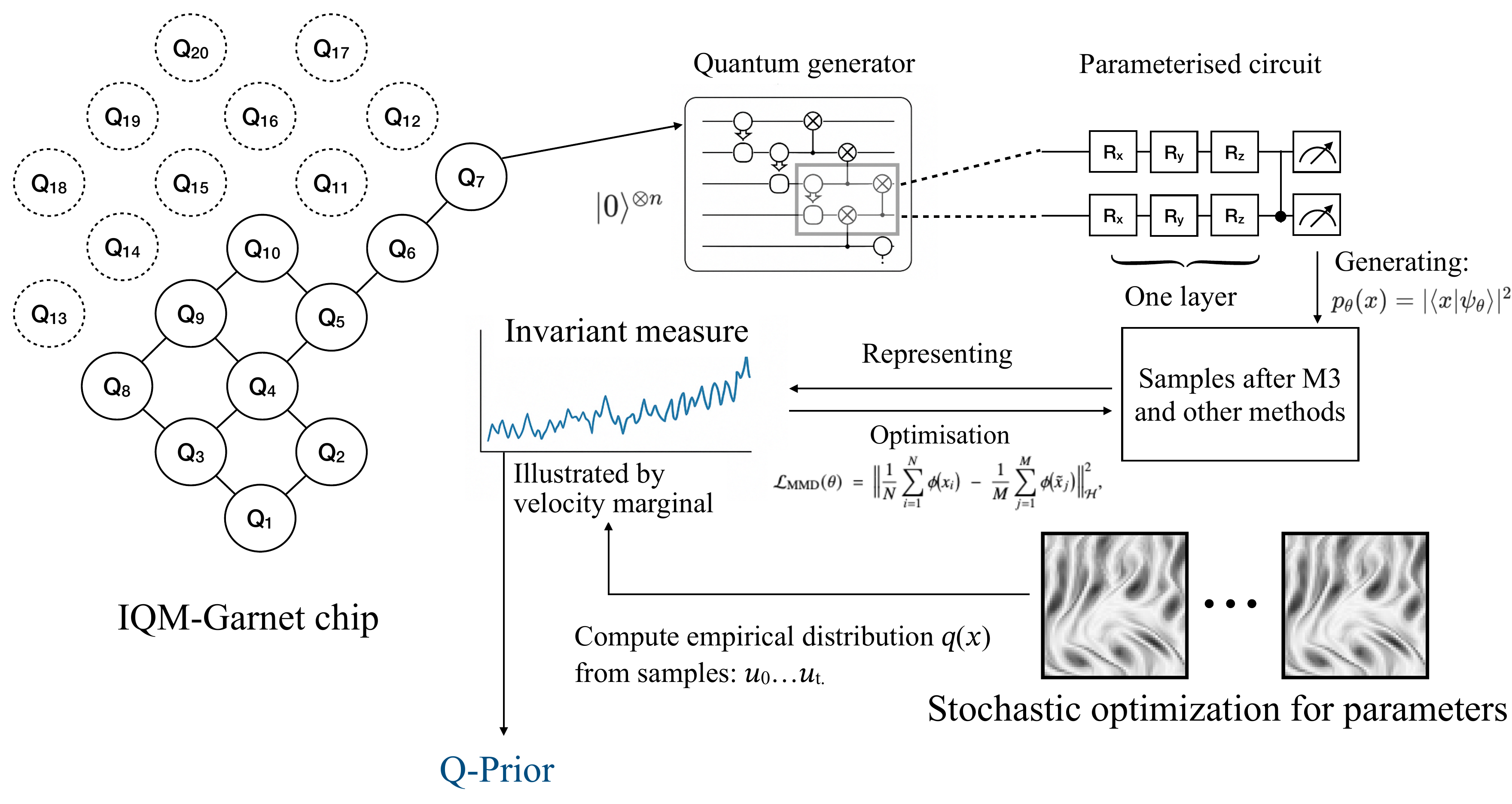}
\caption{\R{\textbf{Schematic illustration of the Q-Prior generation within the QIML framework.} The quantum circuit functions as a Born machine, initialized in a fixed vacuum state $|0\rangle^{\otimes n}$. No classical data is encoded into the input state amplitudes. The circuit $U(\theta)$ evolves the state to generate a wavefunction $|\psi_\theta\rangle$, and the measurement statistics in the computational basis, $p_\theta(x) = |\langle x|\psi_\theta\rangle|^2$, form the model distribution. The classical turbulent data samples serve solely as the target for calculating the loss, driving the optimisation of parameters $\theta$ to $\theta^*$. }}
\label{fig:circuit}

\end{figure}

We develop a quantum-informed learning component in which a quantum generator is used to directly learn invariant statistical properties from high-dimensional physical observations. The schematic illustration of the Q-Prior generation is shown in Fig.~\ref{fig:circuit}. \R{The generator defines a parameterized quantum state $\ket{\psi_\theta}$ on an $n$-qubit register, which serves as a generative model over a discretised spatial domain. It does not encode a specific state (e.g., a velocity field) into the circuit. Instead, by observing many snapshots of the system, it learns to reproduce its underlying invariant probability distribution.
Specifically, each computational basis state $x \in {\{0,1\}}^n$ here is mapped bijectively to a spatial coordinate in the simulation grid. The corresponding amplitude $\langle x | \psi_\theta \rangle$ is optimized so that its squared magnitude approximates the normalized velocity magnitude at that point.
Although the circuit does not encode classical inputs directly via amplitude encoding, it constructs a quantum state whose Born distribution approximates the invariant measure Q-Prior of the system. The resulting Born distribution is given by:
\begin{equation}
p_\theta(x) = \left| \left\langle x \middle| U(\theta) \middle| 0 \right\rangle^{\otimes n} \right|^2,
\end{equation}
where $U(\theta)$ is a parameterized unitary circuit composed of L layers, each consisting of single-qubit rotations and entangling operations.  Each computational basis state represents a discrete spatial point, and the quantum amplitudes are trained to reflect the empirical distribution of local velocity magnitudes. This provides a compact, data-driven mapping from physical observables into the Hilbert space. The exponential size of this space, combined with the use of entangling gates, provides a natural mechanism to encode the complex, non-local correlations inherent in chaotic systems. This feature provides a quantum advantage rooted in the ability of quantum systems to represent high-dimensional classical data with exponential efficiency~\cite{huang2025vast}. This enables it to learn the required statistical properties using exponentially fewer parameters than would be required by a classical model~\cite{wang2025parameter}. Furthermore, by leveraging the vast Hilbert space and entanglement, the quantum circuit can capture the complex, non-local correlations inherent in such systems efficiently~\cite{huang2022quantum}.
Further details can be found in section~\ref{sec:advantage} and Supplementary S1.}

As shown in Algorithm~\ref{alg:qcbm}, the quantum circuit is trained as a quantum generative model.  Each computational basis state
$x \in \{0,1\}^{n}$ is mapped uniquely to a spatial location
$\mathbf{r}_{x}$ in the discretised velocity field, and its target value
is the corresponding velocity magnitude $u(\mathbf{r}_{x})$.
Because the underlying probability distribution over spatial locations
is unknown, we adopt a sample–based training strategy that minimises the MMD loss:

\begin{equation}
  \mathcal{L}_{\text{MMD}}(\theta) \;=\;
  \Bigl\|
    \frac{1}{N}\sum_{i=1}^{N}\phi\!\bigl(x_{i}\bigr)
    \;-\;
    \frac{1}{M}\sum_{j=1}^{M}\phi\!\bigl(\tilde{x}_{j}\bigr)
  \Bigr\|_{\mathcal{H}}^{2},
  \label{eq:mmd_loss}
\end{equation}

\vspace{0.5em}
\noindent
Here, $\{x_i\}_{i=1}^{N}$ denotes the batch of empirical samples
drawn from the reference velocity field, while
$\{\tilde{x}_j\}_{j=1}^{M}\!\sim p_{\theta}(x)$ are samples generated by the
quantum circuits.  Because the MMD objective is expressed solely through kernel evaluations on finite sample sets, optimisation of the generator requires no closed-form expression for the target probability density. This sample-based objective is particularly advantageous for high-dimensional chaotic flows, where the invariant measure is available only as a time-ordered set of simulation snapshots rather than an explicit probability density.  By aligning the quantum generator’s samples with these numerical trajectories, the model infers the underlying measure without ever needing a closed-form expression.

The trained quantum distribution $p_\theta(x)$ thus can serve as a data-driven prior capturing the invariant measure of the underlying physical process. Unlike classical regularizers based on fixed heuristics, the quantum generator constructs this Q-Prior through optimisation on empirical data. 

All quantum circuits are implemented using hardware-efficient structures, which are detailed in Supplementary S2. Training is carried out on noiseless simulators and on superconducting quantum hardware, with circuit transpilation, error mitigation, and device-specific calibration performed using the Qiskit runtime. Further details and the optimisation methods are provided in Supplementary~S1 and S2.

\begin{algorithm}[H]
  \caption{Training of the quantum generator for invariant measure}
  \label{alg:qcbm}
  \begin{algorithmic}[1]
    \Require Velocity snapshots $\{{u}_{t}\}$; circuit depth $L$; qubits $n$
    \Ensure  optimized parameters $\theta^{\ast}$ such that
             $p_{\theta^{\ast}}(x)$ matches the empirical distribution
    \State Initialise $\theta \sim \mathcal{U}[-\pi,\pi]$
    \For{\textbf{each} training epoch}
        \State Build unitary $U(\theta)$ (depth $L$, $n$ qubits)
        \State Generate $M$ samples
              $\{\tilde{x}_{j}\}_{j=1}^{M}\!\sim p_{\theta}(x)$
        \State Extract $N$ target samples
              $\{x_{i}\}_{i=1}^{N}$ from $\{{u}_{t}\}$
        \State Compute
              $\mathcal{L}_{\text{MMD}}\!\bigl(\{x_{i}\},\{\tilde{x}_{j}\}\bigr)$
        \State $\theta \;\leftarrow\; \theta
                - \eta\,\nabla_{\theta}\mathcal{L}_{\text{MMD}}$
    \EndFor
    \State \Return $\theta^{\ast}$
  \end{algorithmic}
\end{algorithm}


\subsection{Quantum hardware implementation }
The sample-based quantum generator model was trained using both emulators on classical computers and real superconducting quantum processors.  A classical emulator here is a program running on a conventional computer that simulates the mathematical operations of an ideal, noise-free quantum device. In this work, these emulations were supported by the PennyLane and Qiskit software packages. The practical constraints of current NISQ-era hardware, which require cryogenic operating temperatures and are subject to frequent maintenance and recalibration, make continuous, large-scale training across numerous datasets challenging. For this reason, we adopted a strategic approach here. The first two systems (Kuramoto-Sivashinsky and Kolmogorov flow) were studied on the classical emulator to establish the Q-Prior's efficacy in an ideal, noise-free setting. We then reserved the use of the real quantum hardware for the most complex and challenging dataset, which is the turbulent channel flow, precisely because this is the case where leading classical methods fail, providing the most stringent test for the QIML framework's practical utility and robustness under realistic noise conditions. To ensure effective execution on near-term quantum hardware and to manage the impact of noise on the quality of Q-Priors, this study constrains the total number of trainable parameters and adopts conservative circuit configurations. Specifically, the number of qubits is 10–15 on the emulator and 10 on the hardware, with circuit depths ranging from 3 to 8 layers, resulting in fewer than 300 trainable parameters in total (see Table S3 in Supplementary Information). Hardware experiments were mainly conducted on IQM’s 20-qubit superconducting chip (Garnet)~\cite{abdurakhimov2024technology}, using qubits 1–10 as the active subset, supported by Qiskit and IQM-Qiskit. Further discussion on hardware implementation and sensitivity is provided in Supplementary S2. To ensure scalability, the quantum emulation and hardware backend were integrated into a GPU-accelerated classical training loop on NVIDIA A100 nodes provided by the BEAST GPU cluster (Leibniz Supercomputing Centre). Gradient evaluations, loss computations, and optimizer updates were performed classically and synchronised with the quantum hardware at each iteration. This setup constitutes an applied demonstration of embedding quantum devices into scientific machine learning pipelines. 
The employed quantum circuit architecture consisted of alternating layers of parameterized $R_x$ rotations and hardware-native entangling gates. Importantly, the same circuit structure was used across all experiments without dataset-specific tailoring, ensuring general applicability. Although performance could potentially be improved through customised ansatz design, our results demonstrate that the proposed quantum-informed framework is compatible with any hardware-efficient quantum circuit. This adaptability enables the integration of diverse circuit architectures without modifying the overall training pipeline, suggesting broad utility across problem domains and quantum backends.

\R{The Q-prior is generated via a one-time, offline training process. This training consists of a total of 50 epochs. Following the completion of this training, the quantum circuit is not utilized during the main classical model's training loop. For each of the 50 training epochs, the circuit was sampled with 20,000 shots to ensure a high-fidelity statistical distribution. Readout errors were mitigated using IQM’s native transpiler and the M3 measurement mitigation toolkit~\cite{nation2021scalable}. While we used this number of shots to ensure maximum stability and demonstrate the framework's potential, this is not a strict requirement for its effectiveness. In separate tests, we reduced the measurement count to 8,000 shots and found the impact on the final model's performance was less than 5\%, indicating that the Q-Prior remains effective. Nevertheless, to present the best possible outcomes in this work, all results and figures are based on data obtained from 20,000 shots per iteration. optimisation was performed using L-BFGS~\cite{zhu1997algorithm}, Adam~\cite{zhang2018improved}, and COBYLA~\cite{regis2011stochastic}, selected based on convergence stability and noise robustness. To prevent mode collapse, such as excessive sampling concentration on the $\ket{0}$ state, runtime monitoring and redundancy checks were incorporated into the training loop.  Despite operating with only 10 to 15 qubits, less than 20 circuit depth, and 300 quantum gates, the Q-Prior successfully captured the structure of invariant distributions over a $1000$- to $60,000$-dimensional Hilbert space, demonstrating the effectiveness of quantum priors in modelling high-dimensional systems. To contextualize the practical feasibility of our approach, Table \ref{tab:quantum_resources} compares the resource budget of the QIML framework against established quantum algorithms such as VQE and HHL, highlighting its significantly lower requirements for circuit depth and measurement overhead. Further discussions of the error mitigations, experiment results, and ablations are provided in Supplementary S2 and S7.}

\clearpage
\bibliography{science_qiml} 

@article{zhong2020jiuzhang,
author = {Han-Sen Zhong  and Hui Wang  and Yu-Hao Deng  and Ming-Cheng Chen  and Li-Chao Peng  and Yi-Han Luo  and Jian Qin  and Dian Wu  and Xing Ding  and Yi Hu  and Peng Hu  and Xiao-Yan Yang  and Wei-Jun Zhang  and Hao Li  and Yuxuan Li  and Xiao Jiang  and Lin Gan  and Guangwen Yang  and Lixing You  and Zhen Wang  and Li Li  and Nai-Le Liu  and Chao-Yang Lu  and Jian-Wei Pan },
title = {Quantum computational advantage using photons},
journal = {Science},
volume = {370},
number = {6523},
pages = {1460-1463},
year = {2020},
doi = {https://doi.org/10.1126/science.abe8770},
eprint = {https://www.science.org/doi/pdf/10.1126/science.abe8770},
}

@article{kubavr2023hybrid,
  title={Hybrid quantum mechanical/molecular mechanical methods for studying energy transduction in biomolecular machines},
  author={Kuba{\v{r}}, T and Elstner, Marcus and Cui, Qiang},
  journal={Annu. Rev. Biophys.},
  volume={52},
  number={1},
  pages={525--551},
  year={2023},
  publisher={Annual Reviews}
}

@article{barron2002universal,
  title={Universal approximation bounds for superpositions of a sigmoidal function},
  author={Barron, Andrew R},
  journal={{IEEE Trans. Inf. Theory}},
  volume={39},
  number={3},
  pages={930--945},
  year={2002},
  publisher={IEEE}
}

@book{Goodfellow2016,
  author    = {Goodfellow, I. and Bengio, Y. and Courville, A.},
  title     = {Deep Learning},
  publisher = {MIT Press},
  year      = {2016},
  chapter   = {6}
}

@article{bickley2025extending,
  title={Extending quantum computing through subspace{,} embedding and classical molecular dynamics techniques},
  author={Bickley, Thomas M. and Mingare, Angus and Weaving, Tim and Williams de la Bastida, Michael and Wan, Shunzhou and Nibbi, Martina and Seitz, Philipp and Ralli, Alexis and Love, Peter J. and Chung, Minh and Hernández Vera, Mario and Schulz, Laura and Coveney, Peter V.},
  journal={Digit. Discov.},
  year={2025},
  volume  ="4",
  issue  ="12",
  pages  ="3427-3444",
  publisher  ="RSC",
  doi="https://doi.org/10.1039/D5DD00225G"
}

@article{Preskill_2018,

	year = {2018},
  
	publisher = {Verein zur Forderung des Open Access Publizierens in den Quantenwissenschaften},
  
	volume = {2},
  
	pages = {79},
  
	author = {John Preskill},
  
	title = {Quantum Computing in the {NISQ} era and beyond},
  
	journal = {Quantum}
}

@article{abdurakhimov2024technology,
  title={Technology and Performance Benchmarks of IQM's 20-Qubit Quantum Computer},
  author={Abdurakhimov, Leonid and Adam, Janos and Ahmad, Hasnain and Ahonen, Olli and Algaba, Manuel and Alonso, Guillermo and Bergholm, Ville and Beriwal, Rohit and Beuerle, Matthias and Bockstiegel, Clinton and others},
  journal={arXiv:2408.12433},
  year={2024},
  doi="https://doi.org/10.48550/arXiv.2408.12433"
}

@article{regis2011stochastic,
  title={Stochastic radial basis function algorithms for large-scale optimization involving expensive black-box objective and constraint functions},
  author={Regis, Rommel G},
  journal={Comput. Oper. Res.},
  volume={38},
  number={5},
  pages={837--853},
  year={2011},
  publisher={Elsevier}
}

@article{gao2022enhancing,
  title={Enhancing generative models via quantum correlations},
  author={Gao, Xun and Anschuetz, Eric R and Wang, Sheng-Tao and Cirac, J Ignacio and Lukin, Mikhail D},
  journal={Phys. Rev. X},
  volume={12},
  number={2},
  pages={021037},
  year={2022},
  publisher={APS}
}

@article{arute2019quantum,
  title={Quantum supremacy using a programmable superconducting processor},
  author={Arute, Frank and Arya, Kunal and Babbush, Ryan and Bacon, Dave and Bardin, Joseph C and Barends, Rami and Biswas, Rupak and Boixo, Sergio and Brandao, Fernando GSL and Buell, David A and others},
  journal={Nature},
  volume={574},
  number={7779},
  pages={505--510},
  year={2019},
  publisher={Nature Publishing Group UK London},
  doi="https://doi.org/10.1038/s41586-019-1666-5"
}

@article{cao2019quantum,
  title={Quantum chemistry in the age of quantum computing},
  author={Cao, Yudong and Romero, Jonathan and Olson, Jonathan P and Degroote, Matthias and Johnson, Peter D and Kieferov{\'a}, M{\'a}ria and Kivlichan, Ian D and Menke, Tim and Peropadre, Borja and Sawaya, Nicolas PD and others},
  journal={Chem. Rev.},
  volume={119},
  number={19},
  pages={10856--10915},
  year={2019},
  publisher={ACS Publications},
  doi="https://doi.org/10.1021/acs.chemrev.8b00803"
}

@article{bauer2020quantum,
  title={Quantum algorithms for quantum chemistry and quantum materials science},
  author={Bauer, Bela and Bravyi, Sergey and Motta, Mario and Chan, Garnet Kin-Lic},
  journal={Chem. Rev.},
  volume={120},
  number={22},
  pages={12685--12717},
  year={2020},
  publisher={ACS Publications}
}

@article{sanavio2024lattice,
  title={Lattice Boltzmann--Carleman quantum algorithm and circuit for fluid flows at moderate Reynolds number},
  author={Sanavio, Claudio and Succi, Sauro},
  journal={AVS Quantum Sci.},
  volume={6},
  number={2},
  year={2024},
  publisher={AIP Publishing}
}

@article{sanavio2024three,
  title={Three Carleman routes to the quantum simulation of classical fluids},
  author={Sanavio, Claudio and Scatamacchia, Riccardo and De Falco, Carlo and Succi, Sauro},
  journal={Phys. Fluids},
  volume={36},
  number={5},
  year={2024},
  publisher={AIP Publishing}
}

@article{succi2025physical,
  title={On the physical interpretation of neural PDEs},
  author={Succi, Sauro},
  journal={Math. Mech. Complex Syst.},
  volume={13},
  number={3},
  pages={275--286},
  year={2025},
  publisher={Mathematical Sciences Publishers}
}

@article{kingma2013auto,
  title={Auto-encoding variational bayes},
  author={Kingma, Diederik P and Welling, Max},
  journal={arXiv:1312.6114},
  year={2013}
}

@article{birkhoff1931proof,
  title={Proof of the ergodic theorem},
  author={Birkhoff, George D},
  journal={Proc. Natl. Acad. Sci. U.S.A.},
  volume={17},
  number={12},
  pages={656--660},
  year={1931},
  publisher={National Academy of Sciences}
}

@article{eckmann1985ergodic,
  title={Ergodic theory of chaos and strange attractors},
  author={Eckmann, J-P and Ruelle, David},
  journal={Rev. Mod. Phys.},
  volume={57},
  number={3},
  pages={617},
  year={1985},
  publisher={APS}
}

@article{farmer1983dimension,
  title={The dimension of chaotic attractors},
  author={Farmer, J Doyne and Ott, Edward and Yorke, James A},
  journal={Physica D},
  volume={7},
  number={1-3},
  pages={153--180},
  year={1983},
  publisher={Elsevier}
}

@article{neumann1932proof,
  title={Proof of the quasi-ergodic hypothesis},
  author={Neumann, J v},
  journal={Proc. Natl. Acad. Sci. U.S.A.},
  volume={18},
  number={1},
  pages={70--82},
  year={1932},
  publisher={National Academy of Sciences}
}

@article{romero2017quantum16,
  title={Quantum autoencoders for efficient compression of quantum data},
  author={Romero, Jonathan and Olson, Jonathan P and Aspuru-Guzik, Alan},
  journal={Quantum Sci. Technol.},
  volume={2},
  number={4},
  pages={045001},
  year={2017},
  publisher={IOP Publishing}
}

@article{lamata2018quantum,
  title={Quantum autoencoders via quantum adders with genetic algorithms},
  author={Lamata, Lucas and Alvarez-Rodriguez, Unai and Mart{\'\i}n-Guerrero, Jos{\'e} David and Sanz, Mikel and Solano, Enrique},
  journal={Quantum Sci. Technol.},
  volume={4},
  number={1},
  pages={014007},
  year={2018},
  publisher={IOP Publishing}
}

@article{18ding2019experimental,
  title={Experimental implementation of a quantum autoencoder via quantum adders},
  author={Ding, Yongcheng and Lamata, Lucas and Sanz, Mikel and Chen, Xi and Solano, Enrique},
  journal={Adv. Quantum Technol.},
  volume={2},
  number={7-8},
  pages={1800065},
  year={2019},
  publisher={Wiley Online Library}
}

@article{19kieferova2017tomography,
  title={Tomography and generative training with quantum Boltzmann machines},
  author={Kieferov{\'a}, M{\'a}ria and Wiebe, Nathan},
  journal={Phys. Rev. A},
  volume={96},
  number={6},
  pages={062327},
  year={2017},
  publisher={APS}
}

@article{20jain2020quantum,
  title={Quantum and classical machine learning for the classification of non-small-cell lung cancer patients},
  author={Jain, Siddhant and Ziauddin, Jalal and Leonchyk, Paul and Yenkanchi, Shashibushan and Geraci, Joseph},
  journal={SN Appl. Sci.},
  volume={2},
  number={6},
  pages={1--10},
  year={2020},
  publisher={Springer}
}

@article{21dallaire2018quantum,
  title={Quantum generative adversarial networks},
  author={Dallaire-Demers, Pierre-Luc and Killoran, Nathan},
  journal={Phys. Rev. A},
  volume={98},
  number={1},
  pages={012324},
  year={2018},
  publisher={APS}
}

@article{23romero2021variational,
  title={Variational quantum generators: Generative adversarial quantum machine learning for continuous distributions},
  author={Romero, Jonathan and Aspuru-Guzik, Al{\'a}n},
  journal={Adv. Quantum Technol.},
  volume={4},
  number={1},
  pages={2000003},
  year={2021},
  publisher={Wiley Online Library}
}

@article{24zeng2019learning,
  title={Learning and inference on generative adversarial quantum circuits},
  author={Zeng, Jinfeng and Wu, Yufeng and Liu, Jin-Guo and Wang, Lei and Hu, Jiangping},
  journal={Phys. Rev. A},
  volume={99},
  number={5},
  pages={052306},
  year={2019},
  publisher={APS}
}

@article{25rebentrost2014quantum,
  title={Quantum support vector machine for big data classification},
  author={Rebentrost, Patrick and Mohseni, Masoud and Lloyd, Seth},
  journal={Phys. Rev. Lett.},
  volume={113},
  number={13},
  pages={130503},
  year={2014},
  publisher={APS}
}

@article{27mengoni2019kernel,
  title={Kernel methods in quantum machine learning},
  author={Mengoni, Riccardo and Di Pierro, Alessandra},
  journal={Quantum Mach. Intell.},
  volume={1},
  number={3},
  pages={65--71},
  year={2019},
  publisher={Springer}
}

@article{gretton2012kernel,
  title={A kernel two-sample test},
  author={Gretton, Arthur and Borgwardt, Karsten M and Rasch, Malte J and Sch{\"o}lkopf, Bernhard and Smola, Alexander},
  journal={J. Mach. Learn. Res.},
  volume={13},
  number={1},
  pages={723--773},
  year={2012},
  publisher={JMLR. org}
}

@article{huang2022quantum,
  title={Quantum advantage in learning from experiments},
  author={Huang, Hsin-Yuan and Broughton, Michael and Cotler, Jordan and Chen, Sitan and Li, Jerry and Mohseni, Masoud and Neven, Hartmut and Babbush, Ryan and Kueng, Richard and Preskill, John and others},
  journal={Science},
  volume={376},
  number={6598},
  pages={1182--1186},
  year={2022},
  publisher={American Association for the Advancement of Science},
  doi="https://doi.org/10.1126/science.abn7293"
}

@article{28huang2021power,
  title={Power of data in quantum machine learning},
  author={Huang, Hsin-Yuan and Broughton, Michael and Mohseni, Masoud and Babbush, Ryan and Boixo, Sergio and Neven, Hartmut and McClean, Jarrod R},
  journal={{Nat. Commun.}},
  volume={12},
  number={1},
  pages={1--9},
  year={2021},
  publisher={Nature Publishing Group},
  doi="https://doi.org/10.1038/s41467-021-22539-9"
}

@article{gao2018quantum,
  title={A quantum machine learning algorithm based on generative models},
  author={Gao, Xun and Zhang, Z-Y and Duan, L-M},
  journal={Sci. Adv.},
  volume={4},
  number={12},
  pages={eaat9004},
  year={2018},
  publisher={American Association for the Advancement of Science}
}

@article{tennie2025quantum,
  title={Quantum computing for nonlinear differential equations and turbulence},
  author={Tennie, Felix and Laizet, Sylvain and Lloyd, Seth and Magri, Luca},
  journal={Nat. Rev. Phys.},
  volume={7},
  number={4},
  pages={220--230},
  year={2025},
  publisher={Nature Publishing Group UK London}
}

@article{carrasquilla2019reconstructing,
  title={Reconstructing quantum states with generative models},
  author={Carrasquilla, Juan and Torlai, Giacomo and Melko, Roger G and Aolita, Leandro},
  journal={Nat. Mach. Intell.},
  volume={1},
  number={3},
  pages={155--161},
  year={2019},
  publisher={Nature Publishing Group UK London}
}

@article{zhu1997algorithm,
  title={Algorithm 778: L-BFGS-B: Fortran subroutines for large-scale bound-constrained optimization},
  author={Zhu, Ciyou and Byrd, Richard H and Lu, Peihuang and Nocedal, Jorge},
  journal={ACM Trans. Math. Softw.},
  volume={23},
  number={4},
  pages={550--560},
  year={1997},
  publisher={ACM New York, NY, USA}
}

@article{edeling2024global,
  title={Global ranking of the sensitivity of interaction potential contributions within classical molecular dynamics force fields},
  author={Edeling, Wouter and Vassaux, Maxime and Yang, Yiming and Wan, Shunzhou and Guillas, Serge and Coveney, Peter V},
  journal={npj Comput. Mater.},
  volume={10},
  number={1},
  pages={87},
  year={2024},
  publisher={Nature Publishing Group UK London},
  doi="https://doi.org/10.1038/s41524-024-01272-z"
}

@article{nation2021scalable,
  title={Scalable mitigation of measurement errors on quantum computers},
  author={Nation, Paul D and Kang, Hwajung and Sundaresan, Neereja and Gambetta, Jay M},
  journal={PRX Quantum},
  volume={2},
  number={4},
  pages={040326},
  year={2021},
  publisher={APS}
}

@article{liu2018differentiable,
  title={Differentiable learning of quantum circuit born machines},
  author={Liu, Jin-Guo and Wang, Lei},
  journal={Phys. Rev. A},
  volume={98},
  number={6},
  pages={062324},
  year={2018},
  publisher={APS}
}

@article{cerezo2022challenges,
  title={Challenges and opportunities in quantum machine learning},
  author={Cerezo, Marco and Verdon, Guillaume and Huang, Hsin-Yuan and Cincio, Lukasz and Coles, Patrick J},
  journal={Nat. Comput. Sci.},
  volume={2},
  number={9},
  pages={567--576},
  year={2022},
  publisher={Nature Publishing Group US New York}
}

@article{wang2025parameter,
  title = {Parameter-efficient quantum anomaly detection method on a superconducting quantum processor},
  author = {Wang, Maida and Jiang, Jinyang and Coveney, Peter V.},
  journal = {Phys. Rev. Res.},
  volume = {7},
  issue = {4},
  pages = {043094},
  numpages = {24},
  year = {2025},
  month = {Oct},
  publisher = {American Physical Society},
  
}

@article{Farmer1983,
title = {Information Dimension and the Probabilistic Structure of Chaos},
author = {J. Doyne Farmer},
pages = {1304--1326},
volume = {37},
number = {11},
journal = {Z. Naturforsch. A},
doi = {https://doi.org/10.1515/zna-1982-1117},
year = {1982},
}

@ARTICLE{Vapnik1998,
  author={Vapnik, V.N.},
  journal={IEEE Trans. Neural Netw.}, 
  title={An overview of statistical learning theory}, 
  year={1999},
  volume={10},
  number={5},
  pages={988-999},
}

@article{Gretton2012,
author = {Gretton, Arthur and Borgwardt, Karsten M. and Rasch, Malte J. and Sch\"{o}lkopf, Bernhard and Smola, Alexander},
title = {A kernel two-sample test},
year = {2012},
issue_date = {3/1/2012},
publisher = {JMLR.org},
volume = {13},
number = {1},
issn = {1532-4435},
journal = {J. Mach. Learn. Res.},
pages = {723–773},
numpages = {51},
}

@article{stokes2020quantum,
  title={Quantum natural gradient},
  author={Stokes, James and Izaac, Josh and Killoran, Nathan and Carleo, Giuseppe},
  journal={Quantum},
  volume={4},
  pages={269},
  year={2020},
  publisher={Verein zur F{\"o}rderung des Open Access Publizierens in den Quantenwissenschaften}
}

@article{kandala2017hardware,
  title={Hardware-efficient variational quantum eigensolver for small molecules and quantum magnets},
  author={Kandala, Abhinav and Mezzacapo, Antonio and Temme, Kristan and Takita, Maika and Brink, Markus and Chow, Jerry M and Gambetta, Jay M},
  journal={Nature},
  volume={549},
  number={7671},
  pages={242--246},
  year={2017},
  publisher={Nature Publishing Group},
  doi="https://doi.org/10.1038/nature23879"
}

@book{coveney2025molecular,
  title={Molecular Dynamics: Probability and Uncertainty},
  author={Coveney, Peter V and Wan, Shunzhou},
  year={2025},
  publisher={Oxford University Press}
}

@article{calude2017deluge,
  title={The deluge of spurious correlations in big data},
  author={Calude, Cristian S and Longo, Giuseppe},
  journal={Found. Sci.},
  volume={22},
  number={3},
  pages={595--612},
  year={2017},
  publisher={Springer}
}

@article{benedetti2019generative,
  title={A generative modeling approach for benchmarking and training shallow quantum circuits},
  author={Benedetti, Marcello and Garcia-Pintos, Delfina and Perdomo, Oscar and Leyton-Ortega, Vicente and Nam, Yunseong and Perdomo-Ortiz, Alejandro},
  journal={npj Quantum Inf.},
  volume={5},
  number={1},
  pages={45},
  year={2019},
  publisher={Nature Publishing Group UK London},
  doi="https://doi.org/10.1038/s41534-019-0157-8"
}

@article{böselt2021machine,
  title={Machine learning in {QM/MM} molecular dynamics simulations of condensed-phase systems},
  author={Böselt, Lennard and Thürlemann, Moritz and Riniker, Sereina},
  journal={J. Chem. Theory Comput.},
  volume={17},
  number={5},
  pages={2641--2658},
  year={2021},
  publisher={ACS Publications}
}

@article{o2016scalable,
  title={Scalable quantum simulation of molecular energies},
  author={O’Malley, Peter JJ and Babbush, Ryan and Kivlichan, Ian D and Romero, Jonathan and McClean, Jarrod R and Barends, Rami and Kelly, Julian and Roushan, Pedram and Tranter, Andrew and Ding, Nan and others},
  journal={Phys. Rev. X},
  volume={6},
  number={3},
  pages={031007},
  year={2016},
  publisher={APS},
  doi="https://doi.org/10.1103/PhysRevX.6.031007"
}

@article{sanyal2022neuro,
  title={{Neuro-Ising: Accelerating large-scale traveling salesman problems via graph neural network guided localized Ising solvers}},
  author={Sanyal, Sourav and Roy, Kaushik},
  journal={{IEEE Trans. Comput.-Aided Des. Integr. Circuits Syst.}},
  volume={41},
  number={12},
  pages={5408--5420},
  year={2022},
  publisher={IEEE}
}

@article{benedetti2019parameterized,
  title={Parameterized quantum circuits as machine learning models},
  author={Benedetti, Marcello and Lloyd, Erika and Sack, Stefan and Fiorentini, Mattia},
  journal={Quantum Sci. Technol.},
  volume={4},
  number={4},
  pages={043001},
  year={2019},
  publisher={IOP Publishing}
}

@article{holevo1973bounds,
  title={Bounds for the quantity of information transmitted by a quantum communication channel},
  author={Holevo, Alexander Semenovich},
  journal={Probl. Peredachi Inf.},
  volume={9},
  number={3},
  pages={3--11},
  year={1973},
  publisher={Russian Academy of Sciences, Branch of Informatics, Computer Equipment and~…}
}

@article{gilboa2024exponential,
  title={Exponential quantum communication advantage in distributed inference and learning},
  author={Gilboa, Dar and Michaeli, Hagay and Soudry, Daniel and McClean, Jarrod},
  journal={Advances in Neural Information Processing Systems},
  volume={37},
  pages={30425--30473},
  year={2024}
}

@article{shor1999polynomial,
  title={Polynomial-time algorithms for prime factorization and discrete logarithms on a quantum computer},
  author={Shor, Peter W},
  journal={SIAM Rev.},
  volume={41},
  number={2},
  pages={303--332},
  year={1999},
  publisher={SIAM}
}

@article{harrow2009quantum,
  title={Quantum algorithm for linear systems of equations},
  author={Harrow, Aram W and Hassidim, Avinatan and Lloyd, Seth},
  journal={Phys. Rev. Lett.},
  volume={103},
  number={15},
  pages={150502},
  year={2009},
  publisher={APS}
}

@article{smagorinsky1963general,
  title={General circulation experiments with the primitive equations: I. The basic experiment},
  author={Smagorinsky, Joseph},
  journal={Mon. Weather Rev.},
  volume={91},
  number={3},
  pages={99--164},
  year={1963},
  publisher={American Meteorological Society}
}

@article{biferale2004multifractal,
  title={Multifractal statistics of Lagrangian velocity and acceleration in turbulence},
  author={Biferale, L and Boffetta, Guido and Celani, Antonio and Devenish, BJ and Lanotte, Alessandra and Toschi, Federico},
  journal={Phys. Rev. Lett.},
  volume={93},
  number={6},
  pages={064502},
  year={2004},
  publisher={APS},
  doi = "https://doi.org/10.1103/PhysRevLett.93.064502"
}

@Preamble{ {\hyphenation{Post-Script Sprin-ger}}
}

@book{succi2001lattice,
  title={The {L}attice {B}oltzmann {E}quation for {F}luid {D}ynamics and {B}eyond},
  author={Succi, Sauro},
  year={2001},
  publisher={Oxford University Press}
}

@article{liu2012three,
  title={Three-dimensional lattice {B}oltzmann model for immiscible two-phase flow simulations},
  author={Liu, Haihu and Valocchi, Albert J and Kang, Qinjun},
  journal={Phys. Rev. E},
  volume={85},
  number={4},
  pages={046309},
  year={2012},
  publisher={APS}
}

@article{chen1998lattice,
  title={Lattice {B}oltzmann method for fluid flows},
  author={Chen, Shiyi and Doolen, Gary D},
  journal={Annu. Rev. Fluid Mech.},
  volume={30},
  number={1},
  pages={329--364},
  year={1998},
  publisher={Annual Reviews 4139 El Camino Way, PO Box 10139, Palo Alto, CA 94303-0139, USA}
}

@article{d2002multiple,
  title={Multiple--relaxation--time lattice {B}oltzmann models in three dimensions},
  author={d'Humieres, Dominique},
  journal={Philos. Trans. R. Soc. A},
  volume={360},
  number={1792},
  pages={437--451},
  year={2002},
  publisher={The Royal Society}
}

@article{Moser1999,
   author = {Moser, Robert D. and Kim, John and Mansour, Nagi N.},
   title = {Direct numerical simulation of turbulent channel flow up to $Re_{\tau}=590$},
   journal = {Phys. Fluids},
   volume = {11},
   number = {4},
   pages = {943-945},
   ISSN = {1070-6631
1089-7666},
   DOI = {https://doi.org/10.1063/1.869966},
   year = {1999},
   type = {Journal Article}
}

@article{hou1994lattice,
  title={A lattice {B}oltzmann subgrid model for high {R}eynolds number flows},
  author={Hou, Shuling and Sterling, James and Chen, Shiyi and Doolen, Gary},
  journal={Pattern Formation and Lattice Gas Automata},
  pages={151--166},
  year={1995},
  publisher={American Mathematical Society}
}

@article{xue2022synthetic,
  title={Synthetic turbulence generator for lattice {B}oltzmann method at the interface between RANS and {LES}},
  author={Xue, Xiao and Yao, Hua-Dong and Davidson, Lars},
  journal={Phys. Fluids},
  volume={34},
  number={5},
  pages={055118},
  year={2022},
  publisher={AIP Publishing LLC}
}

@article{latt2008straight,
  title={Straight velocity boundaries in the lattice {B}oltzmann method},
  author={Latt, Jonas and Chopard, Bastien and Malaspinas, Orestis and Deville, Michel and Michler, Andreas},
  journal={Phys. Rev. E},
  volume={77},
  number={5},
  pages={056703},
  year={2008},
  publisher={APS}
}

@article{koda2015lattice,
  title={{The lattice Boltzmann method implemented on the GPU to simulate the turbulent flow over a square cylinder confined in a channel}},
  author={Koda, Yusuke and Lien, Fue-Sang},
  journal={Flow Turbul. Combust.},
  volume={94},
  number={3},
  pages={495--512},
  year={2015},
  publisher={Springer}
}

@article{yang2019predictive,
  title={Predictive large-eddy-simulation wall modeling via physics-informed neural networks},
  author={Yang, XIA and Zafar, Suhaib and Wang, J-X and Xiao, Heng},
  journal={Phys. Rev. Fluids},
  volume={4},
  number={3},
  pages={034602},
  year={2019},
  publisher={APS}
}

@article{bae2022scientific,
  title={Scientific multi-agent reinforcement learning for wall-models of turbulent flows},
  author={Bae, H Jane and Koumoutsakos, Petros},
  journal={Nat. Commun.},
  volume={13},
  number={1},
  pages={1443},
  year={2022},
  publisher={Nature Publishing Group UK London}
}

@article{raissi2019physics,
  title={Physics-informed neural networks: A deep learning framework for solving forward and inverse problems involving nonlinear partial differential equations},
  author={Raissi, Maziar and Perdikaris, Paris and Karniadakis, George E},
  journal={J. Comput. Phys.},
  volume={378},
  pages={686--707},
  year={2019},
  publisher={Elsevier}
}

@inproceedings{long2018pde,
  title={Pde-net: Learning pdes from data},
  author={Long, Zichao and Lu, Yiping and Ma, Xianzhong and Dong, Bin},
  booktitle={International conference on machine learning},
  pages={3208--3216},
  year={2018},
  organization={PMLR}
}

@book{galaktionov2012stability,
  title={A stability technique for evolution partial differential equations: a dynamical systems approach},
  author={Galaktionov, Victor A and V{\'a}zquez, Juan Luis},
  volume={56},
  year={2012},
  publisher={Springer Science \& Business Media}
}

@article{price2025probabilistic,
  title={Probabilistic weather forecasting with machine learning},
  author={Price, Ilan and Sanchez-Gonzalez, Alvaro and Alet, Ferran and Andersson, Tom R and El-Kadi, Andrew and Masters, Dominic and Ewalds, Timo and Stott, Jacklynn and Mohamed, Shakir and Battaglia, Peter and others},
  journal={Nature},
  volume={637},
  number={8044},
  pages={84--90},
  year={2025},
  publisher={Nature Publishing Group},
  doi="https://doi.org/10.1038/s41586-024-08252-9"
}

@article{li2020fourier,
  title={Fourier neural operator for parametric partial differential equations},
  author={Li, Zongyi and Kovachki, Nikola and Azizzadenesheli, Kamyar and Liu, Burigede and Bhattacharya, Kaushik and Stuart, Andrew and Anandkumar, Anima},
  journal={arXiv:2010.08895},
  year={2020},
  doi="https://doi.org/10.48550/arXiv.2010.08895"
}

@article{lu2021learning,
  title={Learning nonlinear operators via DeepONet based on the universal approximation theorem of operators},
  author={Lu, Lu and Jin, Pengzhan and Pang, Guofei and Zhang, Zhongqiang and Karniadakis, George Em},
  journal={Nat. Mach. Intell.},
  volume={3},
  number={3},
  pages={218--229},
  year={2021},
  publisher={Nature Publishing Group UK London}
}

@article{schiff2024dyslim,
  title={Dyslim: Dynamics {S}table {L}earning by {I}nvariant {M}easure for {C}haotic {S}ystems},
  author={Schiff, Yair and Wan, Zhong Yi and Parker, Jeffrey B and Hoyer, Stephan and Kuleshov, Volodymyr and Sha, Fei and Zepeda-N{\'u}{\~n}ez, Leonardo},
  journal={arXiv:2402.04467},
  year={2024},
  doi="https://doi.org/10.48550/arXiv.2402.04467"
}

@article{goodfellow2020generative,
  title={Generative adversarial networks},
  author={Goodfellow, Ian and Pouget-Abadie, Jean and Mirza, Mehdi and Xu, Bing and Warde-Farley, David and Ozair, Sherjil and Courville, Aaron and Bengio, Yoshua},
  journal={Commun. ACM},
  volume={63},
  number={11},
  pages={139--144},
  year={2020},
  publisher={ACM New York, NY, USA},
  doi="https://doi.org/10.1145/3422622"
}

@article{xue2024physics,
  title={Physics informed data-driven near-wall modelling for lattice {B}oltzmann simulation of high Reynolds number turbulent flows},
  author={Xue, Xiao and Wang, Shuo and Yao, Hua-Dong and Davidson, Lars and Coveney, Peter V},
  journal={Commun. {P}hys.},
  volume={7},
  number={1},
  pages={338},
  year={2024},
  publisher={Nature Publishing Group UK London}
}

@article{maulik2019sub,
  title={Sub-grid scale model classification and blending through deep learning},
  author={Maulik, Romit and San, Omer and Jacob, Jamey D and Crick, Christopher},
  journal={J. Fluid Mech.},
  volume={870},
  pages={784--812},
  year={2019},
  publisher={Cambridge University Press}
}

@article{pal2020deep,
  title={Deep learning emulation of subgrid-scale processes in turbulent shear flows},
  author={Pal, Anikesh},
  journal={Geophys. Res. Lett.},
  volume={47},
  number={12},
  pages={e2020GL087005},
  year={2020},
  publisher={Wiley Online Library}
}

@article{fukami2019synthetic,
  title={Synthetic turbulent inflow generator using machine learning},
  author={Fukami, Kai and Nabae, Yusuke and Kawai, Ken and Fukagata, Koji},
  journal={Phys. Rev. Fluids},
  volume={4},
  number={6},
  pages={064603},
  year={2019},
  publisher={APS}
}

@article{yousif2022physics,
  title={Physics-guided deep learning for generating turbulent inflow conditions},
  author={Yousif, Mustafa Z and Yu, Linqi and Lim, HeeChang},
  journal={J. Fluid Mech.},
  volume={936},
  pages={A21},
  year={2022},
  publisher={Cambridge University Press}
}

@article{vanchurin2021toward,
  title={Toward a theory of machine learning},
  author={Vanchurin, Vitaly},
  journal={Mach. Learn.: Sci. Technol.},
  volume={2},
  number={3},
  pages={035012},
  year={2021},
  publisher={IOP Publishing}
}

@article{carleo2019machine,
  title={Machine learning and the physical sciences},
  author={Carleo, Giuseppe and Cirac, Ignacio and Cranmer, Kyle and Daudet, Laurent and Schuld, Maria and Tishby, Naftali and Vogt-Maranto, Leslie and Zdeborov{\'a}, Lenka},
  journal={Rev. Mod. Phys.},
  volume={91},
  number={4},
  pages={045002},
  year={2019},
  publisher={APS},
  doi="https://doi.org/10.1103/RevModPhys.91.045002"
}

@article{Kochkov2021-ML-CFD,
  author = {Kochkov, Dmitrii and Smith, Jamie A. and Alieva, Ayya and Wang, Qing and Brenner, Michael P. and Hoyer, Stephan},
  title = {Machine learning{\textendash}accelerated computational fluid dynamics},
  volume = {118},
  number = {21},
  elocation-id = {e2101784118},
  year = {2021},
  doi = {https://doi.org/10.1073/pnas.2101784118},
  publisher = {National Academy of Sciences},
  issn = {0027-8424},
  journal = {Proc. Natl. Acad. Sci. U.S.A.}
}

@article{jerbi2024shadows,
  title={Shadows of quantum machine learning},
  author={Jerbi, Sofiene and Gyurik, Casper and Marshall, Simon C and Molteni, Riccardo and Dunjko, Vedran},
  journal={Nat. Commun.},
  volume={15},
  number={1},
  pages={5676},
  year={2024},
  publisher={Nature Publishing Group UK London}
}

@article{caro2023out,
  title={Out-of-distribution generalization for learning quantum dynamics},
  author={Caro, Matthias C and Huang, Hsin-Yuan and Ezzell, Nicholas and Gibbs, Joe and Sornborger, Andrew T and Cincio, Lukasz and Coles, Patrick J and Holmes, Zo{\"e}},
  journal={Nat. Commun.},
  volume={14},
  number={1},
  pages={3751},
  year={2023},
  publisher={Nature Publishing Group UK London},
  doi="https://doi.org/10.1038/s41467-023-39381-w"
}

@article{mezic2021koopman,
  title={Koopman operator, geometry, and learning of dynamical systems},
  author={Mezi{\'c}, Igor},
  journal={Notices Am. Math. Soc.},
  volume={68},
  number={7},
  pages={1087--1105},
  year={2021}
}

@article{brunton2021modern,
  title={{Modern Koopman theory for dynamical systems}},
  author={Brunton, Steven L and Budi{\v{s}}i{\'c}, Marko and Kaiser, Eurika and Kutz, J Nathan},
  journal={arXiv:2102.12086},
  year={2021}
}

@article{jerrum1996markov,
  title={The Markov chain Monte Carlo method: an approach to approximate counting and integration},
  author={Jerrum, Mark and Sinclair, Alistair},
  journal={Approximation Algorithms for NP-hard problems, PWS Publishing},
  year={1996}
}

@inproceedings{bengio2013better,
  title={Better mixing via deep representations},
  author={Bengio, Yoshua and Mesnil, Gr{\'e}goire and Dauphin, Yann and Rifai, Salah},
  booktitle={International conference on machine learning},
  pages={552--560},
  year={2013},
  organization={PMLR}
}

@article{huang2025vast,
  title={The vast world of quantum advantage},
  author={Huang, Hsin-Yuan and Choi, Soonwon and McClean, Jarrod R and Preskill, John},
  journal={arXiv:2508.05720},
  year={2025},
}

@inproceedings{zhang2018improved,
  title={Improved {Adam} optimizer for deep neural networks},
  author={Zhang, Zijun},
  booktitle={2018 IEEE/ACM 26th international symposium on quality of service (IWQoS)},
  pages={1--2},
  year={2018},
  organization={Ieee}
}

@article{chandler2013invariant,
  title={Invariant recurrent solutions embedded in a turbulent two-dimensional Kolmogorov flow},
  author={Chandler, Gary J and Kerswell, Rich R},
  journal={J. Fluid Mech.},
  volume={722},
  pages={554--595},
  year={2013},
  publisher={Cambridge University Press}
}

@article{coveney2024sharkovskii,
  title={Sharkovskii’s theorem and the limits of digital computers for the simulation of chaotic dynamical systems},
  author={Coveney, Peter V},
  journal={J. Comput. Sci.},
  volume={83},
  pages={102449},
  year={2024},
  publisher={Elsevier}
}

@article{klower2023periodic,
  title={Periodic orbits in chaotic systems simulated at low precision},
  author={Kl{\"o}wer, Milan and Coveney, Peter V and Paxton, E Adam and Palmer, Tim N},
  journal={Sci. Rep.},
  volume={13},
  number={1},
  pages={11410},
  year={2023},
  publisher={Nature Publishing Group UK London}
}

@article{boghosian2019new,
  title={A new pathology in the simulation of chaotic dynamical systems on digital computers},
  author={Boghosian, Bruce M and Coveney, Peter V and Wang, Hongyan},
  journal={Adv. Theory Simul.},
  volume={2},
  number={12},
  pages={1900125},
  year={2019},
  publisher={Wiley Online Library}
}

@article{mezic2005spectral,
  title={Spectral properties of dynamical systems, model reduction and decompositions},
  author={Mezi{\'c}, Igor},
  journal={Nonlinear Dyn.},
  volume={41},
  number={1},
  pages={309--325},
  year={2005},
  publisher={Springer}
}

@book{petersen1989ergodic,
  title={Ergodic theory},
  author={Petersen, Karl E and Petersen, Karl},
  year={1989},
  publisher={Cambridge university press}
}

@article{budivsic2012applied,
  title={Applied Koopmanism},
  author={Budi{\v{s}}i{\'c}, Marko and Mohr, Ryan and Mezi{\'c}, Igor},
  journal={Chaos},
  volume={22},
  number={4},
  year={2012},
  publisher={AIP Publishing}
}

@inproceedings{pascanu2013difficulty,
  title={On the difficulty of training recurrent neural networks},
  author={Pascanu, Razvan and Mikolov, Tomas and Bengio, Yoshua},
  booktitle={International conference on machine learning},
  pages={1310--1318},
  year={2013}
}

@article{bi2023accurate,
  title={Accurate medium-range global weather forecasting with 3D neural networks},
  author={Bi, Kaifeng and Xie, Lingxi and Zhang, Hengheng and Chen, Xin and Gu, Xiaotao and Tian, Qi},
  journal={Nature},
  volume={619},
  number={7970},
  pages={533--538},
  year={2023},
  publisher={Nature Publishing Group UK London},
  doi="https://doi.org/10.1038/s41586-023-06185-3"
}

@article{lam2023learning,
  title={Learning skillful medium-range global weather forecasting},
  author={Lam, Remi and Sanchez-Gonzalez, Alvaro and Willson, Matthew and Wirnsberger, Peter and Fortunato, Meire and Alet, Ferran and Ravuri, Suman and Ewalds, Timo and Eaton-Rosen, Zach and Hu, Weihua and others},
  journal={Science},
  volume={382},
  number={6677},
  pages={1416--1421},
  year={2023},
  publisher={American Association for the Advancement of Science},
  doi="https://doi.org/10.1126/science.adi2336"
}

@article{cavaiola2024hybrid,
  title={Hybrid {ai}-enhanced lightning flash prediction in the medium-range forecast horizon},
  author={Cavaiola, Mattia and Cassola, Federico and Sacchetti, Davide and Ferrari, Francesco and Mazzino, Andrea},
  journal={Nat. Commun.},
  volume={15},
  number={1},
  pages={1188},
  year={2024},
  publisher={Nature Publishing Group UK London}
}

@article{chang2024survey,
  title={A survey on evaluation of large language models},
  author={Chang, Yupeng and Wang, Xu and Wang, Jindong and Wu, Yuan and Yang, Linyi and Zhu, Kaijie and Chen, Hao and Yi, Xiaoyuan and Wang, Cunxiang and Wang, Yidong and others},
  journal={ACM Trans. Intell. Syst. Technol.},
  volume={15},
  number={3},
  pages={1--45},
  year={2024},
  publisher={ACM New York, NY},
  doi="https://doi.org/10.1145/3641289"
}

@article{thirunavukarasu2023large,
  title={Large language models in medicine},
  author={Thirunavukarasu, Arun James and Ting, Darren Shu Jeng and Elangovan, Kabilan and Gutierrez, Laura and Tan, Ting Fang and Ting, Daniel Shu Wei},
  journal={Nat. Med.},
  volume={29},
  number={8},
  pages={1930--1940},
  year={2023},
  publisher={Nature Publishing Group US New York},
  doi="https://doi.org/10.1038/s41591-023-02448-8"
}

@article{zhang2024vision,
  title={Vision-language models for vision tasks: A survey},
  author={Zhang, Jingyi and Huang, Jiaxing and Jin, Sheng and Lu, Shijian},
  journal={IEEE Trans. Pattern Anal. Mach. Intell.},
  volume={46},
  number={8},
  pages={5625--5644},
  year={2024},
  publisher={IEEE}
}

@article{zhou2022learning,
  title={Learning to prompt for vision-language models},
  author={Zhou, Kaiyang and Yang, Jingkang and Loy, Chen Change and Liu, Ziwei},
  journal={Int. J. Comput. Vis.},
  volume={130},
  number={9},
  pages={2337--2348},
  year={2022},
  publisher={Springer}
}

@article{cheng2025learning,
  title={Learning chaos in a linear way},
  author={Cheng, Xiaoyuan and He, Yi and Yang, Yiming and Xue, Xiao and Cheng, Sibo and Giles, Daniel and Tang, Xiaohang and Hu, Yukun},
  journal={arXiv:2503.14702},
  year={2025},
  doi="https://doi.org/10.48550/arXiv.2503.14702"
}

@article{ghazi2025quantum,
  title={Quantum-computing-enhanced algorithm unveils potential KRAS inhibitors},
  author={Ghazi Vakili, Mohammad and Gorgulla, Christoph and Snider, Jamie and Nigam, AkshatKumar and Bezrukov, Dmitry and Varoli, Daniel and Aliper, Alex and Polykovsky, Daniil and Padmanabha Das, Krishna M and Cox Iii, Huel and others},
  journal={Nat. Biotechnol.},
  pages={1--6},
  year={2025},
  publisher={Nature Publishing Group US New York},
  doi="https://doi.org/10.1038/s41587-024-02526-3"
}

@article{li2021learning,
  title={Learning dissipative dynamics in chaotic systems},
  author={Li, Zongyi and Liu-Schiaffini, Miguel and Kovachki, Nikola and Liu, Burigede and Azizzadenesheli, Kamyar and Bhattacharya, Kaushik and Stuart, Andrew and Anandkumar, Anima},
  journal={arXiv:2106.06898},
  year={2021},
  doi="https://doi.org/10.48550/arXiv.2106.06898"
}

@book{kuramoto2003chemical,
  title={Chemical oscillations, waves, and turbulence},
  author={Kuramoto, Yoshiki},
  year={2003},
  publisher={Courier Corporation}
}

@article{sivashinsky1980flame,
  title={On flame propagation under conditions of stoichiometry},
  author={Sivashinsky, Gregory I},
  journal={SIAM J. Appl. Math.},
  volume={39},
  number={1},
  pages={67--82},
  year={1980},
  publisher={SIAM}
}

@book{cornfeld2012ergodic,
  title={Ergodic theory},
  author={Cornfeld, Isaac P and Fomin, Sergei Vasilevich and Sinai, Yakov Grigor'evǐc},
  volume={245},
  year={2012},
  publisher={Springer Science \& Business Media}
}

@article{cheng2025machine,
  title={Machine learning for modelling unstructured grid data in computational physics: a review},
  author={Cheng, Sibo and Bocquet, Marc and Ding, Weiping and Finn, Tobias Sebastian and Fu, Rui and Fu, Jinlong and Guo, Yike and Johnson, Eleda and Li, Siyi and Liu, Che and others},
  journal={Inf. Fusion},
  pages={103255},
  year={2025},
  publisher={Elsevier},
  doi="https://doi.org/10.1016/j.inffus.2025.103255"
}

@article{mccabe2023towards,
  title={Towards stability of autoregressive neural operators},
  author={McCabe, Michael and Harrington, Peter and Subramanian, Shashank and Brown, Jed},
  journal={arXiv:2306.10619},
  year={2023}
}

@article{lippe2023pde,
  title={{PDE-refiner: Achieving accurate long rollouts with neural PDE solvers}},
  author={Lippe, Phillip and Veeling, Bas and Perdikaris, Paris and Turner, Richard and Brandstetter, Johannes},
  journal={Advances in Neural Information Processing Systems},
  volume={36},
  pages={67398--67433},
  year={2023}
}
\bibliographystyle{sciencemag}
\clearpage

\section*{Acknowledgements}
We would like to thank Professor Igor Mezic for his valuable comments on an earlier version of this paper and Dr. Marcello Benedetti for his valuable feedback. We are also grateful to Thomas M. Bickley and Angus Mingare for their careful reading of the manuscript and insightful comments. We also gratefully acknowledge IQM Quantum Computers for providing access to superconducting quantum processors used in hardware benchmarking, and the Leibniz Supercomputing Centre (LRZ) for access to the BEAST GPU cluster, which supported the training of classical models and quantum circuit simulations.

\paragraph*{Funding: }Peter V. Coveney acknowledges funding support from the European Commission CompBioMed Centre of Excellence (Grant No. 675451 and 823712) and from the UK Engineering and Physical Sciences Research Council through the projects “UK Consortium on Mesoscale Engineering Sciences (UKCOMES)” (Grant No. EP/R029598/1) and “Software Environment for Actionable and VVUQ-evaluated Exascale Applications (SEAVEA)” (Grant No. EP/W007711/1). He also acknowledges support from the 2024–2025 DOE INCITE award for computational resources at the Oak Ridge and Argonne Leadership Computing Facilities under the “COMPBIO3” project.

\subsection*{Author contributions} 
M.W. and X.X. jointly conceived the central idea of the study and collaboratively designed the quantum-informed machine learning framework. Both authors conducted the quantum hardware experiments and contributed equally to the classical modelling and validation. M.G. performed the numerical simulations for the machine learning baselines and curated the results for the comparative analysis. M.W. led the writing of the quantum algorithm and hardware implementation sections, while X.X. was responsible for drafting the classical modelling and chaotic system descriptions. Data analysis, hyperparameter tuning, and visualization were carried out jointly by M.W. and X.X. P.V.C. provided guidance throughout the evolution of the project, including software and hardware execution and optimisation processes. All authors discussed the results and contributed to the overall manuscript.

\subsection*{Competing interests}
The authors declare that they have no competing interests.

\subsection*{Data, Code, and Materials Availability}\label{sec:data}

All data and code needed to evaluate and reproduce the results in the paper are present in the paper and/or the Supplementary Materials. This study did not generate new materials. The training/validation/testing datasets for all three cases are available on Zenodo: 16419085. The complete training code for the quantum generator module and the QIML is openly available at https://github.com/UCL-CCS/QIML.git.

\subsection*{Supplementary materials}
Supplementary Materials.



\clearpage

\setcounter{page}{1}
\begin{center}
    \textbf{\huge Supplementary Materials}\\
    \vspace{0.4cm}
    \textbf{\large Quantum-Informed Machine Learning for Predicting Spatiotemporal Chaos with Practical Quantum Advantage}\\
    \vspace{0.4cm}
    \text{Maida Wang\textsuperscript{1,\dag}, Xiao Xue\textsuperscript{1,\dag}, Mingyang Gao\textsuperscript{1}, Peter V. Coveney\textsuperscript{1,2,3,*}}\\
    \vspace{0.4cm}
    \parbox{\textwidth}{%
        \begin{center}
        \textsuperscript{1} Centre for Computational Science, Department of Chemistry, University College London, London, UK\\
        \textsuperscript{2} Informatics Institute, University of Amsterdam, Amsterdam, The Netherlands\\
        \textsuperscript{3} Centre for Advanced Research Computing, University College London, UK
        \end{center}
    }
    \text{\dag These authors contributed equally to this work}\\
    \text{*Email: p.v.coveney@ucl.ac.uk}
    \vspace{0.2cm}
\end{center}
\textbf{The PDF file includes:}\\
  Materials and Methods\\
  Supplementary Text\\
  Figs. S1 to S9\\
  Table S1 to S5\\
  Sections S1 to S10\\
  References (90–120)


\clearpage

\noindent \textbf{Table of Contents}\\

S1. Quantum Computing and Quantum Generator \dotfill 3

Quantum computing $\cdot$ Quantum machine learning $\cdot$ Sample-based quantum generator

S2. Implementation of QIML on the emulator and IQM Devices \dotfill 7

Quantum circuit architecture and parameterization $\cdot$ Implementation details $\cdot$ Optimisation strategy and training protocol $\cdot$ Hardware sensitivity analysis $\cdot$ Chip selection and error mitigation

S3. Kuramoto--Sivashinsky Equation \dotfill 15

Governing equation $\cdot$ Boundary conditions $\cdot$ Data source

S4. Kolmogorov Flow Governing Equation \dotfill 16

Navier--Stokes equations with forcing $\cdot$ Kolmogorov forcing $\cdot$ Data source

S5. LBM-based 3D Turbulent Channel Flow Simulation \dotfill 18

Lattice Boltzmann method $\cdot$ BGK collision kernel $\cdot$ Smagorinsky subgrid-scale modelling $\cdot$ Simulation setup $\cdot$ Dataset validation $\cdot$ MRT moment transformation matrix

S6. Unit Conversion between LBM Units and Physical Units \dotfill 22

Velocity conversion $\cdot$ Spatial conversion $\cdot$ Dimensional analysis

S7. Neural Networks Architecture Details \dotfill 23

Quantum model architecture $\cdot$ Koopman-based model $\cdot$ Fourier neural operator $\cdot$ Markov neural operator $\cdot$ Variational autoencoder baseline $\cdot$ Integration of classical model and Q-Prior

S8. Quantum Parameter Efficiency and Quantum Memory Advantage \dotfill 33

Theoretical argument for memory advantage $\cdot$ KS compression analysis $\cdot$ Kolmogorov compression analysis $\cdot$ TCF compression analysis

S9. Performance and Statistical Metrics \dotfill 36

Temporal autocorrelation $\cdot$ Error metrics

S10. Argument for the Quantum Advantage in Representing Chaotic Measures \dotfill 37

Supplementary Tables \dotfill 41

\clearpage

\setcounter{section}{0}
\setcounter{figure}{0}
\setcounter{table}{0}
\setcounter{equation}{0}
\renewcommand{\thesection}{S\arabic{section}}
\renewcommand{\thefigure}{S\arabic{figure}}
\renewcommand{\thetable}{S\arabic{table}}
\renewcommand{\theequation}{S\arabic{equation}}

\section{Quantum Computing and Quantum Generator}

\subsection{Quantum Computing}
Quantum computing operates on the principles of quantum mechanics, allowing for a novel class of information processing paradigms. The fundamental unit of quantum information is the \emph{qubit}, which exists in a superposition of classical states. Formally, a single qubit can be represented as
\begin{equation}
|\psi\rangle = \alpha |0\rangle + \beta |1\rangle,
\end{equation}
where \( \alpha, \beta \in \mathbb{C} \) and \( |\alpha|^2 + |\beta|^2 = 1 \). Upon measurement in the computational basis, the qubit collapses to state \( |0\rangle \) or \( |1\rangle \) with probabilities \( |\alpha|^2 \) and \( |\beta|^2 \), respectively. In an \( n \)-qubit system, the global state resides in a \( 2^n \)-dimensional Hilbert space, allowing the compact representation of complex, high-dimensional distributions.

\subsection{Quantum Machine Learning}
Recent progress in quantum computing has spurred the rapid emergence of QML as a promising interdisciplinary field at the intersection of quantum information science and statistical learning. QML algorithms are designed to exploit quantum phenomena, such as entanglement, interference, and superposition, to augment learning efficiency or representational capacity in ways that may surpass traditional counterparts. A variety of model classes have been developed within this paradigm, including quantum autoencoders~\cite{romero2017quantum16,lamata2018quantum,18ding2019experimental}, which perform lossy compression and denoising on quantum data using a reduced number of qubits, and quantum Boltzmann machines~\cite{19kieferova2017tomography,20jain2020quantum}, which offer a framework for generative modelling and state tomography through quantum sampling of thermal distributions.

Another line of work explores quantum generative adversarial networks (QGANs)~\cite{21dallaire2018quantum,23romero2021variational,24zeng2019learning}, which have been employed to emulate quantum systems and approximate entangled states through adversarial training mechanisms. In the context of supervised learning, quantum kernel methods~\cite{25rebentrost2014quantum,27mengoni2019kernel,28huang2021power} have been proposed to enhance the expressive power of traditional classifiers by mapping data into high-dimensional Hilbert spaces via quantum feature maps.

Despite the theoretical potential of QML models, their application to real-world, non-quantum datasets remains in its infancy. Many benchmark results have been demonstrated, primarily in controlled settings with synthetic or quantum-native inputs. Notably, Huang et al.~\cite{huang2022quantum} recently showed that a 40-qubit quantum processor could infer global properties of a data distribution using exponentially fewer samples than traditional learners, albeit in a tightly constrained regime. While such results mark a significant step forward, the challenge of applying QML to traditional scientific tasks with high-dimensional structure and dynamical complexity remains open.

Such matters motivate us to develop a quantum-informed machine learning framework (QIML) that incorporates quantum components as functional modules within traditional workflows. In particular, demonstrating the ability of QML models to learn nontrivial statistical priors or invariant structures from real data, such as those arising in fluid dynamics or nonlinear PDEs, would provide a critical proof-of-concept for their practical relevance. To this end, and to circumvent the significant challenge of encoding high-dimensional classical data into quantum states, our QIML framework utilizes a sample-based quantum circuit as its generative module.

\subsection{Sample-based Quantum Generator}
Quantum generative models have shown great potential recently~\cite{carrasquilla2019reconstructing,gao2018quantum}, where quantum circuits are trained to learn quantum states or model traditional probability distributions. The quantum generator employed in this work is a sample-based model, with an architecture based on the quantum circuit Born machine. Given a parameterized quantum circuit \( U(\boldsymbol{\theta}) \), initialized from the all-zero state \( |0\rangle^{\otimes n} \), the probability of observing a bitstring \( x \in \{0,1\}^n \) upon measurement is governed by the Born rule:
\begin{equation}
p_{\boldsymbol{\theta}}(x) = |\langle x | U(\boldsymbol{\theta}) | 0\rangle^{\otimes n}|^2.
\end{equation}
These probabilities define an implicit generative model from which samples can be drawn directly via quantum measurement, without requiring explicit likelihoods or tractable gradients.

Quantum generators have demonstrated utility in modelling structured datasets in domains such as quantum chemistry, generative learning, and combinatorial optimisation. However, their application to traditional physical systems, particularly those governed by nonlinear PDEs, remains limited. This is despite the deep structural parallels between quantum mechanics and classical dynamical systems: both evolve within high-dimensional Hilbert spaces, exhibit conservation laws, and are constrained
by symmetry principles. These analogies motivate the integration of quantum-generated priors into classical scientific machine learning pipelines.

The sample-based quantum generator implementation in this work is distinguished by its likelihood-free training approach: a parameterized quantum circuit encodes a probability distribution through measurement statistics, without ever writing down an explicit probability mass function.  We describe the full circuit ansatz and its Born-rule output distribution in Methods C; the Supplementary material therefore omits the detailed equations and simply points readers to that section for implementation specifics.

\textbf{Empirical and model sample sets.}
A batch of empirical samples $\{x_{i}\}_{i=1}^{N}$ is obtained by converting the normalized velocity magnitudes $v(\mathbf{r})$ into a categorical distribution over spatial indices and drawing $N$ indices.  Model samples $\{\tilde{x}_{j}\}_{j=1}^{M}\!\sim p_{\boldsymbol{\theta}}$ are produced by executing the circuit $M$ times on the quantum processor.  These two finite sample sets constitute the sole inputs to the training loss.

\textbf{MMD training objective.}
Using a characteristic kernel $k(x,x')=\langle\phi(x),\phi(x')\rangle$ the maximum-mean-discrepancy (MMD) between the empirical and model samples is
\begin{equation}
  \mathcal{L}_{\mathrm{MMD}}(\boldsymbol{\theta})=
  \frac{1}{N(N-1)}
  \sum_{i\neq i'} k(x_i,x_{i'})
  +\frac{1}{M(M-1)}
  \sum_{j\neq j'} k(\tilde{x}_j,\tilde{x}_{j'})
  -\frac{2}{NM}
  \sum_{i,j} k(x_i,\tilde{x}_j),
  \label{eq:mmd_loss_clean}
\end{equation}
which vanishes if and only if $p_{\boldsymbol{\theta}}$ matches the empirical distribution (N denotes the size of the empirical sample set $\{x_i\}$ and M denotes the size of the model sample set $\{\tilde{x}_j\}$). The loss depends solely on kernel evaluations of finite samples and therefore requires no closed-form density—an essential feature in high-dimensional chaotic flows.

\textbf{Parameter-update strategy.}
On hardware, we employ a mixed optimisation scheme. When analytic or simulator-based gradients are available (e.g.\ during pre-training), we use BFGS~\cite{zhu1997algorithm} or Adam~\cite{zhang2018improved};   on a noisy quantum processor, we switch to derivative-free methods such as COBYLA~\cite{regis2011stochastic} or BFGS. In both cases, a mini-batch of $M$ circuit samples is generated per update, with $20\,000$ shots and \textsc{M3} read-out mitigation (see Supplementary section S2.2) to suppress measurement noise on the quantum hardware.

\textbf{Why the quantum generator is efficient in this setting.}
With $10$ or $15$ qubits, the Hilbert space already spans
$2^{10}=1024$ or $2^{15}=32\,768$ computational basis states—matching the spatial
resolution of our coarsest flow field.  
Entanglement permits the implicit encoding of multi-point velocity
correlations without enumerating them explicitly;  
the sample-based MMD objective circumvents intractable likelihoods
and normalization constants;  
and the adopted optimisation schemes require only additional
forward shots, avoiding ancillary qubits or deep state
tomography.  
These properties allow the quantum generator to provide an informative,
low-parameter prior that complements the classical Koopman
machine learning model discussed in the main text.

In this study, we propose a hybrid quantum–classical architecture in which quantum generators are trained to learn invariant velocity distributions that characterise the long-term statistical structure of chaotic fluid systems. These learned quantum distributions serve as data-driven priors to regularize the predictions of a classical machine learning model trained on PDE dynamics. Specifically, let \( p_{\theta}(x) \) denote the generated prior distribution and \( \hat{q}(x) \) the empirical velocity distribution obtained from the prediction. We design a composite regularization strategy combining both a Kullback–Leibler (KL) divergence loss and an MMD loss:

\begin{align}
\mathcal{L}_{\text{KL}} &= D_{\text{KL}}\left( \hat{q}(x) \| p_{\theta}(x) \right), \\
\mathcal{L}_{\text{MMD}} &= \left\| \mathbb{E}_{x \sim \hat{q}(x)}[\phi(x)] - \mathbb{E}_{x \sim p_{\theta}(x)}[\phi(x)] \right\|^2_{\mathcal{H}}.
\end{align}

Here, \( \phi(x) \) denotes a feature mapping into a reproducing kernel Hilbert space \( \mathcal{H} \), and \( \hat{q}(x) \) is derived from the predicted velocity field. The KL term in equation $(A4)$ captures first-order alignment between predicted and reference distributions in information-theoretic terms, while the MMD term measures higher-order statistical discrepancies under a kernel embedding. This dual-objective regularization is inspired by recent developments in physics-informed learning frameworks such as DysLIM~\cite{schiff2024dyslim}, which have shown improved robustness in chaotic and high-dimensional regimes.

The total training objective combines the standard reconstruction loss with these quantum-informed priors:
\begin{equation}
\mathcal{L}_{\text{total}} = \mathcal{L}_{\text{recon}} + \lambda_{\text{KL}} \mathcal{L}_{\text{KL}} + \lambda_{\text{MMD}} \mathcal{L}_{\text{MMD}},
\end{equation}
where \( \mathcal{L}_{\text{recon}} = \| \hat{{u}}_{t+1} - {u}_{t+1} \|^2 \) and the hyperparameters \( \lambda_{\text{KL}}, \lambda_{\text{MMD}} \) are selected empirically to balance fidelity and long-term physical consistency.

By anchoring machine learning model predictions to quantum-learned invariant structures, the framework improves stability during roll-out and reduces drift caused by chaotic sensitivity. \R{The quantum modules are implemented using 10–15 qubits and 3–8 layers of parameterized rotation gates, with a total of 30–300 trainable parameters. Execution is performed on the IQM Garnet superconducting quantum processor—a 20-qubit near-term quantum device with gate fidelities reported in~\cite{abdurakhimov2024technology} and Table~\ref{tab:qpu_specs}.} Circuit details, ansatz choices, and calibration parameters are provided in the next section.

These results provide an early but concrete demonstration of how quantum components may be meaningfully integrated into classical PDE solvers, offering scalable hybrid strategies even within the noise and qubit limitations of current quantum hardware.

\section{Implementation of QIML on the emulator and IQM Devices}
\begin{figure}[t]
  \centering
  \includegraphics[width=0.9\linewidth]{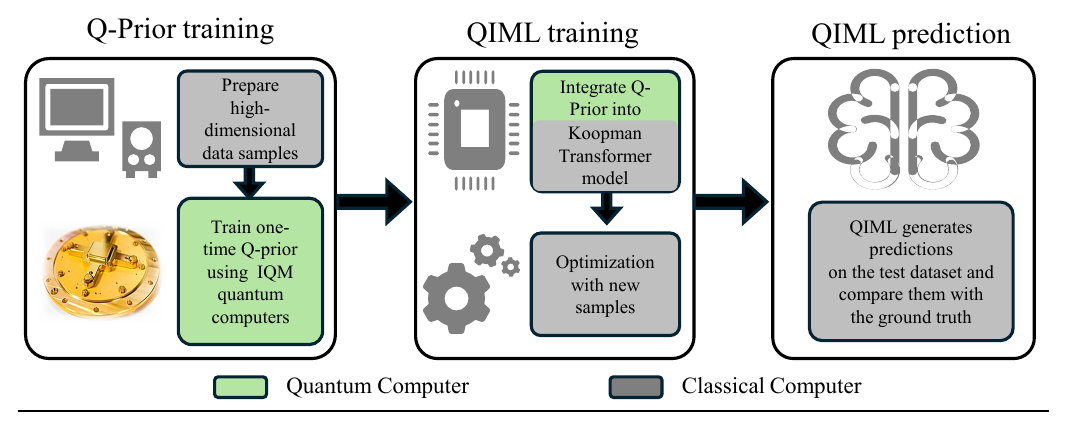}
  \caption{
    \R{\textbf{Workflow of the QIML framework.} 
    The overall process consists of three sequential stages: 
    (left)~Using high-dimensional chaotic data, a one-time offline training is performed on a quantum processor to learn a compressed Q-Prior. 
    (middle)~The pretrained and fixed Q-Prior is then integrated into the training loop of a classical machine learning model (such as a Koopman Transformer model) as a physical constraint that guides optimisation. 
    (right)~ Finally, the trained QIML model performs autoregressive prediction on new data.}
  }
  \label{fig:qiml_trainingloop}
\end{figure}

In the QIML framework proposed in this work, the quantum component is realized using a parameterized quantum circuit as a quantum generator, as shown in Fig.~\ref{fig:qiml_trainingloop}. Its core function, learning invariant distributions from observational data, has been outlined in the preceding section. Here, we detail the implementation of this quantum model on the classical emulator and superconducting quantum hardware provided by IQM~\cite{abdurakhimov2024technology}. A classical emulator is a conventional computer program that simulates the mathematical operations of an ideal, noise-free quantum device.  Due to experimental resource constraints, real quantum hardware implementation was conducted only on the most challenging benchmark—the turbulent channel flow dataset.

\subsection{Quantum Circuit Architecture and Parameterization}

We construct our quantum circuit using a layered structure composed of parameterized single-qubit rotations followed by entangling operations. The circuit uses Qiskit and consists of $L = 3$ to $8$ alternating layers. Each layer comprises three rotation gates per qubit: $R_y(\theta_i)$, $R_z(\phi_i)$, and $R_x(\psi_i)$ which are all transpiled to local rotation gates, followed by a sequence of controlled-Z (CZ) gates between adjacent qubits. The number of qubits $n$ ranges from $10$ to $15$, depending on the resolution of the discretised target distribution.

\begin{equation}
U(\boldsymbol{\theta}) \;=\;
\prod_{\ell = 1}^{L}
\Bigl[
  \Bigl(\,
    \prod_{j = 1}^{n}
      R_x^{(j)}\!\bigl(\psi^{(\ell)}_{j}\bigr)\,
      R_z^{(j)}\!\bigl(\phi^{(\ell)}_{j}\bigr)\,
      R_y^{(j)}\!\bigl(\theta^{(\ell)}_{j}\bigr)
  \Bigr)
  \Bigl(\,
    \prod_{(j,k)\in\mathcal{C}}
      \mathrm{CZ}_{j,k}
  \Bigr)
\Bigr],
\label{eq:qcbm_unitary}
\end{equation}

The full quantum state $|\psi_{\theta}\rangle = U({\theta})|0\cdots 0\rangle$ is sampled $N$ times to generate empirical frequencies $\hat{p}_{\theta}(x)$ used to approximate the target distribution.

Notably, we do not inject explicit classical features into the quantum circuit. Instead, the quantum module acts purely as a generative module: its parameters are optimized so that the Born distribution reproduces the empirical statistics of the data. This choice eliminates costly quantum-to-classical data interfacing and highlights the expressive power of comparatively shallow circuits for modelling complex, high-dimensional distributions.

A parameterized quantum state $|\psi_{\theta}\rangle$ lives in a $2^{n}$-dimensional Hilbert space, where the amplitudes $\langle x \lvert \psi_{\theta} \rangle$ define a probability distribution over computational basis states. Each basis string x is mapped bijectively to a spatial grid point, so the quantum state provides an implicit embedding of the target measure into a linear feature space. Unlike explicit kernel expansions, however, this Hilbert-space representation is generated by a quantum circuit whose entangling gates can capture non-local correlations—potentially including quantum entanglement—without the need for an exponential number of classical parameters.

This construction allows for efficient sampling and generalization from complex distributions. Classical networks often require thousands of parameters to represent distributions with long-range correlations or multi-modal structure. In contrast, our Q-Prior uses fewer parameters and leverages the quantum state’s exponential support to encode richer structure in the generated samples. This trade-off between circuit expressivity and parameter efficiency is particularly valuable under the constraints of quantum hardware.

\begin{figure}
\centering
\includegraphics[width=1\textwidth]{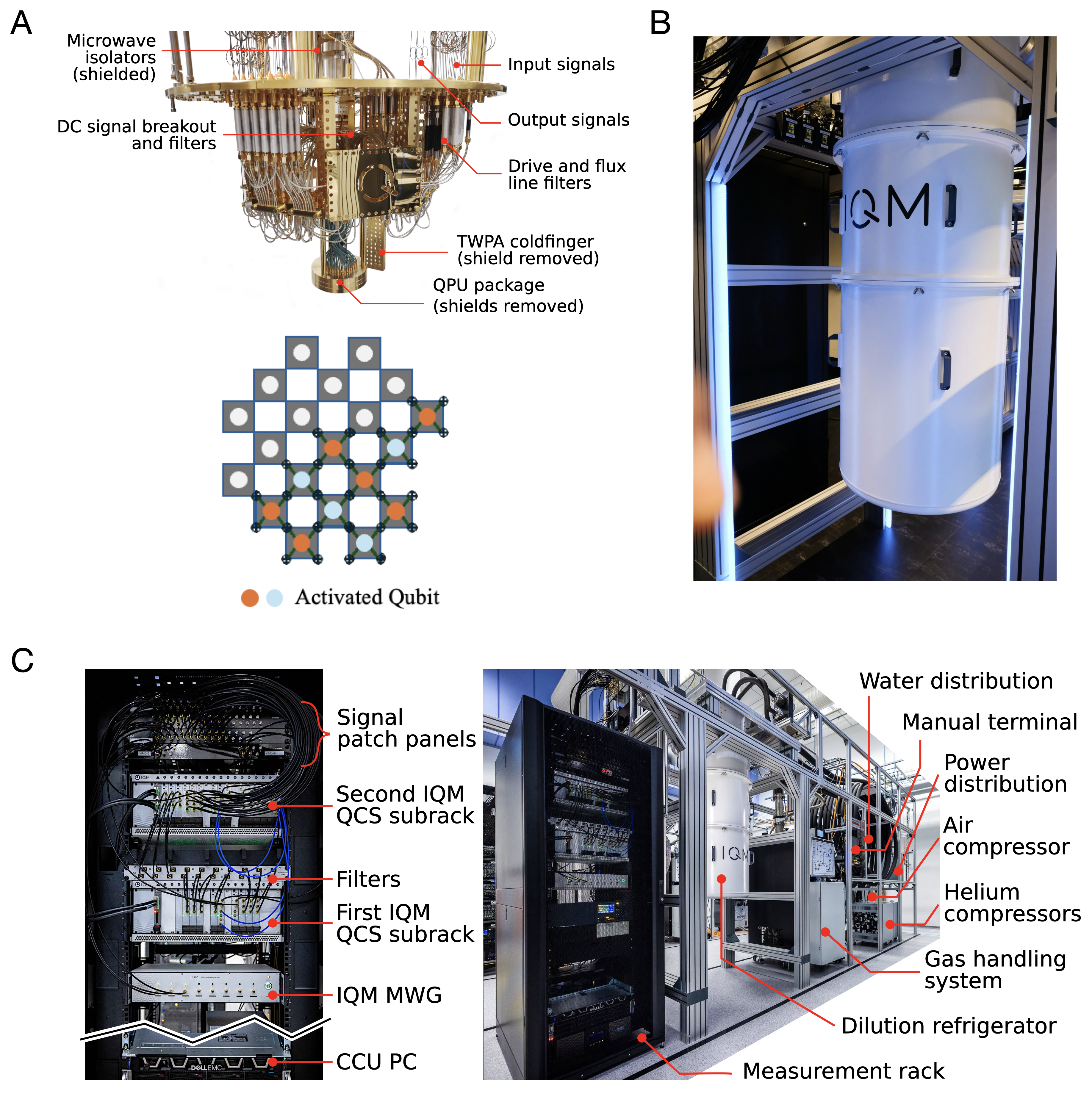}
\caption{\textbf{Representative images of the IQM Radiance 20 quantum computer (Garnet).} Copyright permission from IQM. A. Cryogenic components inside the host dilution refrigerator. 10 qubits are used of the 20-qubit IQM computer.  B. IQM's quantum computer is housed within an ultra-high vacuum cryostat during operation, maintaining the ultra-low temperatures required for superconducting qubit performance. c. Photograph of the IQM system without covers.
}
\label{fig:device}
\end{figure}

\subsection{Implementation of QIML}
\R{We implemented and validated the classical machine learning and quantum emulator on BEAST GPU cluster from Leibniz Supercomputing Centre and the quantum generator on superconducting quantum hardware provided by IQM, mainly using the 20-qubit \textit{Garnet} chip. A subset of 10-15 qubits was selected based on individual coherence performance and gate error metrics. 
The above circuit is compiled to superconducting hardware using IQM's transpiler stack. Transpilation includes single- and two-qubit gate fusion, qubit remapping, and hardware-specific gate decompositions. A representative transpiled circuit for the 10-qubit Garnet chip is shown in Supplementary Fig.~\ref{fig:device}, illustrating the translation of $R_y$/$R_z$/$R_x$ blocks into native $R(\theta,\phi)$ rotations with optimized layout and connectivity. The total number of parameters is approximately bounded by 300, as each qubit hosts 2–3 rotation gates per layer (each with a trainable parameter), and the circuit involves up to 10 qubits and 10 layers.}

\R{Each quantum circuit is executed with $N = 20{,}000$ measurement shots on the emulator and quantum devices, yielding samples $\{x_i\}_{i=1}^{N}$ distributed according to the Born rule, i.e., with probabilities $|\langle x_i | \psi_{\theta}\rangle|^2$. These are post-processed into traditional histograms and compared to the empirical fluid velocity distribution via KL divergence, MMD loss, and peak structure preservation losses. Sampling outcomes typically yield $8000$--$20{,}000$ distinct outcomes per shot batch, corresponding to $10$--$15$-bit strings depending on circuit size. All remaining traditional computation, including forward simulation and loss evaluation, was executed on NVIDIA A100 GPUs. To provide context on the practical feasibility of our approach, Table~\ref{tab:quantum_resources} compares the resource budget of the QIML framework with that of established quantum algorithms, such as VQE and HHL. It highlights that the QIML framework has significantly lower requirements for circuit depth and measurement overhead. }

\begin{sidewaystable}[t]
\centering
\small
\caption{
Representative quantum resource requirements reported for selected quantum algorithms and application settings.
The comparison contextualizes the practical scale of quantum resources used in this work, rather than implying direct performance superiority across fundamentally different computational tasks.
}
\label{tab:quantum_resources}
\renewcommand{\arraystretch}{1.3}
\begin{tabular}{@{}p{4.8cm} p{5.8cm} p{1.6cm} p{2.2cm} p{2.6cm} p{4.5cm}@{}}
\toprule
Framework & Task type & Qubits & Circuit depth & Shots per eval. & Reported quantum runtime \\
\midrule
QIML (this work) 
& Offline generative prior for turbulence 
& $<15$
& $<20$ 
& $<2\times10^{4}$ 
& $<11$ h (one-time) \\
VQE~\cite{cao2019quantum}
& Molecular ground-state energy 
& 10--50 
& $10^{2}$--$10^{3}$ 
& $10^{5}$--$10^{7}$ 
& Days to weeks \\
QMMM/Embedding~\cite{bauer2020quantum}
& Electronic structure embedding 
& 20--50 
& $>10^{2}$ 
& $>10^{5}$ 
& Days to weeks \\
HHL~\cite{harrow2009quantum}
& Linear systems solving 
& $>10^{3}$ (logical) 
& Deep (FTQC) 
& N/A 
& Not NISQ-feasible \\
Random circuit sampling~\cite{arute2019quantum} 
& Sampling benchmark 
& 53 
& $\sim20$ 
& $\sim10^{6}$ 
& Seconds \\
\bottomrule
\end{tabular}
\end{sidewaystable}

\subsection{Optimisation Strategy and Training Protocol}\label{SI: optimization Strategy and Training Protocol}

\begin{figure}[htbp!]
\centering
\includegraphics[width=1\textwidth]{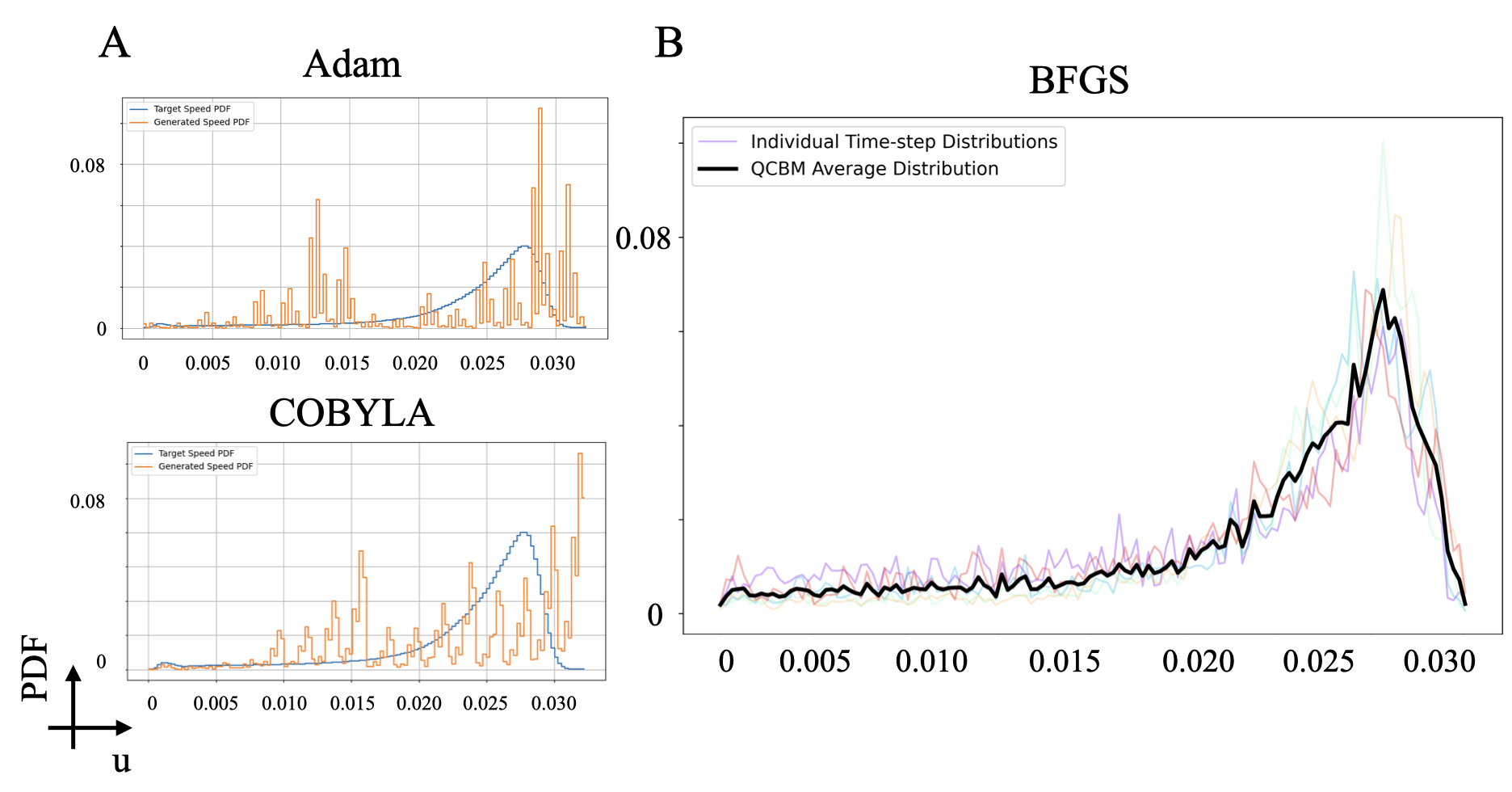}
\caption{\textbf{Quantum generator results obtained using different optimisation methods.} A. Training with Adam and COBYLA optimizers results in poor convergence and noisy distributions. B. Applying a single-shot L-BFGS optimizer in combination with the M3 measurement mitigation technique yields significantly improved performance.}
\label{fig:distributions using different methods}
\end{figure}

The training for the Kuramoto-Sivashinsky and Kolmogorov flow systems was conducted on a quantum circuit emulator, which is a classical simulation of an ideal quantum device supported by the PennyLane and Qiskit libraries, optimized by Adam. Given its higher complexity and the failure of classical models to learn its dynamics, the TCF system was used for validation on real quantum hardware, accessed via Qiskit and the IQM-Qiskit provider. As shown in Fig.~\ref{fig:distributions using different methods}, for the training on the hardware, our optimisation of quantum generator parameters was carried out using several traditional algorithms, with L-BFGS yielding the most stable convergence under hardware noise. Across all simulations and experiments, and depending on dataset complexity and resolution, training epochs varied between 400 and 500. The number of learnable parameters, corresponding to the angles of the parameterized rotation gates within the quantum circuit as shown in Fig.~\ref{fig:circuit}, ranged from 120 to 300. For the specific parameter count for each system, we refer the reader to Table~\ref{tab:param_counts} and Table~\ref{tab:compression} in our discussion on parameter efficiency and memory advantage.  Target distributions were generated by binning velocity fields from turbulent and Kolmogorov flow datasets into one-dimensional histograms with support ranging from $2^{10} = 1024$ to $2^{15} = 32,768$ bins.

\subsection{Sensitivity of Hardware Quantum Generator to Qubit Number and Circuit Depth}

To better understand the scalability limitations of Q-priors on real quantum hardware, we conducted an ablation study by varying both the number of active qubits (from 4 to 15) and the circuit depth (from 2 to 12 layers) on the IQM superconducting processors. As shown on the IQM official website, increasing either the qubit count or the number of layers leads to a marked rise in total circuit error, which in turn degrades the quality of the sampled distributions—particularly their ability to reproduce accurate velocity statistics.



This degradation is primarily due to the accumulation of readout noise and gate errors, which increase non-linearly with circuit size. In contrast to noiseless simulators or emulators, real hardware suffers from decoherence, sampling noise, and hardware-specific imperfections that compound as more gates and qubits are used. These effects place practical constraints on how expressive a quantum circuit can be without error correction. Such challenges are hallmarks of the present NISQ era, where the accumulation of errors fundamentally limits computational power.

To address these limitations, we apply hardware-aware strategies including targeted chip and qubit selection, suitable circuit depth, and error mitigation, which are detailed in the following section.

\subsection{Chip Selection, Error Mitigation and Robustness Analysis}

\begin{figure}[htbp!]
\centering
\includegraphics[width=1\textwidth]{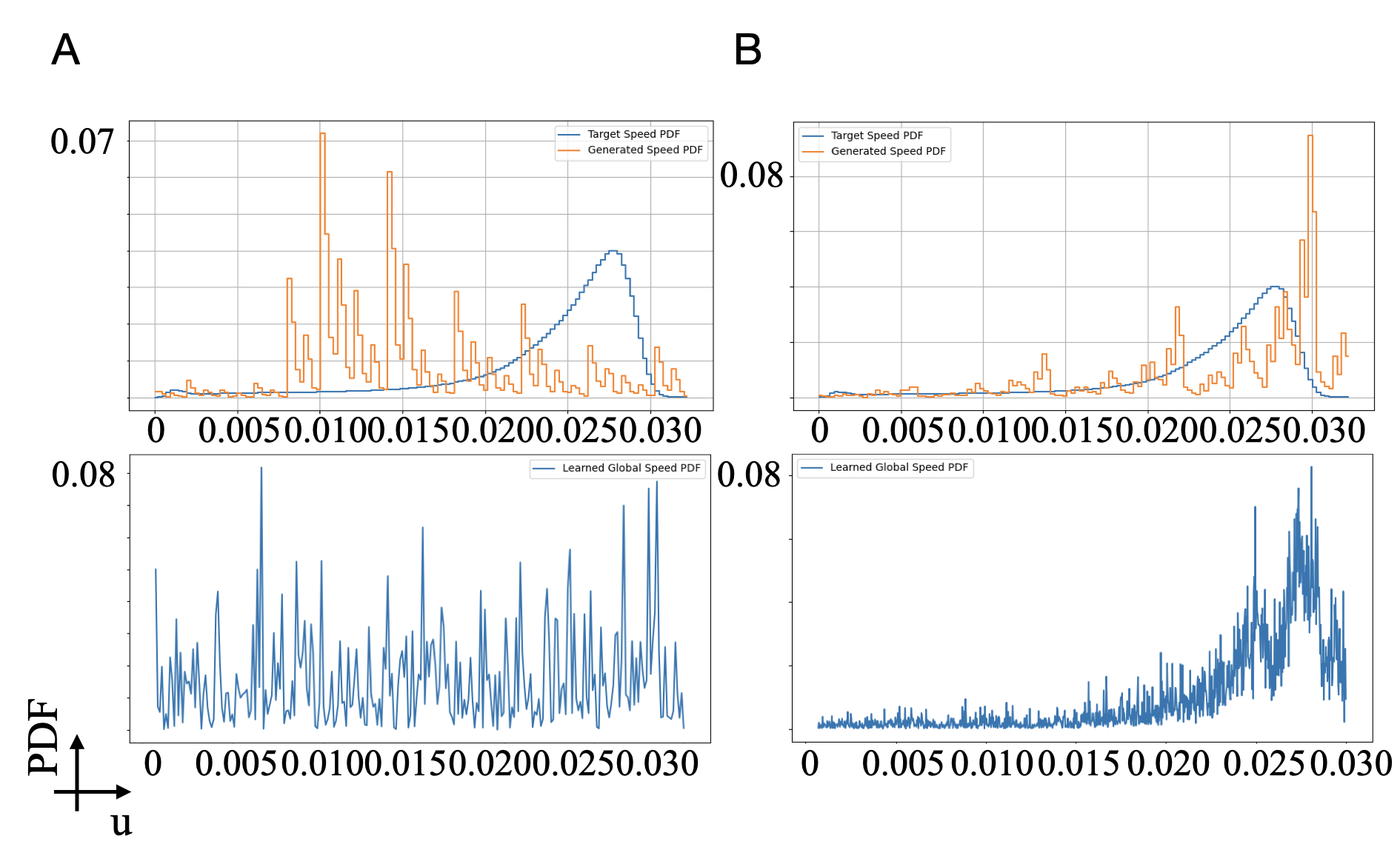}
\caption{\textbf{Quantum generator results obtained on different IQM quantum devices.} A. Training on IQM-Sirius results in poorer distributions. B. Quantum Generator achieves better results on IQM-Garnet. 
}
\label{fig:distributions on different iqm devices}
\end{figure}

As shown on IQM's official website and Fig. \ref{fig:distributions on different iqm devices}, our initial experiments using IQM's \textit{Sirius} chip yielded not good convergence due to fidelity limitations, two-bit gate connection limitations due to hardware topology, and the inherent sensitivity of high-resolution distributions to noise. \R{The technical specifications and performance metrics, including coherence times and gate fidelities, for the Sirius and Garnet quantum processors are summarized in Table~\ref{tab:qpu_specs}.} To mitigate these issues, we first migrated our algorithm to the \textit{Garnet} high-fidelity architecture and reduced histogram binning granularity from 1024 to 256. To mitigate the effect of readout errors during quantum sampling, we implemented a post-processing strategy based on the matrix-free measurement mitigation (\texttt{M3}) protocol \cite{nation2021scalable} shown in Fig. \ref{fig:circuit}. This technique calibrates the readout noise by learning a probabilistic response model from the device’s native measurement behaviour. Then it applies Bayesian corrections to raw bitstring outputs without explicitly inverting a response matrix, thereby preserving numerical stability at scale. The use of \texttt{M3} proved particularly valuable given the multi-qubit readout complexity and the high measurement resolution required to fit fine-grained distributional features.

\R{Additionally, in the present work, practical experiments were conducted at a reduced resolution (see last paragraph). Scaling the QIML framework to a high-resolution geophysical grid of $1024 \times 1024$ points (approx. $10^6$ degrees of freedom) requires encoding a Hilbert space of dimension $2^{20}$. Assuming a compact generative ansatz, this necessitates a minimum of $n \approx 20$ logical qubits for spatial addressing, plus auxiliary qubits for feature channels, placing the requirement in the range of 30-50 qubits.
Regarding coherence, capturing the multi-scale correlations of such a large system likely requires increased circuit depth, estimated at $D \sim 50-100$ layers. To maintain a meaningful signal-to-noise ratio without full error correction, the two-qubit gate fidelity would need to exceed 99.9\%, a target approachable by next-generation superconducting or trapped-ion processors combined with advanced error mitigation strategies.}

\begin{table}[h!]
\centering
\caption{Comparison of Quantum Processor Specifications for Sirius and Garnet.}
\label{tab:qpu_specs}
\begin{tabular}{lcc}
\toprule
\textbf{Property} & \textbf{Sirius} & \textbf{Garnet} \\
\midrule
Topology & STAR 24 & CRYSTAL 20 \\
Qubits & 16 & 20 \\
Pulse-Level Access & Available & Not available \\
Native Gates & barrier, cz, measure, move, rx & barrier, cz, measure, rx \\
Max Circuits / Shots & 500 / 20000 & 500 / 20000 \\
Median T1 & 29.33 $\mu$s & 38.96 $\mu$s \\
Median T2 (Ramsey) & 23.30 $\mu$s & 7.78 $\mu$s \\
Median T2 (Echo) & 30.20 $\mu$s & 16.95 $\mu$s \\
Median Rotation Gate Fidelity & 99.89 \% & 99.89 \% \\
Median CZ Gate Fidelity & 98.27 \% & 99.29 \% \\
Median Move-Move Fidelity & 98.96 \% & Not listed \\
\bottomrule
\end{tabular}
\end{table}

In addition to \texttt{M3}, we employed a combination of large-sample averaging and outlier rejection to suppress noise further. One key feature of our framework is the one-time, offline training of the Q-prior. The quantum generator is trained for a total of 50 epochs. Once this initial training process is complete, the quantum circuit is no longer needed for the main training loop. Within each of the 50 training epochs, each quantum circuit was executed with up to 20,000 shots, and the resulting histograms were averaged across repeated trials to stabilize statistical fluctuations. Outlier detection routines were applied to eliminate anomalous measurement rounds—specifically, instances where a disproportionate number of outcomes collapsed into trivial all-zero or all-one configurations. These spurious distributions, often symptomatic of transient decoherence or control drift, were discarded from the final ensemble to avoid biasing the learned quantum distribution. Together, these techniques enabled robust extraction of high-dimensional statistical structure from quantum circuits on quantum hardware. Circuit transpilation was performed via IQM’s native compiler stack, which supports qubit mapping, single-gate merging, and readout-aware optimisation. These improvements allowed us to stabilize the training loss and achieve consistent sampling performance across multiple runs. We tested numerous qubit configurations on both 10-qubit and 15-qubit hardware layouts. Results show that expressive distributions can be reliably approximated using $\sim$120 parameters, with deeper circuits providing better capture of multimodal or heavy-tailed distributions. The absence of input encoding enables this model to generalize across different datasets (e.g., Kolmogorov flow and shear-driven turbulence) by retraining only on the output distribution.
\R{We observed consistent convergence behaviours across independent runs, with less than 5\% variance in MMD divergence across 5 seeds.}

Furthermore, we found that higher-resolution distributions exhibit increased robustness against circuit-level noise, while low-resolution cases suffer from distributional collapse and mode imbalance. The experimental evidence suggests that quantum circuits are particularly well-suited to modelling high-dimensional dynamical systems. These findings support the feasibility of embedding small-scale quantum devices within traditional learning pipelines. The successful deployment of quantum generators on real quantum hardware, together with GPU-based optimisation and software differentiation, illustrates a viable hybrid computing paradigm for physics-informed machine learning.


\section{Kuramoto--Sivashinsky Equation}
The Kuramoto-Sivashinsky (KS) equation is a fourth-order nonlinear partial differential equation that models spatiotemporal instabilities in a range of physical systems~\cite{kuramoto2003chemical,sivashinsky1980flame}. It is especially important in the study of pattern formation and chaos. The governing equation is described as follows.

\subsection{Governing Equation}

In one spatial dimension, the KS equation is written as:

\begin{equation}
\frac{\partial u}{\partial t} + u \frac{\partial u}{\partial x} + \frac{\partial^2 u}{\partial x^2} + \nu \frac{\partial^4 u}{\partial x^4} = 0,
\end{equation}
where $u(x,t)$ is a scalar field, $x$ is the spatial coordinate, $t$ is time, and $\nu$ is a positive parameter controlling the strength of the fourth-order dissipation term. The term $\frac{\partial u}{\partial t}$ describes the temporal evolution of the field. The nonlinear term $ u\frac {\partial u}{\partial x}$ represents convective transport and introduces nonlinearity into the system. The second derivative term $\frac{\partial^2 u}{\partial x^2}$ acts as a linear destabilizing mechanism, similar to anti-diffusion, amplifying short-wavelength perturbations. The fourth derivative term $\nu \frac{\partial^4 u}{\partial x^4}$ provides a stabilizing effect by damping high-frequency modes, thereby preventing blow-up and enabling bounded chaotic behaviour.

\subsection{Boundary Conditions}

In this study, the KS equation is performed under periodic boundary conditions of the form: $u(x + L, t) = u(x, t),$ where $L$ is the spatial period of the domain. These conditions reflect the translational symmetry of many physical systems and simplify the analysis of chaotic dynamics.

\subsection{Data Source}
In the first application, we performed KS equation dataset with the help of CFD jax community code~\cite{Kochkov2021-ML-CFD}. The dataset has also been used and validated in Ref.~\cite{schiff2024dyslim}. 

\section{Kolmogorov Flow Governing Equation}
The Kolmogorov flow is governed by the incompressible Navier-Stokes (NS) equations with a sinusoidal forcing term. In this paper, we apply a 2D Kolmogorov flow as an example to examine our QIML framework. Below, we will describe our system in 2D NS equations with forcing.

\subsection{Navier-Stokes Equations with Forcing}
The flow is described by the incompressible Navier-Stokes equations with an external forcing term, which is denoted as 

\begin{equation}
\frac{\partial \mathbf{u}}{\partial t} + (\mathbf{u} \cdot \nabla)\mathbf{u} = -\nabla p + \nu \nabla^2 \mathbf{u} + \mathbf{F},
\end{equation}

\begin{equation}
\nabla \cdot \mathbf{u} = 0,
\end{equation}
where $\mathbf{u} = (u(x,y,t), v(x,y,t))$ is the velocity vector field of the fluid, $p(x,y,t)$ is the scalar pressure field, $\nu$ is the kinematic viscosity, and $\mathbf{F}$ represents the external body force applied to the fluid.

\subsection{Kolmogorov Forcing}

In the Kolmogorov setup, the force is applied only in the $x$-direction and varies sinusoidally in the $y$-direction. The forcing term is defined as:
\begin{equation}
\mathbf{F} = \left(F_0 \sin(k y), 0\right),
\end{equation}
where $F_0$ is the amplitude of the forcing and $k$ is the wavenumber determining the periodicity in the $y$-direction. 

\subsection{Data Source}

In the second application, we utilized the high-fidelity Kolmogorov dataset derived from~\cite{chandler2013invariant}, which offers a comprehensive and statistically rich representation of turbulent flow fields. This dataset was critical for evaluating the robustness and generalization of our proposed model, particularly under complex conditions characterized by a high Reynolds number of $Re = 1000$. It comprises 40 trajectories, each containing 320 temporal snapshots, with a spatial resolution of $256 \times 256$ grid points. To establish a high-fidelity ground truth, all numerical simulations for data generation were performed using double-precision (FP64) floating-point arithmetic. For the machine learning stage, however, our QIML framework operates entirely in single-precision (FP32) during both training and inference, a standard practice to balance computational efficiency and numerical stability. Consequently, while the models are trained on FP64 data, their predictive outputs are themselves of FP32 precision. The extent to which training the QIML framework in double-precision would alter these predictive outcomes remains a subject for future study.

\section{LBM-based 3D turbulent channel flow simulation}
\subsection{The Lattice Boltzmann Method}
The Lattice Boltzmann Method is known as an alternative computational fluid dynamics (CFD) framework that models the evolution of single particle distribution functions at the kinetic-level. It is based on the discrete form of the Boltzmann equation and operates on a lattice grid in space and time. The governing equation for the probability distribution function, or populations $\mathbf{f}$, located at position $\mathbf{x}$ at time $t$, accounting for both collisions and external forces, is given by:
\begin{equation}
\label{eq:lbe}
\mathbf{f}(\mathbf{x}+\mathbf{c}_{i}\Delta t,t+\Delta t) =\mathbf{f}(\mathbf{x}, t) + \Omega\left(\mathbf{f}(\mathbf{x},t )\right)+\mathbf{F}(\mathbf{x},t),
\end{equation}
where $\mathbf{c}_i$ are the discrete lattice velocities, $\Delta t$ is the simulation time step which is set to unity, $\Omega$ is the collision operator for the probability distribution function. $\mathbf{F}$ is the external volume force.

The LBE has gained popularity due to its simplicity, ease of implementation on parallel architectures, and its ability to naturally handle complex boundary conditions. Macroscopic fluid quantities such as density and velocity are obtained by taking moments of the distribution function. Specifically, the fluid density $\rho$ and momentum density $\rho\mathbf{u}$ are computed as follows:
\begin{equation}\label{eq:density}
\rho(\mathbf{x}, t) = \sum_{i=0}^{Q-1} f_i(\mathbf{x}, t), \\
\end{equation} 
\begin{equation}
\label{eq:momentum}
\rho(\mathbf{x}, t)\mathbf{u}(\mathbf{x}, t) = \sum_{i=0}^{Q-1} f_i(\mathbf{x}, t)\mathbf{c}_{i},
\end{equation}
where $f_i(\mathbf{x}, t)$ is the particle distribution function in the $i$-th discrete velocity direction at position $\mathbf{x}$ and time $t$.

\subsection{Bhatnagar–Gross–Krook Collision Kernel}\label{sec:method-bgk}
We define BGK collision kernel $\Omega$ as follow:
\begin{equation}
    \Omega\left(\mathbf{f}(\mathbf{x},t )\right)=-\frac{1}{\tau}(\mathbf{f}(\mathbf{x},t )-\mathbf{f}^{\mbox{ eq}}(\mathbf{x},t )),
\end{equation}
where $\mathbf{f}^{\mbox{ eq}}(\mathbf{x},t )$ denoted as the equilibrium distribution function, $\tau$ is correlated with the kinematic viscosity $\nu$:
\begin{equation}
\label{eq:nu}
\nu = c_s^2\left(\tau-\frac{1}{2}\right)\Delta t_{coll},
\end{equation}
where $\Delta t_{coll}$ is set to identity in the simulation.

\subsection{Smagorinsky Subgrid-Scale Modelling}\label{sec:method-sgs}
In this part, we summarize the lattice-Boltzmann-based Smagorinsky Subgrid Scale (SGS) LES techniques. Within the LBM framework, the effective viscosity $\nu _{\mathrm{eff}}$ \cite{smagorinsky1963general, hou1994lattice, koda2015lattice} is modeled as the sum of the molecular viscosity, $\nu_0$, and the turbulent viscosity, $\nu_t$:

\begin{equation}
\nu _{\mathrm{eff}}=\nu _0+\nu_t, \hspace{.6in} \nu_t = C_{\mathrm{smag}}\Delta ^2\left |\bar{\mathbf{S}}\right |,
\label{smogrinsky_model}
\end{equation}
where $\left |\mathbf{\bar{S}} \right |$ is the filtered strain rate tensor, $C_{\mathrm{smag}}$ is the Smagorinsky constant, $\Delta$ represents the filter size, which is set to 1 LBU. The Smagorinsky constant for this study is set to $C_{\mathrm{smag}}=0.01$.

\subsection{Simulation Set up}

In this study, the computational domain for the turbulent channel flow simulation is defined with dimensions $ L_x \times L_y \times L_z = 1024\times192 \times 192   $, where $x$, $y$, and $z$ denote the streamwise, vertical, and spanwise directions, respectively. The friction Reynolds number is set to $Re_{\tau}=180$ which is equivalent to $Re=3250$. Periodic boundary conditions are applied in the streamwise and spanwise directions, while the vertical direction is governed by no-slip boundary conditions~\cite{latt2008straight}. This configuration distinguishes our dataset from existing studies that primarily rely on 2D turbulent cases, as our dataset is based on fully three-dimensional simulations. The fully developed 3D turbulent channel flow simulation follows the configuration outlined in reference~\cite{xue2022synthetic}. The simulation begins from an initial zero-velocity field, with a square block of size $20 \times 20 \times 100$ grid points positioned at $x = 192$. A volumetric force is applied uniformly across the domain to drive the flow. The simulation is run for 50 domain-through times to establish initial flow characteristics.
After this initial phase, the block is removed, and the simulation is continued for an additional 50 domain-through times, allowing the flow to fully develop into a turbulent state. Data sampling begins after this stage, focusing on the 2D cross-section at $x = 512$. Data is collected over a further 100 turnover times, ensuring that the samples represent fully developed turbulence. The simulation timestep is set to $\Delta t = 0.02 \, \mathrm{s}$, and it operates in dimensionless units (800 timesteps), as is typical for LBM simulations. The detailed transformation from dimensionless units to physical units can be found in section S6.
To ensure robustness and generality, we conducted three independent 3D turbulent channel flow simulations under these conditions, generating a comprehensive dataset for analysis.

\subsection{Dataset Periodic Turbulent Channel Flow Validation}

\begin{figure}[htbp!]
\centering
\includegraphics[width=0.65\textwidth]{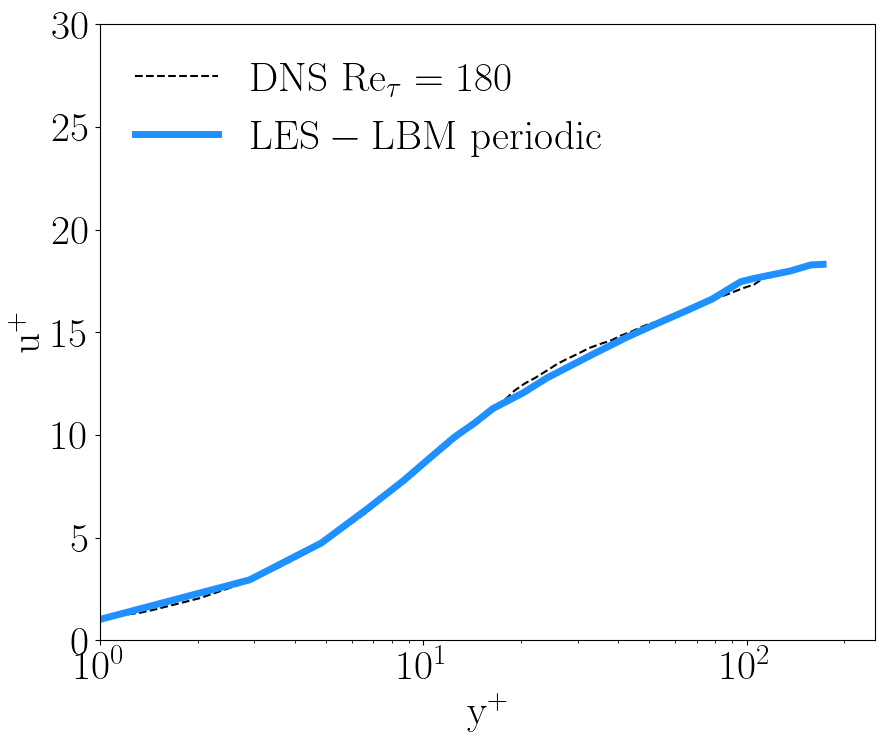}
\caption{\textbf{Comparison of dimensionless mean streamwise velocity profiles in wall units ($u^+$ vs. $y^+$) between the LES–LBM simulation with periodic boundary conditions and DNS data at $\text{Re}_\tau = 180$.} The LES–LBM result (solid blue) closely follows the reference DNS profile (dashed black), validating the accuracy of the LBM in capturing the near-wall behaviour and log-layer scaling of turbulent channel flow.}
\label{fig:lbm}
\end{figure}

Fig.~\ref{fig:lbm} presents a comparison between the streamwise mean velocity profile obtained from our large eddy simulation using the lattice Boltzmann method (LES–LBM) and benchmark direct numerical simulation (DNS) data at a friction Reynolds number of $\text{Re}\tau = 180$. The velocity is normalized in wall units, where $u^+ = u/u_\tau$ and $y^+ = y u_\tau / \nu$.  The LES–LBM result shows good agreement with the DNS reference across both the viscous sublayer and the logarithmic region, indicating that the model successfully captures key features of wall-bounded turbulence under periodic boundary conditions.

\begin{figure}[htbp!]
\centering
\includegraphics[width=0.65\textwidth]{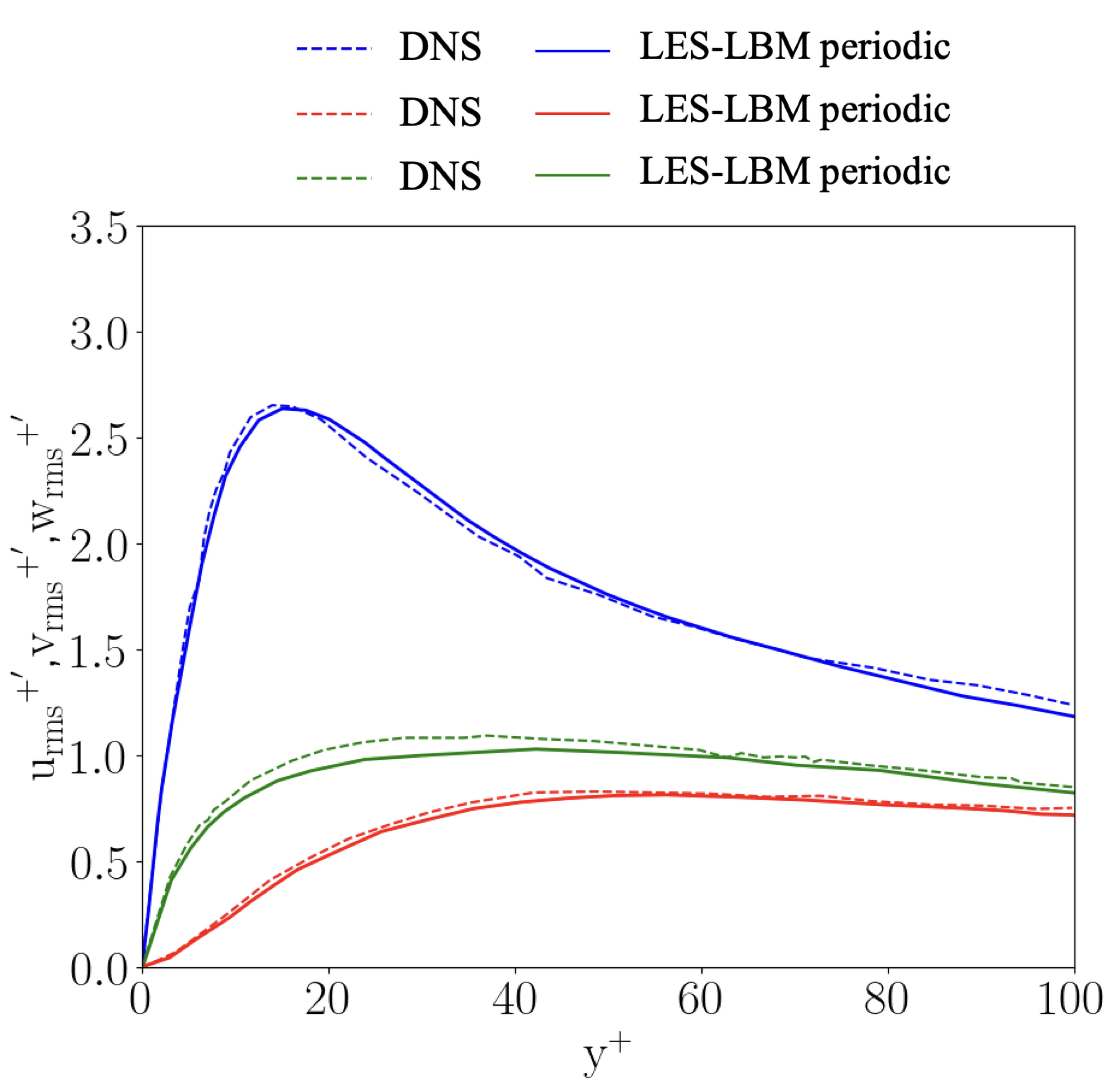}
\caption{\textbf{Comparison of root-mean-square (RMS) velocity fluctuations in dimensionless wall units between LES–LBM simulations with periodic boundary conditions and DNS data at $\text{Re}_\tau = 180$.} The streamwise (blue), wall-normal (red), and spanwise (green) RMS components are shown. The LES–LBM results (solid lines) closely match the DNS reference (dashed lines), particularly in the near-wall region, capturing the peak in $u_{\mathrm{rms}}^{+'}$ and maintaining the correct anisotropy among components.}
\label{fig:lbm_uvw}
\end{figure}

~\ref{fig:lbm_uvw} shows the comparison of RMS velocity fluctuations between our LES–LBM simulations and DNS data at $\text{Re}_\tau = 180$. The LES–LBM results (solid lines) exhibit good agreement with DNS (dashed lines), particularly in reproducing the characteristic peak in $u_{\mathrm{rms}}^{+'}$ near $y^+ \approx 15$ and capturing the general anisotropic distribution of turbulent fluctuations across the channel height. 

\subsection{The Multiple-Relaxation Time Lattice Boltzmann Moment Space Transformation Matrix}\label{sec:matrix}
Let's revisit the evolution equation for the distribution functions expressed as:
\begin{equation}
\label{eq:mrtlbe}
\mathbf{f}(\mathbf{x}+\mathbf{c}_{i}\Delta t,t+\Delta t) =\mathbf{f}(\mathbf{x}, t) - \mathbf{M}^{-1}\mathbf{S}\mathbf{M} \left[\mathbf{f}(\mathbf{x},t )-\mathbf{f}^{\mbox{ eq}}(\mathbf{x},t )\right] + \mathbf{F}(\mathbf{x}, t) \Delta t,
\end{equation}
where $\mathbf{M}$ denotes the moment space transformation matrix for the multiple relaxation time (MRT) collision kernel\cite{d2002multiple}. The explicit matrix is shown in the supplementary Table \ref{tab:matrix_M}.
\section{Unit Conversion between LBM Units and Physical Units}
\label{sec:unit}
In this subsection, we demonstrate the unit conversion between LBU and physical units in both velocity and distance. Within the LBM simulation, we can obtain the LBM velocity $\mathbf{u}_{\text{LB}}(\mathbf{x}, t)$ at location $\mathbf{x}$. The physical unit of the velocity, $\mathbf{u}_{\text{phys}}(\mathbf{x}, t)$, can be written as
\begin{equation}
\label{eq:u_transfer}
{u}_{\text{phys}}(\mathbf{x}, t) = {u}_{\text{LB}}(\mathbf{x}, t) \frac{c_x}{c_t},
\end{equation}
where $c_x$ and  $c_t$ are the conversion factors from the lattice Boltzmann simulation to the physical system, which is defined as
\begin{equation}
\label{eq:c_factor}
c_x = \frac{L_{\text{phys}}}{L_{\text{LB}}},\hspace{.6in} c_t = \frac{t_{\text{phys}}}{t_{\text{LB}}}, 
\end{equation}
where $L_{\text{phys}}$ and $t_{\text{phys}}$ represent the space and time units from the physical system, while $L_{\text{LB}}$ and $t_{\text{LB}}$ are the space and time lattice Boltzmann units from the lattice Boltzmann simulation. Accordingly, we can also obtained the physical distance $y_{\text{phys}}$ from the LBM length scale: 
\begin{equation}
\label{eq:y_transfer}
y_{\text{phys}} = y_{\text{LB}} c_x.
\end{equation}
For example, in Eq. \ref{eq:c_factor}, $L_{\text{phys}}$ may denote the height of the channel as $L_{\text{phys}}=2 \mathrm{m}$ so that $L_{\text{LB}} = 20$ LBU. Then, $c_x = 0.1\mathrm{m}$. Similar calculations can be performed to determine the time scale conversion factor $c_t$.

\section{Neural Networks Architecture Details}
\subsection{Quantum Machine Learning Model Architecture}
The quantum machine learning model consists of parameterized quantum circuits trained to approximate target Q-Priors via a quantum generator; the optimisation is performed traditionally through gradient-based updates, as detailed in Supplementary S1 and illustrated in Fig. 1 of the main text.

In our QIML architecture, quantum and traditional computations are optimized independently but operate in a tightly coupled feedback loop. The outputs of traditional modules dynamically guide quantum circuit objectives, establishing a synergistic interaction that underscores the relevance and practicality of hybrid quantum–traditional integration.

\subsection{Model Architecture for 2D chaotic flows}
The Koopman-based model begins with a patch embedding layer implemented using a single 2D convolutional layer. This layer transforms the input from 1 channel to 32 channels while maintaining a spatial resolution of $(64, 64)$. No activation function is specified for this layer.

Following the embedding, the encoder consists of three transformer blocks. The first block contains 2 layers and maps features from 32 channels to 64 channels, reducing the spatial dimensions to $(32, 32)$, with ReLU activations. The second block has 3 layers and increases the channels from 64 to 128 while downsampling to $(8, 8)$, again using ReLU activations. The third block also comprises 3 layers, maintaining 128 channels and compressing the spatial dimensions to $(2, 2)$, with ReLU activation applied.

After encoding, the representation is passed through a fully connected layer that operates on a flattened feature vector of size $128 \times 2 \times 2$. The activation function is not specified for this operator. Next, the model includes a fully connected layer applied to the same flattened $128 \times 2 \times 2$ feature vector, without a specified activation. 

A third component, the backward component, is similarly implemented using a fully connected layer acting on the $128 \times 2 \times 2$ vector, with no activation mentioned.

The decoder mirrors the encoder architecture in reverse. The first decoder block uses 3 transformer layers to maintain 128 channels and upsample to $(8, 8)$ with ReLU activation. The second block also has 3 layers, reducing the channels from 128 to 64 and increasing the spatial resolution to $(32, 32)$, again with ReLU. The third block consists of 2 layers, mapping 64 channels to 32 with output size $(64, 64)$ and ReLU activations. Finally, a single 2D convolutional layer reduces the channels from 32 back to 1 while preserving the $(64, 64)$ spatial resolution, with no activation function specified.

Parameter comparison across different models (QIML, Koopman, FNO and MNO) can be found in Table~\ref{tab:param_counts}. 

\subsection{Fourier Neural Operator Architecture}
The Fourier Neural Operator is a neural network architecture designed to learn mappings between function spaces, with a particular focus on solving partial differential equations and modelling spatial-temporal systems. Unlike traditional convolutional networks, it operates in the frequency domain, enabling efficient learning of long-range dependencies.

The processing pipeline in this paper consists of three main stages: 

\textbf{Lifting}, which maps the input tensor from $C_{\text{in}}=1$ to $C_{\text{hidden}}=128$ channels using a $1 \times 1$ convolution:
\begin{equation}
    \mathbf{X}_1 = \text{Conv}_{1\times 1}(\mathbf{X}_0) \,,
\end{equation}
where $\mathbf{X}_0 \in \mathbb{R}^{B,1,H,W}$ and $\mathbf{X_1} \in \mathbb{R}^{B,128,H,W}$.

\textbf{Four Fourier layers}, each of which is formed by spectral convolutional blocks with residual connections. The spectral convolution block first applies a 2D Fast Fourier Transform to the input tensor along its spatial dimensions. Then, a mode truncation step retains only the lowest $n_{\text{mode}} = (32,32)$ frequency components. These retained modes are acted upon by complex-valued spectral weights $\mathbf{W}_{\text{spec}} \in \mathbb{C}^{C_{\text{out}}\times C_{\text{in}}\times 32\times 32}$. To reduce storage and computational cost, these spectral weights are factorized using Tucker factorization as:
\begin{equation}
    \mathbf{W}_{\text{spec}} \approx \mathbf{U}_{\text{out}} \times \mathbf{G} \times \mathbf{U}_{\text{in}} \,,
\end{equation}
where $\mathbf{U}_{\text{in}} \in \mathbb{R}^{C_{\text{in}}\times r_{\text{in}}}$, $\mathbf{U}_{\text{out}} \in \mathbb{R}^{C_{\text{out}}\times r_{\text{out}}}$ and $\mathbf{G} \in \mathbb{C}^{r_{\text{in}}\times r_{\text{out}}\times 32\times 32}$. The modified spectrum is then transformed back to the spatial domain via an inverse FFT. In parallel, a pointwise convolution path applies a $1\times 1$ convolution to the original input tensor to retain local information. Finally, a residual connection combines these paths with the original input, followed by a GELU non-linear activation function:
\begin{equation}
    \mathbf{X_{\mathcal{L}+1}} = \sigma(\text{Spectral}(\mathbf{X}_\mathcal{L})+\text{Pointwise}(\mathbf{X}_\mathcal{L})+\mathbf{X}_\mathcal{L}) \,.
\end{equation}

\textbf{Projection}, the final stage, which reduces the hidden representation from $C_{\text{hidden}}=128$ to $C_{\text{out}}=1$ channels using a 2-layer $1\times 1$ convolutional MLP:
\begin{equation}
    \mathbf{Y} = \text{Conv}_{1\times 1}(\sigma(\text{Conv}_{1\times 1}(\mathbf{X}_\mathcal{L}))) \,,
\end{equation}
with intermediate channels set to $r_{\text{proj}}\times C_{\text{hidden}} = 1 \times 128 = 128$.

See Table \ref{tab:param_counts} for the total parameters of the FNO model.

\subsection{Markov Neural Operator Architecture}

The Markov Neural Operator is a framework to learn discretization-independent mappings between function spaces. In this approach, the learned operator is expressed as a finite composition of Markov kernel layers, each representing a local operator that evolves the state over a discrete step in ``operator depth.”

The architecture setting in this paper consists of three main stages:

\textbf{Lifting}, where a $1\times 1$ convolution projects the input tensor from $C_{\text{in}}=1$ to $C_{\text{hidden}}=128$:
\begin{equation}
    \mathbf{X}_1 = \text{Conv}_{1\times 1}(\mathbf{X}_0) \,,
\end{equation}
where $\mathbf{X}_0 \in \mathbb{R}^{B,1,H,W}$ and $\mathbf{X}_1 \in \mathbb{R}^{B,128,H,W}$.

\textbf{Single Markov kernel block}, which consists of four sequential Markov kernel layers. Each layer is implemented via a spectral convolution in Fourier space combined with local pointwise updates, following a similar setting as the FNO architecture described previously. A residual connection combines these paths, but the non-linear activation function used is SELU. This composition ensures that each layer satisfies the Markov property, where the updated state depends only on the previous state.

\textbf{Projection}, the final stage, which reduces the hidden representation from $C_{\text{hidden}}=128$ back to $C_{\text{out}}=1$ channels.

Another key part of the MNO is its explicit stability features, which are realized by incorporating dissipative target mappings and partition of unity post-processing in the training loop. First, training inputs are drawn using spherical shell sampling:
\begin{equation}
    \mathbf{x} \sim \mathcal{U}(\mathcal{S}(R_{\text{inner}}, R_{\text{outer}}))
\end{equation}
where $\mathcal{S}(r_1,r_2)=\{\mathbf{v}\in \mathbb{R}^n \,|\, r_1 \leq \lVert \mathbf{v}\rVert_2 \leq r_2 \}$. This ensures diverse energy levels are present in the training set. Second, a baseline dissipative target mapping is defined as:
\begin{equation}
    \mathcal{D}(\mathbf{x}) = s \cdot \mathbf{x} \,, \quad 0<s<1
\end{equation}
where we use $s=0.8$. This map serves as a stable fallback for states outside the data manifold. Finally, a partition of unity correction blends the learned mapping $f_\theta$ with the dissipative baseline $\mathcal{D}$ using a smooth partition of unity function $\rho$:
\begin{equation}
    \hat{\mathbf{y}}(\mathbf{x}) = \rho(\lVert \mathbf{x} \rVert) \cdot f_\theta(\mathbf{x}) + (1-\rho(\lVert \mathbf{x}\rVert))\cdot\mathcal{D}(\mathbf{x})
\end{equation}
with $\rho$ defined as a sigmoid function $\rho = (1+e^{s(r-c)})^{-1}$, where $c$ is the shift and $s$ controls the transition steepness. This ensures a high-fidelity learned output when $\rho \approx 1$ (inside the training region) and a stable dissipative fallback when $\rho \approx 0$ (outside the training region).

See Table \ref{tab:param_counts} for the total parameters of the MNO model.

\begin{table}[h!]
\centering
\caption{Total parameter counts for the compared models.}
\label{tab:param_counts}
\begin{tabular}{l r}
\hline
\textbf{Model} & \textbf{Total Parameters} \\
\hline
Q-Prior (in the QIML) & $<$300  \\
C-Prior (VAE comparable) & 124,418  \\
C-Prior (VAE fail) & 300  \\
Koopman  & 4,534,872  \\
FNO      & 35,971,333 \\
MNO      & 45,823,681 \\
\hline
\end{tabular}
\end{table}

\subsection{Variational Autoencoder Baseline for the Classical Prior}

\R{To provide a classical generative baseline aligned with the Q-Prior, we consider a standard VAE architecture to construct a C-Prior. The VAE is employed purely as an unconditional generative model for learning the invariant marginal distribution of local velocity magnitudes, rather than for reconstructing full spatial fields or performing conditional prediction.}

\begin{figure*}[htbp!]
\centering
\includegraphics[width=0.95\textwidth]{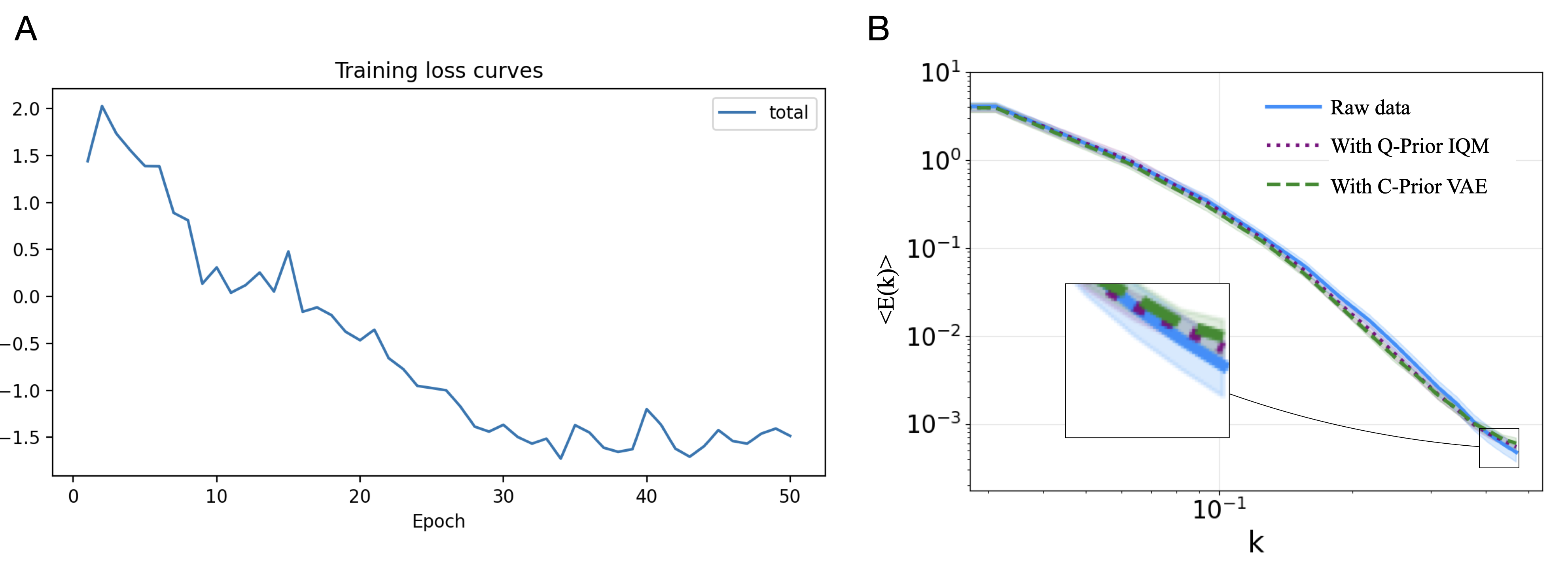}
\caption{\R{\textbf{Training behaviour and spectral comparison for Q-Prior and C-Prior in turbulent channel flow.} A. Representative training loss curve for the classical generative prior, illustrating stable convergence over optimisation epochs.
B. Time-averaged kinetic energy spectra E(k) for the turbulent channel inflow, comparing the ground-truth data with predictions obtained using the Q-Prior ($300$ parameters) and the VAE-based C-Prior (more than $10^5$ parameters). While both priors reproduce the large-scale spectral behaviour, deviations become apparent in the middle and high-wavenumber regime, where the Q-Prior more faithfully preserves the energy content at small scales. Insets highlight the differences in the spectral tail.}}
\label{fig:ciml_comparison}
\end{figure*}

\R{The VAE architecture follows a canonical encoder--decoder design operating on scalar inputs~\cite{kingma2013auto}. The encoder maps a one-dimensional velocity sample $v \in \mathbb{R}$ to a latent representation $z \in \mathbb{R}^{d_z}$ via a multilayer perceptron, producing the parameters of a Gaussian variational posterior $q_\phi(z|v) = \mathcal{N}(\mu(v), \sigma^2(v))$. The decoder is a symmetric MLP that maps samples drawn from the latent prior $p(z)=\mathcal{N}(0,I)$ back to the velocity space. Training is performed by maximising the standard evidence lower bound, consisting of a reconstruction term and a Kullback--Leibler regularisation term.
To ensure a fair and controlled comparison, the learning procedure for the C-Prior is designed to be statistically identical to that of the Q-Prior. Specifically, the VAE is trained on the same dataset and target quantity. Consequently, the only distinction between the C-Prior and Q-Prior lies in the representational mechanism, classical latent variables versus a parameterized quantum circuit, rather than in the data, objective, or training protocol.}

\R{Using a sufficiently high-capacity VAE (more than $10^5$ trainable parameters), the resulting C-Prior is able to reproduce the coarse-grained energy distribution and large-scale statistics at a level comparable to the Q-Prior (see Fig.~\ref{fig:comparison}). However, beyond intermediate prediction windows, the corresponding CIML rollouts exhibit a tendency toward dynamical saturation, with the predicted flow becoming progressively less variable in time and approaching a quasi-stationary configuration.
Consistent with this observation, Fig.~\ref{fig:ciml_comparison} shows modest but systematic deviations in the high-wavenumber regime, where the classical prior slightly underestimates the tail of the energy spectrum relative to QIML. While these differences are small at the level of marginal distributions, they can accumulate during long-term autoregressive prediction and contribute to a gradual loss of dynamical richness when the C-Prior is incorporated into the Koopman-based predictor.}

\R{We further note that achieving this level of performance with a C-Prior requires a substantially larger number of parameters. In our experiments, the VAE-based C-Prior employs on the order of more than $10^5$ trainable parameters, compared to approximately $300$ parameters in the quantum prior. For completeness and fairness, we also tested a VAE constrained to a comparable parameter budget of approximately $300$ parameters; in this regime, the model was unable to train stably and collapsed to an overly smooth, mean-field representation, failing to capture meaningful fine-scale statistics. This contrast highlights the parameter efficiency of the Q-Prior in representing multiscale statistical features relevant for chaotic dynamics. A summary of parameter counts for all models considered is provided in Table~\ref{tab:param_counts}.}

\subsection{Integration of the Traditional Model and the Quantum Prior Guidance}
In our QIML framework, the integration of quantum and traditional components is achieved by embedding a quantum-learned invariant distribution into the loss function of a traditional machine learning model tasked with forecasting high-dimensional dynamical systems. This coupling is designed to address a fundamental limitation in traditional neural PDE solvers: while they can capture short-term evolution, their predictive accuracy typically deteriorates over longer horizons due to the accumulation of numerical error and sensitivity to initial conditions.

The traditional machine learning model, constructed using convolutional or transformer-based architectures, is trained to predict the next-step velocity field $\hat{u}_{t+1}$ from a history of past states $\{u_t, u_{t-1}, \dots\}$. To enhance robustness in long-term rollouts, we incorporate a distributional Q-Prior $p_\theta(x)$ learned by the quantum generator, which approximates the system’s invariant measure.

The initial loss formulation includes a reconstruction term and a Kullback–Leibler divergence between the predicted and quantum-learned distributions:

\begin{equation}
\mathcal{L}_{\text{total}} = \mathcal{L}_{\text{recon}} + \mathcal{L}_{\text{unitary}} + \lambda_{\text{KL}} \mathcal{L}_{\text{KL}},
\end{equation}
where

\begin{equation}
\mathcal{L}_{\text{unitary}} = |K^\top K - I|_F^2,
\text{ for details, see the main text, }\end{equation}

\begin{equation}
\mathcal{L}_{\text{recon}} = | \hat{u}_{t+1} - u_{t+1} |^2,
\end{equation}

and

\begin{equation}
\mathcal{L}_{\text{KL}} = D_{\text{KL}}\left( \hat{q}(x) | p_\theta(x) \right),
\end{equation}
with $\hat{q}(x)$ denoting the empirical distribution derived from the predicted field $\hat{u}_{t+1}$.

While this formulation successfully captures the global distributional shape, it often converges to a low-variance mean-field solution, insufficient to preserve local or high-frequency structures critical for short-term dynamics. To mitigate this, we introduce two additional constraints.

First, we add a MMD term to better match higher-order statistics:

\begin{equation}
\mathcal{L}_{\text{MMD}} = \left| \mathbb{E}{x \sim \hat{q}(x)}[\phi(x)] - \mathbb{E}{x \sim p_\theta(x)}[\phi(x)] \right|^2_{\mathcal{H}},
\end{equation}
where $\phi(x)$ is a feature mapping into a reproducing kernel Hilbert space.

Second, in some cases (TCF), we also incorporate a peak loss to explicitly align dominant modes of the predicted and target distributions:

\begin{equation}
\mathcal{L}_{\text{peak}} = \left| \text{TopK}(\hat{q}(x)) - \text{TopK}(p_\theta(x)) \right|^2,
\end{equation}
where \text{TopK} selects the highest-probability bins from both histograms.

The final training loss is a weighted combination of all four components:

\begin{equation}
\mathcal{L}_{\text{total}} = \mathcal{L}_{\text{recon}} + \lambda_{\text{KL}} \mathcal{L}_{\text{KL}} + \lambda_{\text{MMD}} \mathcal{L}_{\text{MMD}} + \lambda_{\text{peak}} \mathcal{L}_{\text{peak}}.
\end{equation}

This composite objective ensures that predictions remain consistent with the invariant statistics of the underlying physical system, while also preserving critical local features necessary for short- and mid-term forecasting. Our results demonstrate that Q-Priors can significantly enhance the stability and accuracy of machine learning models in high-dimensional chaotic regimes.

\begin{figure*}[htbp!]
\centering
\includegraphics[width=0.75\textwidth]{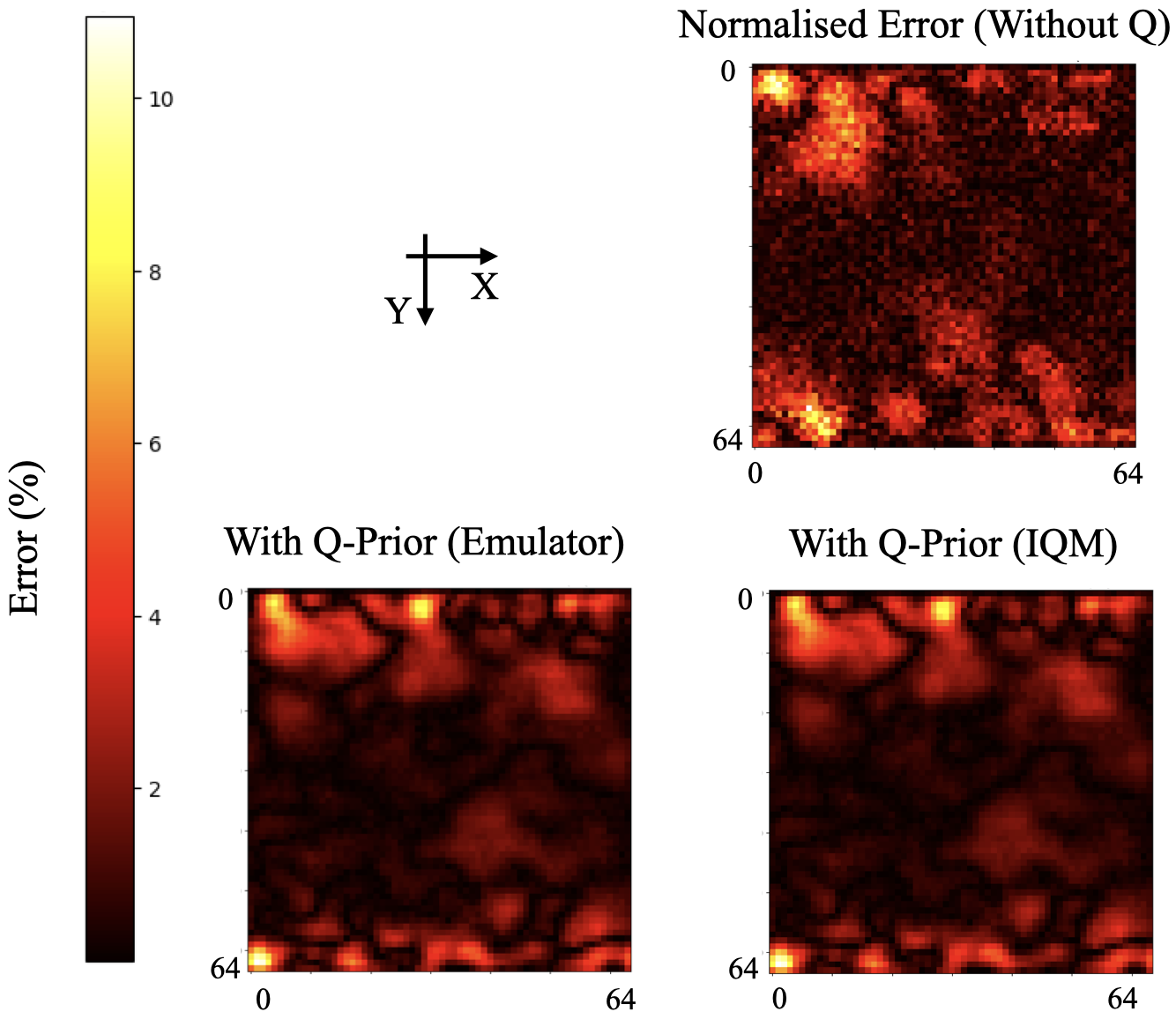}
\caption{\textbf{Diagram of the normalized error of the turbulent channel flow.} The baseline model without the Q-Prior (top right) is compared against the QIML model on a classical emulator (bottom left) and on IQM quantum hardware (bottom right). The QIML model demonstrates a marked reduction in error compared to its classical counterparts.}
\label{fig:tcf_ne}
\end{figure*}

\begin{figure}[htbp]
\centering
\includegraphics[width=0.62\textwidth]{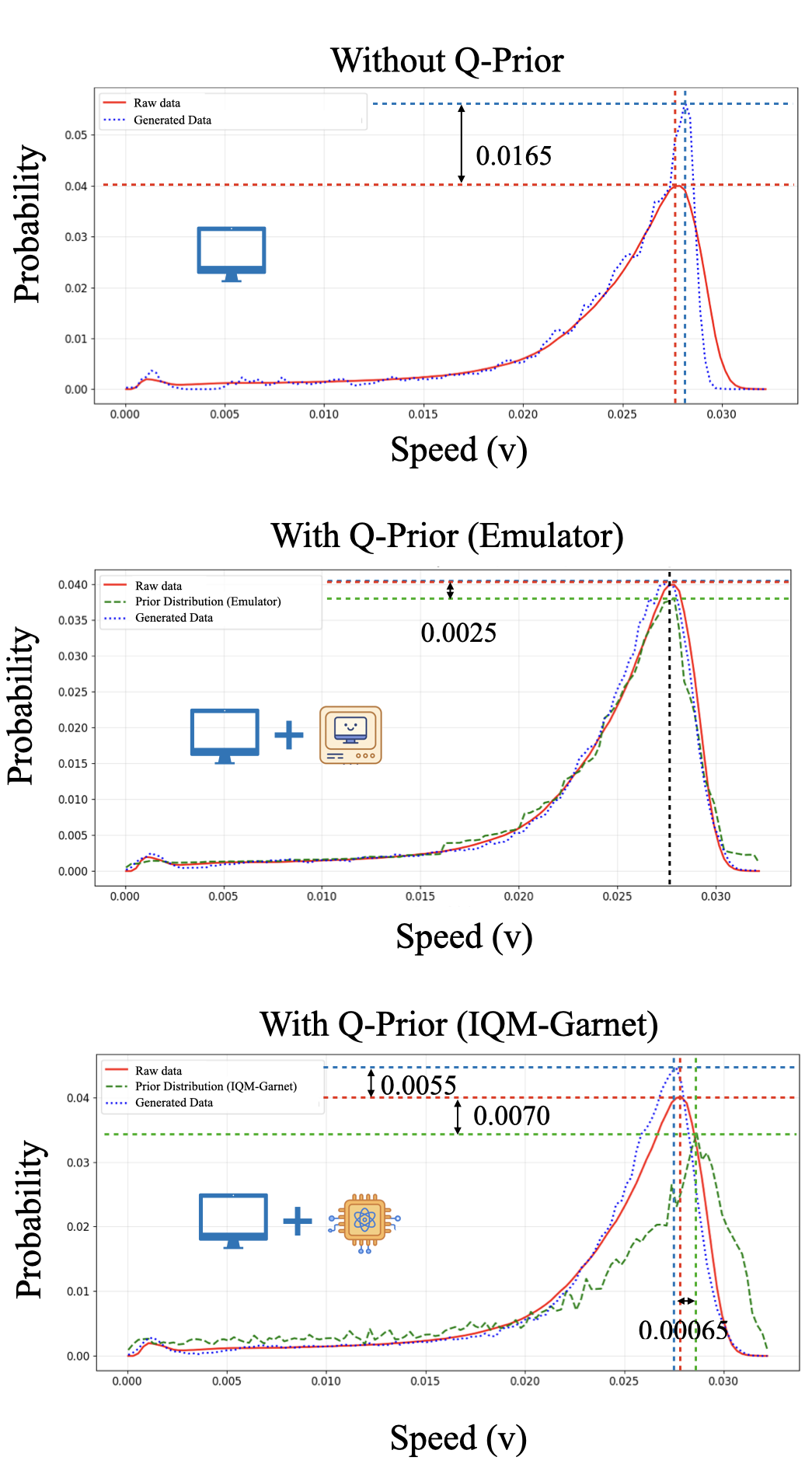}
\caption{\textbf{Diagram of the velocity distribution of the turbulent channel flow predicted under the same three learning regimes (no Q-Prior, emulator-based Q-Prior, and IQM hardware-based Q-Prior), compared to the target distribution.} Without quantum guidance, the model fails to capture the dominant statistical mode—both the modal peak and its amplitude are significantly misaligned. With Q-Prior, the predicted distribution nearly perfectly matches the raw data.}
\label{fig:tcf_dis}
\end{figure}
Fig.~\ref{fig:tcf_ne}  presents the spatial distribution of the normalized velocity error across the domain under three modelling regimes. The top panel displays error patterns obtained without Q-Priors, revealing widespread, spatially incoherent deviations with peak errors exceeding 10\%. Incorporating a Q-Prior from a classical emulator (middle panel) leads to a pronounced reduction in error magnitude and spatial variance, with the majority of regions falling below 4\%. Further improvement is observed when employing the Q-Prior trained on real quantum hardware (bottom panel), where error distributions are both quantitatively reduced and exhibit greater spatial smoothness, indicating enhanced generalization and stability in chaotic regime reconstruction.
Fig. \ref{fig:tcf_dis} shows the TCF velocity magnitude distribution $\hat{q}(x)$ predicted under the same three learning regimes (no Q-Prior, emulator-based Q-Prior, and hardware-based Q-Prior), compared to the empirical target distribution. Without quantum guidance, the model fails to capture the dominant statistical mode—both the modal peak and its amplitude are significantly misaligned. With simulator-based Q-Prior, the predicted distribution nearly perfectly matches the raw data in both location and shape, confirming the Q-Prior’s capacity to encode high-dimensional invariant measures with minimal parameterization. Hardware-based Q-Prior introduces moderate deviations, including a downward shift in peak probability and a small modal displacement; however, the distribution remains substantially closer to the ground truth than the classical result.

\begin{table}[h!]
\centering
\setlength{\tabcolsep}{3pt}
\caption{Quantum resources and storage compression achieved by the Q-Prior.}
\label{tab:compression}
\begin{tabular}{lcccccc}
\toprule
\textbf{System} &
\textbf{Qubits} &
\textbf{Trainable} \(\boldsymbol{\theta}\) &
\textbf{Raw data} &
\textbf{Q-Prior file} &
\textbf{Compression} &
\textbf{Device}\\[-2pt]
& & \textbf{parameters} & (full set) & (full set) & (ratio)\\
\midrule
Kuramoto–Sivashinsky & 10 & \(\sim120\) & \(300\;\mathrm{MB}\) & \(0.25\;\mathrm{MB}\) & \(\approx 1.2\times10^{3}\!:\!1\) & Emulator \\
2D Kolmogorov flow   & 10 & \(\sim180\) & \(400\;\mathrm{MB}\) & \(0.40\;\mathrm{MB}\) & \(\approx   10^{3}\!:\!1\) & Emulator \\
Turbulent channel flow & 15 & \(\sim300\) & \(500\;\mathrm{MB}\) & \(2.3\;\mathrm{MB}\) & \(\approx        2.2\times10^{2}\!:\!1\) & Emulator \\
Turbulent channel flow & 10 & \(\sim240\) & \(500\;\mathrm{MB}\) & \(2.0\;\mathrm{MB}\) & \(\approx        2.5\times10^{2}\!:\!1\) & IQM-Garnet \\
\bottomrule
\end{tabular}
\end{table}

\section{Quantum Parameter Efficiency and Quantum Memory Advantage}
This section explains the parameter efficiency and memory advantage for the QIML. According to the parameter numbers in this work, as shown in Table~\ref{tab:compression}, we used 120 parameters on the KS system, 180 on the Kolmogorov flows on the emulator, and 240 or 300 for the TCF system on the hardware and emulator. As is evident, our method uses a remarkably small number of parameters to learn the necessary statistical information. We have already shown the orders-of-magnitude difference in parameter counts compared to classical models in Table~\ref{tab:param_counts}. Both the parameter efficiency discussed here and the memory advantage that follows originate from the same foundational quantum advantage~\cite{huang2025vast}: the ability of a quantum circuit to logarithmically compress classical data. This is the fundamental source of the quantum advantage established in our work, a topic we have discussed in detail in section S6. The critical challenge in harnessing this advantage is circumventing the limitations imposed by Holevo's bound. This bound arises because retrieving the complete classical information from the quantum state would necessitate an exponential number of measurements, rendering the approach impractical. This obstacle is overcome because the task of guiding a dynamical model does not necessitate access to the complete classical dataset. Instead, the framework's objective is to extract a statistical summary, the distribution of velocity information, which represents the system's invariant measure. For a chaotic system, such a measure inherently contains far less information than the full dataset from which it is derived. The ability to function with only this compressed distributional information allows the method to bypass the need for an exponential number of measurements. This principle is the foundation for the quantum memory advantage demonstrated empirically for the three systems investigated in this section. \R{Before detailing these quantitative results, we first proceed with a formal theoretical argument for the aforementioned quantum advantage.}

\subsection{Theoretical argument for the quantum memory advantage via QIML}
\label{sec:theory_efficiency}

\R{Our QIML framework employs a Parameterized Quantum Circuit (PQC), denoted \(U(\theta)\), acting on \(n_q\) qubits to learn the invariant measure \(\mu(x)\) associated with a high-dimensional chaotic dynamical system. The quantum state generated by the circuit, \(\ket{\psi_\theta} = U(\theta)\ket{0}^{\otimes n_q}\), resides in a Hilbert space \(\mathcal{H}\) of dimension \(N = 2^{n_q}\). A naive reconstruction of \(\ket{\psi_\theta}\) would indeed require resources that scale exponentially with \(n_q\), apparently contradicting our assertion of efficiency based on a polynomial number of measurements, \(S = \mathrm{poly}(n_q)\). Furthermore, Holevo’s theorem~\cite{holevo1973bounds} limits the classical information obtainable per measurement to \(n_q\) bits. We now formalize why our learning task circumvents these exponential barriers.}

\R{The essential insight arises from the structure of the target distribution \(\mu(x)\) and the nature of the learning objective. For a dissipative chaotic system with a discrete state space \(X \subseteq \{0,1\}^{n_q}\), the invariant measure \(\mu(x)\) is supported on a lower-dimensional strange attractor \(\mathcal{A} \subset X\). The complexity of \(\mu(x)\) is therefore governed not by the ambient dimension \(N\), but by quantities such as its information dimension \(D_1(\mu)\)~\cite{Farmer1983}, which for physical systems typically scales sub-exponentially, often polynomially or even remaining constant with respect to \(n_q\). Let \(C(\mu, \epsilon)\) denote the minimal information required to specify \(\mu(x)\) up to accuracy \(\epsilon\). The compressibility of the invariant measure implies
\begin{equation}
    C(\mu, \epsilon) \ll O(N) = O(2^{n_q}).
    \label{eq:measure_complexity}
\end{equation}}


\R{Furthermore, in simple terms, according to statistical learning theory~\cite{Vapnik1998, Gretton2012}, the number of samples \(S'\) required to estimate the objective or its gradient with accuracy \(\epsilon\) scales polynomially with the number of parameters and inverse precision:
\begin{equation}
    S' = O(\mathrm{poly}(k, 1/\epsilon)).
    \label{eq:sample_complexity}
\end{equation}
Since \(k = \mathrm{poly}(n_q, L_{\text{circ}})\), the required number of quantum measurements \(S'\) scales polynomially with \(n_q\). Holevo’s bound does not impose an exponential limitation here, because we are not attempting full quantum state tomography of \(\ket{\psi_\theta}\). Instead, the quantum device is used solely to generate classical samples from \(p_\theta(x)\), providing stochastic estimates of the loss function for the classical optimizer. In this setting, the quantum circuit acts as an efficient implicit sampler parameterized by \(\theta\), whose complexity is dictated by the number of trainable parameters \(k\). }

\R{In conclusion, the overall efficiency of QIML arises from two concurrent, complementary factors: 
\begin{enumerate}
\item The intrinsic compressibility of the target measure $\mu(x)$, expressed in Eq.~\eqref{eq:measure_complexity}. This factor represents the compression advantage arising from the exponentially large Hilbert space, where a complex classical probability distribution can be efficiently extracted and then compactly encoded within a quantum state.
\item The polynomial sample complexity of learning within the parameterized manifold $\mathcal{P}$, as shown in Eq.~\eqref{eq:sample_complexity}. This represents the sampling advantage offered by quantum computing, allowing us to efficiently extract effective information from the complex quantum state.
\end{enumerate}
We firstly proceed to quantify the resulting memory advantage achieved for the three specific systems investigated in this work and will establish the theoretical basis for this efficiency in section~S10 below.}

\subsection{Kuramoto–Sivashinsky Equation}
The raw KS data comprise
\(1200\) trajectories, each recorded for
\(2000\) time steps on a \(512\)-point spatial grid,
totalling $\sim\!12.3\,$GB.
For training we down-sample to \(N_{\text{traj}}\!=\!1200\) trajectories, each
with \(256\) temporal frames and \(128\) spatial points, leading to a working
tensor of shape \((256,128)\) per trajectory.
With double precision (\(8\) bytes per value) a single trajectory occupies
\begin{equation}
256 \times 128 \times 8 = 2.62 \times 10^{5} \; \text{bytes} \simeq 0.25 \; \text{MB}.
\end{equation}
Thus, the full down-sampled dataset requires
\(0.25\,\text{MB}\times1200 \approx 300\,\text{MB}\).
Each trajectory is paired with an individual quantum generator containing
\(n=10\) qubits, \(L=4\) circuit layers and \(120\) rotation parameters,
for a raw weight file of \(120\times8 = 960\) bytes; including metadata the
stored checkpoint is \(\approx1.0\) kB.
Archiving one checkpoint per trajectory, therefore, costs
\(1200\times1.0\,\text{kB} \approx 1.2\,\text{MB}\).
Consequently, the learned Q-Priors reduce storage from
\(\sim300\,\text{MB}\) to \(\sim1.2\,\text{MB}\), a compression factor of
around \(250{:}1\) while still capturing the essential
statistics used by the classical machine learning model.

\subsection{2D Kolmogorov Flows}
The total dataset comprises $40$ trajectories, each containing
$320$ temporal snapshots on a $256\times256$ grid.
For a single velocity component, one snapshot occupies
\begin{equation}
256^{2} \times 8 = 524,288 \; \text{bytes} \simeq 0.50 \; \text{MB}.
\end{equation}
so that one complete trajectory requires
\(0.50\text{ MB}\times320\approx160\text{ MB}\)
and the full set of $40$ trajectories stores
$\sim6.4\text{ GB}$.
The corresponding quantum generator employs $n=10$ qubits, $L=6$ layers and
$180$ trainable rotation angles.
Saved in double precision, one checkpoint is
\(180\times8=1{,}440\text{ bytes}\simeq1.4\text{ kB}\).
Archiving a separate checkpoint for every snapshot therefore costs  
\begin{equation}
1.4 \; \text{kB} \times 320 = 448 \; \text{kB} \simeq 0.44 \; \text{MB}.
\end{equation}
yielding a compression factor of
\begin{equation}
\frac{160 \; \text{MB}}{0.44 \; \text{MB}} \approx 3.6 \times 10^{2}.
\end{equation}
while faithfully reproducing the empirical invariant measure.

\subsection{Turbulent Channel Flow}

For the TCF benchmark, each snapshot is stored as a $192\times192$ array of double-precision values (8 bytes per entry).
A single velocity component therefore, occupies
\begin{equation}
192^{2} \times 8 = 294,912 \; \text{bytes} \simeq 0.29 \; \text{MB}.
\end{equation}
so that the full sequence of 595 snapshots amounts to
\begin{equation}
0.29 \; \text{MB} \times 595 \approx 170 \; \text{MB}.
\end{equation}
Retaining all three velocity components raises the per-trajectory footprint to roughly 500 MB.

The quantum generators used to reproduce the same invariant statistics each employ $n=15$ qubits, $L=6$ layers, and 240 rotation angles. Each saved quantum generator checkpoint, with minimal metadata, is about 4.3 kB. Given that 595 such quantum generator models are retained (one for each snapshot), the total storage for the entire set of quantum generator checkpoints amounts to $595 \times 4.3\,\text{kB} \approx 2.5\,\text{MB}$. Relative to the raw three-component trajectory ($\sim\!500\,\text{MB}$), this represents a compression factor of approximately 200, which is over two orders of magnitude, while still capturing the statistics required by the classical Koopman machine learning model.

\section{Performance and Statistical Metrics}

This section provides the mathematical definitions for the primary metrics used in this paper to evaluate model performance and analyse the system's statistical properties.

\subsection{Temporal Autocorrelation}

The temporal autocorrelation is used to measure the memory of the dynamical system, quantifying the correlation of a time series with a delayed version of itself. For a discrete time series $u_i$ (the value at a specific spatial point at time step $i$), with mean $\bar{u}$ and total length $N$, the normalized temporal autocorrelation $C(k)$ at a time lag of $k$ steps is defined as:
\begin{equation}
C(k) = \frac{\sum_{i=1}^{N-k} (u_i - \bar{u})(u_{i+k} - \bar{u})}{\sum_{i=1}^{N} (u_i - \bar{u})^2}.
\end{equation}

\subsection{Error Metrics}

To quantify the difference between the ground truth data and the model predictions, the following error metrics are used.

The absolute error provides a direct, point-wise measure of the deviation between a predicted field $\hat{u}(\mathbf{x})$ and the ground truth field $u(\mathbf{x})$:
\begin{equation}
E_{\text{abs}}(\mathbf{x}) = \left|u(\mathbf{x}) - \hat{u}(\mathbf{x})\right|.
\end{equation}

To specifically quantify how well the model reproduces the temporal correlation structure, we define the normalized relative autocorrelation error $E_r$. This measures the relative L2 error between the true autocorrelation function $C_{\text{true}}(k)$ and the predicted one $C_{\text{pred}}(k)$:
\begin{equation}
E_{r} = \frac{\Vert C_{\text{true}}(k) - C_{\text{pred}}(k) \Vert_{2}}{\Vert C_{\text{true}}(k) \Vert_{2}}.
\end{equation}

\section{Argument for the Quantum Advantage in Representing Chaotic Measures}
\label{sec:rigorous_proof}

\R{In section S8.1 and the main text, we identified two theoretical sources of efficiency within the QIML framework: the compressibility of the invariant measure into a polynomial number of quantum parameters, and the polynomial sample complexity associated with learning this representation. In this section, we formalize these arguments using concepts from dynamical systems theory and computational complexity. We argue that the invariant measure of a high-dimensional chaotic system admits an efficient representation by a PQC, whereas classical generative models generically require super-polynomial resources to achieve comparable accuracy.}

\R{Our derivation establishes a logical progression from the physical properties of chaotic dynamics to the geometric structure of the invariant measure, and subsequently to the computational complexity of probabilistic modelling.}

\R{Consider a dissipative chaotic dynamical system governed by a map 
$\Phi : \mathcal{X} \to \mathcal{X}$ on a compact phase space 
$\mathcal{X} \subset \mathbb{R}^N$. The long-time behaviour is characterized by an invariant probability measure $\mu$, satisfying 
$\mu(E) = \mu(\Phi^{-1}(E))$ for any measurable set $E$.  
Classical ergodic theory guarantees the existence of such invariant measures~\cite{birkhoff1931proof,neumann1932proof}, and for high-dimensional chaotic flows such as the KS equation or Navier–Stokes turbulence, $\mu$ is typically a Sinai–Ruelle–Bowen (SRB) measure~\cite{eckmann1985ergodic}. For computational purposes, we consider a coarse-graining and discretization procedure mapping the phase space to bit-strings 
$x \in \{0,1\}^{n}$ with $n = O(N)$. The statements below refer to this discretized representation. At this resolution, chaotic invariant measures display several geometric and statistical features that make them difficult to approximate using standard classical generative models.}

\R{One key property is that $\operatorname{supp}(\mu)$ forms a strange attractor with non-integer (often fractal) dimension~\cite{farmer1983dimension}.  
The combination of nonlinear stretching and dissipative contraction yields a measure that is singular with respect to the ambient Lebesgue measure. Representing such a measure requires specifying an extensive collection of constraints that rule out exponentially many dynamically forbidden regions of phase space.}
\R{A second feature is the effective non-Markovian nature of the symbolic dynamics.  
While correlation functions decay, the conditional distribution of the next state does not collapse to any finite-order Markov approximation at physically relevant resolution.  
Capturing the statistical structure of $\mu$ therefore requires access to information distributed across multiple scales of the attractor.}
\R{A third feature arises from the Koopman description of the dynamics~\cite{mezic2005spectral}. The observable $f$ evolves as $f \circ \Phi^t$, and due to sensitivity to initial conditions, its support typically spreads across all degrees of freedom.  
This spreading is analogous to operator growth or scrambling in quantum many-body systems.  
As a result, the invariant measure involves long-range and high-order correlations that do not factorize over low-order interaction graphs.}

\R{These properties motivate examining the expressivity limitations of classical generative models.  
Architectures such as finite-range Markov Random Fields, Restricted Boltzmann Machines with bounded connectivity, and convolutional networks impose local factorization constraints tied to bounded-treewidth graphs.  
Recent results on contextual and non-local probability distributions~\cite{gao2022enhancing} suggest that such local models may require super-polynomially many parameters to approximate distributions with global correlation structure.  
Although a fully rigorous lower bound for SRB measures remains an open problem, the combination of fractal geometry, non-Markovianity, and non-local structure indicates a significant mismatch between chaotic invariant measures and the inductive biases of classical local models.}

\R{PQCs offer a contrasting representational framework.  
Quantum circuits can generate global entanglement at depths scaling polylogarithmically in system size, allowing correlations across all qubits to be established without imposing local factorization constraints.  
Recent work has shown that multifractal and strongly correlated distributions can be generated efficiently by PQCs~\cite{wang2025parameter}, providing evidence that quantum states can encode structures qualitatively similar to those present in chaotic invariant measures.  
Furthermore, PQCs serve as universal approximators for unitary transformations.  
Since the Koopman operator is itself a linear (unitary on an appropriate function space) transformation, coarse-grained versions of its spectral structure may be represented within families of quantum circuits.  
Although this correspondence is heuristic, it suggests a mechanism by which PQCs can encode dynamical invariants using polynomial resources.}

\R{Sampling considerations further differentiate classical and quantum settings.  
Classical sampling from approximations of $\mu$ often relies on Markov Chain Monte Carlo methods.  
The complex, multimodal, and fractal geometry of chaotic attractors is known to create slow mixing and metastability in high dimensional regimes~\cite{jerrum1996markov,bengio2013better}, making sampling computationally expensive.  
In contrast, a PQC of depth $O(\mathrm{poly}(N))$ enables direct sampling from its Born distribution at a cost proportional to circuit depth, avoiding the need for Markov mixing altogether.  
Moreover, gradients of loss functions involving $p_\theta$ can be estimated via standard quantum differentiation rules, and recent theoretical work indicates potential advantages in learning distributional properties from quantum-prepared samples~\cite{huang2022quantum}.}

\R{Taken together, these considerations provide a conceptual framework for understanding the empirical behaviour observed in the QIML architecture.  
The invariant measures of chaotic systems possess geometric complexity and global correlation structure that challenge classical generative models based on local interactions or finite-order dependencies.  
Quantum circuits, through their ability to generate global entanglement and approximate high-dimensional dynamical structure, offer a plausible and potentially efficient representational basis for these measures.  
From another perspective, we note an intriguing analogy between our framework and the dynamical systems perspective on classical machine learning, as discussed in recent work on the physical interpretation of neural PDEs~\cite{succi2025physical}. In that view, the training process can be conceptualized as a discrete dynamical system evolving towards a local attractor that represents the target truth. Extending this analogy to the quantum domain, our QIML training can be seen as driving the quantum state $|\psi(\theta)\rangle$ within the projective Hilbert space towards a specific quantum attractor—a sub-manifold of states whose Born distributions align with the classical invariant measure. Investigating the convergence properties and geometry of these quantum optimisation landscapes through the lens of discrete dynamical systems offers a fertile ground for future theoretical research.
Furthermore, we note that related notions of effective dimensionality reduction have also been observed in purely classical uncertainty quantification and physics-informed machine learning. In particular, Edeling \textit{et al.}~\cite{edeling2024global} demonstrated that uncertainty arising from thousands of parameters in classical molecular dynamics force fields can, under a kernel-based sensitivity analysis, be reduced to a low-dimensional active subspace. While the resulting reduction does not typically reach the extreme compression levels observed in the present QIML setting, it nonetheless leads to substantial practical gains for uncertainty quantification. These authors also reported closely similar behaviour when using deep active subspace methods~\cite{succi2025physical}.
These classical results provide an important conceptual parallel: despite the apparent high dimensionality of physical models, the uncertainty and statistical variability most relevant for prediction may reside on a significantly lower-dimensional manifold. From this perspective, the Q-Prior in QIML may be viewed as a complementary mechanism for exploiting such effective compressibility, in which a parameterized quantum circuit implicitly encodes a compact statistical representation without requiring explicit kernel construction or sensitivity ranking. This suggests that QIML may also contribute to the broader effort of dimensionality reduction in uncertainty-aware modelling, particularly in regimes where classical methods achieve only partial compression.
While establishing a formal complexity-theoretic separation remains an open direction, the theoretical perspective outlined here offers a coherent explanation for the stability, compression, and robustness properties associated with the Q-Prior in our experiments.}

\clearpage

\begin{table}[h]
\centering
\caption{Matrix definition used in Section~\ref{sec:matrix}.
}
\label{tab:matrix_M}
\renewcommand{\arraystretch}{1.1}
\setlength{\tabcolsep}{3pt}
\begin{tabular}{rrrrrrrrrrrrrrrrrrr}
\toprule
1 & 1 & 1 & 1 & 1 & 1 & 1 & 1 & 1 & 1 & 1 & 1 & 1 & 1 & 1 & 1 & 1 & 1 & 1 \\
8 & 8 & -11 & 8 & 8 & 8 & -11 & 8 & -11 & -30 & -11 & 8 & -11 & 8 & 8 & 8 & -11 & 8 & 8 \\
1 & 1 & -4 & 1 & 1 & 1 & -4 & 1 & -4 & 12 & -4 & 1 & -4 & 1 & 1 & 1 & -4 & 1 & 1 \\
0 & -1 & 0 & 1 & 0 & -1 & 0 & 1 & -1 & 0 & 1 & -1 & 0 & 1 & 0 & -1 & 0 & 1 & 0 \\
0 & -1 & 0 & 1 & 0 & -1 & 0 & 1 & 4 & 0 & -4 & -1 & 0 & 1 & 0 & -1 & 0 & 1 & 0 \\
-1 & 0 & 0 & 0 & 1 & -1 & -1 & -1 & 0 & 0 & 0 & 1 & 1 & 1 & -1 & 0 & 0 & 0 & 1 \\
-1 & 0 & 0 & 0 & 1 & -1 & 4 & -1 & 0 & 0 & 0 & 1 & -4 & 1 & -1 & 0 & 0 & 0 & 1 \\
-1 & -1 & -1 & -1 & -1 & 0 & 0 & 0 & 0 & 0 & 0 & 0 & 0 & 0 & 1 & 1 & 1 & 1 & 1 \\
-1 & -1 & 4 & -1 & -1 & 0 & 0 & 0 & 0 & 0 & 0 & 0 & 0 & 0 & 1 & 1 & -4 & 1 & 1 \\
-2 & 1 & -1 & 1 & -2 & 1 & -1 & 1 & 2 & 0 & 2 & 1 & -1 & 1 & -2 & 1 & -1 & 1 & -2 \\
-2 & 1 & 2 & 1 & -2 & 1 & 2 & 1 & -4 & 0 & -4 & 1 & 2 & 1 & -2 & 1 & 2 & 1 & -2 \\
0 & -1 & -1 & -1 & 0 & 1 & 1 & 1 & 0 & 0 & 0 & 1 & 1 & 1 & 0 & -1 & -1 & -1 & 0 \\
0 & -1 & 2 & -1 & 0 & 1 & -2 & 1 & 0 & 0 & 0 & 1 & -2 & 1 & 0 & -1 & 2 & -1 & 0 \\
0 & 0 & 0 & 0 & 0 & 1 & 0 & -1 & 0 & 0 & 0 & -1 & 0 & 1 & 0 & 0 & 0 & 0 & 0 \\
1 & 0 & 0 & 0 & -1 & 0 & 0 & 0 & 0 & 0 & 0 & 0 & 0 & 0 & -1 & 0 & 0 & 0 & 1 \\
0 & 1 & 0 & -1 & 0 & 0 & 0 & 0 & 0 & 0 & 0 & 0 & 0 & 0 & 0 & -1 & 0 & 1 & 0 \\
0 & 1 & 0 & -1 & 0 & -1 & 0 & 1 & 0 & 0 & 0 & -1 & 0 & 1 & 0 & 1 & 0 & -1 & 0 \\
-1 & 0 & 0 & 0 & 1 & 1 & 0 & 1 & 0 & 0 & 0 & -1 & 0 & -1 & -1 & 0 & 0 & 0 & 1 \\
1 & -1 & 0 & -1 & 1 & 0 & 0 & 0 & 0 & 0 & 0 & 0 & 0 & 0 & -1 & 1 & 0 & 1 & -1 \\
\bottomrule
\end{tabular}
\end{table}

\end{document}